%% file: MSPBinaries_geometry_vaccepted.tex
\documentclass[colorlinks]{aastex61}   
\hypersetup{linkcolor=blue,citecolor=blue,filecolor=cyan,urlcolor=magenta}

\usepackage{graphicx}
\usepackage{amssymb}
\usepackage{amsmath}
\usepackage{ulem}
\usepackage{epstopdf}
\usepackage{color}

\shorttitle{Shocks in MSP Binaries}
\shortauthors{Wadiasingh et al.}

\begin{document}

\title{Constraining Relativistic Bow Shock Properties in Rotation-Powered Millisecond Pulsar Binaries}

\correspondingauthor{Zorawar Wadiasingh}

\author{Zorawar Wadiasingh}
\email{zwadiasingh@gmail.com}
\affiliation{Centre for Space Research, North-West University, Potchefstroom, South Africa}

\author{Alice K. Harding}
\affiliation{Astrophysics Science Division, NASA Goddard Space Flight Center, Greenbelt, MD 20771, USA}

\author{Christo Venter}
\affiliation{Centre for Space Research, North-West University, Potchefstroom, South Africa}


\author{Markus B\"{o}ttcher}
\affiliation{Centre for Space Research, North-West University, Potchefstroom, South Africa}

\author{Matthew G. Baring}
\affiliation{Department of Physics and Astronomy, Rice University, Houston, TX 77251, USA}

\begin{abstract}
Multiwavelength followup of unidentified {\it{Fermi}} sources has vastly expanded the number of known galactic-field ``black widow" and ``redback" millisecond pulsar binaries. Focusing on their rotation-powered state, we interpret the radio to X-ray phenomenology in a consistent framework. We advocate the existence of two distinct modes differing in their intrabinary shock orientation, distinguished by the phase-centering of the double-peaked X-ray orbital modulation originating from mildly-relativistic Doppler boosting. By constructing a geometric model for radio eclipses, we constrain the shock geometry as functions of binary inclination and shock stand-off $R_0$. We develop synthetic X-ray synchrotron orbital light curves and explore the model parameter space allowed by radio eclipse constraints applied on archetypal systems B1957+20 and J1023+0038. For B1957+20, from radio eclipses the stand-off is $R_0 \sim 0.15$ -- $0.3$ fraction of binary separation from the companion center, depending on the orbit inclination. Constructed X-ray light curves for B1957+20 using these values are qualitatively consistent with those observed, and we find occultation of the shock by the companion as a minor influence, demanding significant Doppler factors to yield double peaks. For J1023+0038, radio eclipses imply $R_0 \lesssim 0.4$ while X-ray light curves suggest $0.1\lesssim R_0 \lesssim 0.3$ (from the pulsar). Degeneracies in the model parameter space encourage further development to include transport considerations. Generically, the spatial variation along the shock of the underlying electron power-law index should yield energy-dependence in the shape of light curves motivating future X-ray phase-resolved spectroscopic studies to probe the unknown physics of pulsar winds and relativistic shock acceleration therein.
\end{abstract} 


\input{geo_intro}

\section{CONSTRAINING THE INTRABINARY SHOCK IN BLACK WIDOW AND REDBACK SYSTEMS} 
\label{sec2}

\begin{figure}[t]
\plotone{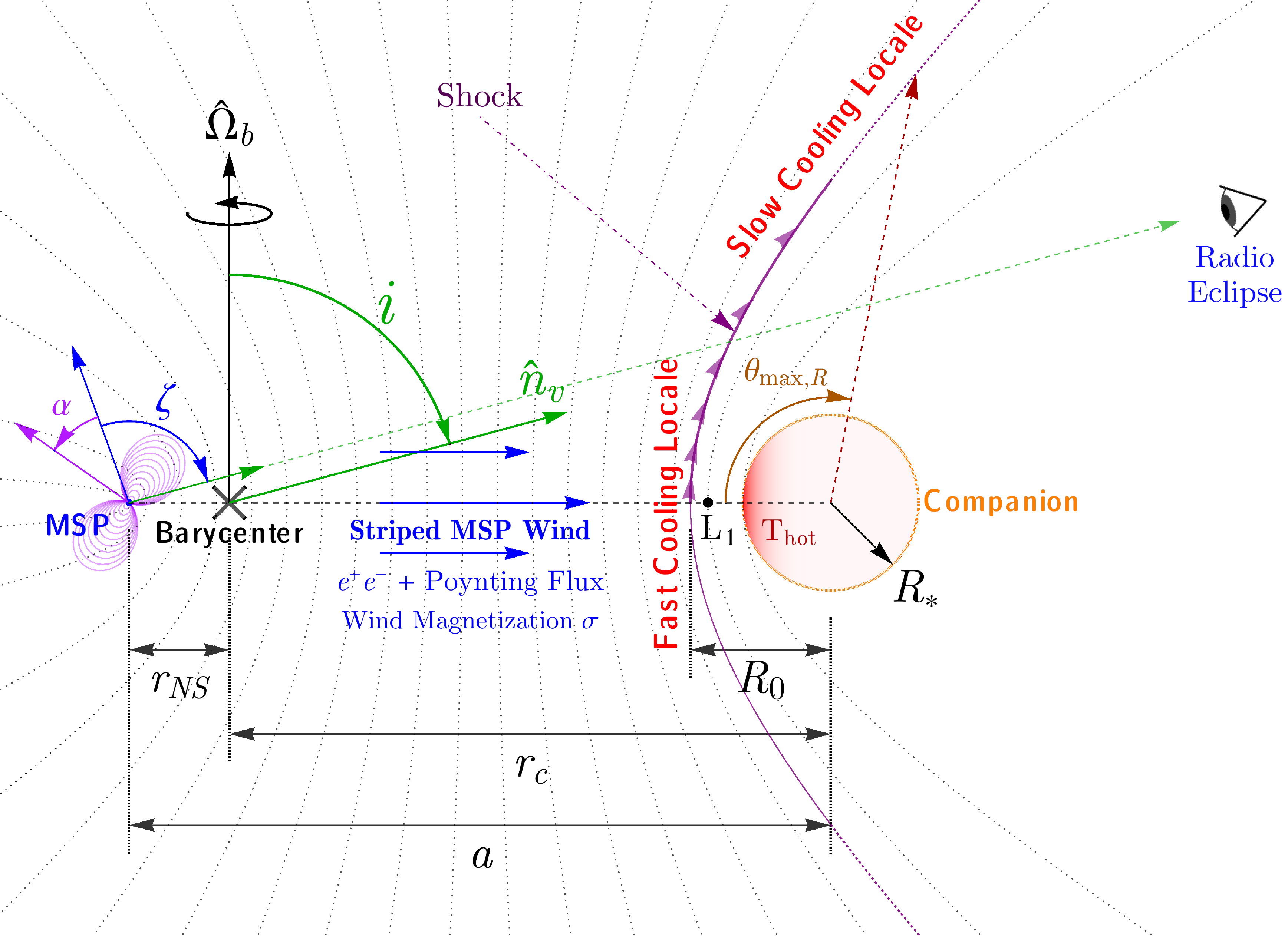}
\caption{Schematic cross-sectional diagram of a \replaced{canonical}{``spider"} MSP binary system, scale exaggerated for clarity, illustrating geometry and defining some variables and parameters used in this paper. \added{The gray dotted curves depict differing thin-shell shock surface realizations for colliding momentum-dominated isotropic winds.} In RBs and transitional objects in the rotation-powered state, the orientation of the shock is reversed such that the shock bows around the pulsar somewhat outside its light cylinder\added{,} with the definition of $R_0$ taken from the MSP \added{ in that case}. \replaced{Symbols are defined throughout the text}{Definitions of symbols are discussed in \S\ref{sec21} and elsewhere}.
\label{geometry_schematic}    }
\end{figure}  

\subsection{Overview and Idealizations}
\label{sec21}
 As is \replaced{the case generically}{generically the case} for collisionless astrophysical shocks, there are two components separated by the contact discontinuity for the intrabinary shock in MSP binaries like B1957+20 and J1023+0038: a relativistic pair shocked pulsar wind and an ionized shocked companion component \citep{1988Natur.333..832P, 1993ApJ...402..271E}. Such a gross structure must exist and is borne out in hydrodynamic \citep{2002A&A...387.1066B,2003A&A...397..913V,2012A&A...544A..59B,2015A&A...581A..27D} and relativistic magnetohydrodynamic \citep[RMHD, e.g.,][]{2005A&A...434..189B} simulations of pulsar wind shocks in various contexts. Relativistic plasma/magnetic turbulence and electron acceleration, likely mediated by magnetic reconnection, DSA, or energization mediated by shear flows \citep{2013ApJ...766L..19L} if mixing between components is low, will occur near the site of the shock contact discontinuity, leading to high-energy emission. For DSA, however, it is widely accepted that oblique relativistic magnetized shocks are less efficient accelerators than parallel shocks and may lead to spatial dependence in the acceleration in the bowed ``head" of the intrabinary shock. Due to the disparate length \deleted{scales}scales of the shock and gyroscale acceleration, such particle acceleration by current kinetic-scale simulations (e.g., particle-in-cell codes) cannot be computed in a self-consistent manner over the large length scales of the shock. Developing an expedient formulation to empirically diagnose the spatial character of such acceleration from high-energy spectroscopically phase-resolved light curves is a planned goal for future work.
  
\added{A simplified structure of the pulsar binary and intrabinary shock, central to this paper, is depicted in Figure~\ref{geometry_schematic}. The circular binary components orbit with radii $r_{\rm c}$ or $r_{\rm NS}$ around a common center-of-mass, with separation $a = r_{\rm c} +r_{\rm NS} $ for a mass ratio $q = M_{\rm MSP}/M_{\rm c} \gg 1$ with a companion that has a characteristic spherical radius $R_*$ that is $\leq L_1$ Lagrange point distance. This spherical approximation for the companion, adopted for expediency, is employed for shadowing and eclipsing calculations in \S\ref{eclipsesB1957} and \S\ref{SR1957}. The orbital momentum vector $\boldsymbol{\hat{\Omega}}_b$ is inclined at angle $i$ with respect to the observer line-of-sight $\boldsymbol{\hat{n}}_v$; if the pulsar spin axis is aligned with $\boldsymbol{\hat{\Omega}}_b$ as one may expect from the recycling evolution, then $\zeta = i$. The MSP's pulsed radio emission, which originates within the light cylinder, is assumed to be point-like. This is a good approximation since the pulsar magnetosphere is small compared to the orbital separation $a$, i.e. $R_{\rm LC}/a \sim 10^{-4}$ for $a \sim 10^{11}$ cm and a typical MSP spin period of $2$ ms. This small distance scale relative to $a$ is also approximately the MSP striped relativistic MHD wind wavelength length scale (all distances hereafter are specified in units where $a=1$ unless otherwise noted). Contours representing the intrabinary shock surface in the thin-shell approximation for two colliding isotropic momentum-dominated winds \citep{1996ApJ...469..729C} are shown, with the purple curve highlighting a particular case at the stagnation point $R_0$. The polar angle $\theta_{\rm max,R}$ defines the region that is optically thick at a particular radio frequency for radio eclipses of the MSP in \S\ref{sec22} and the Appendices. The purple arrows along the shock schematically depict the increasing bulk flow along the shock away from the stagnation point, discussed in \S \ref{mixingsec} and is a necessary ingredient in \S\ref{radiation}. The relevance of red labels referring to fast/slow cooling will be apparent in due course in \S\ref{caveats}.} 

 To furnish insight to the reader and underpin the framework developed in this paper that will be utilized in the future, we begin by briefly discussing select multiwavelength phenomenology that constrain the shock, physics and geometry in BWs and RBs, with a focus on B1957+20 and J1023+0038. Such an assessment of extant empirical conclusions is critical for the development of a self-consistent, unified model of BWs and RBs as we attempt in this paper. \added{In addition to discourse that follows, we note that the orbital inclination $i$ may be constrained using several methods: radio MSP and optical companion mass functions, orbitally-modulated $\gamma$-ray emission and eclipses, radio eclipses, and orbitally-modulated intrabinary X-ray shock emission. The latter two constitute the purview of this paper where $i$ is a critical model parameter. In addition, if $\zeta \approx i$, pulsed radio and $\gamma$-ray light curve models \citep[e.g.,][]{2014MNRAS.439.2033G,2014ApJS..213....6J} and pulsed thermal X-rays from polar-cap hot spots of the MSP inform on the orbital geometry.}
 
 Phase zero of the orbit in most observational contexts for MSP binaries is defined where the pulsar passes the ascending node, with orbital phases $0.25$ and $0.75$ the superior conjunction (SC) and inferior conjunction (IC) of the MSP, respectively --- in this article, phase zero is defined as SC where the MSP is behind the companion for the observer as this is a natural choice for radio eclipses. For some MSP binaries where the pulsar is totally shrouded in the radio and no radio ephemeris is available, the IC phase is associated with the global optical maximum of the irradiated stellar companion such as for J2339.6--0532 \citep{2015ApJ...812L..24R,2015ApJ...814...88S}.
    
\input{doublepeak_list}

\subsubsection{X-ray Phenomenology and Interpretation}
Many rotation-powered BWs and RBs show evidence for orbitally-modulated X-ray emission, likely due to synchrotron cooling of relativistic electrons and positrons at an intrabinary shock in a turbulent and relatively high $\lesssim 50$~G magnetic field anticipated just upstream of the shocked pulsar wind (see Eq.~[\ref{upstreamB}]).  The BW or RB systems, without detectable accretion or disks in the rotation-powered state, have an inherently different origin of X-ray emission than for dipping/eclipsing LMXBs \citep[e.g.,][]{1986ApJ...308..199P} where accretion power may dominate. The emission typically has a strong non-thermal power-law component, with relatively flat photon indices $\Gamma_X \sim 1-1.5$ \citep[e.g.,][]{2015arXiv150207208R} that implies relatively hard underlying electron distributions $p \approx 2 \Gamma_X -1 \lesssim 2$ and efficient acceleration. \added{Typically no additional thermal component is necessary in the power-law fits, and if one is observed it is weak and attributed to the unrelated MSP polar caps.}

 The \replaced{large-scale}{significant} orbital modulation in spiders is generally not strongly energy-dependent \citep[e.g.,][]{2011ApJ...742...97B, 2014ApJ...789...40B, 2014ApJ...791...77T} in phase-resolved spectroscopic studies in the soft X-ray band, thus photoelectric absorption is disfavored as the principal modulating mechanism in most sources \citep[J2339.6--0532 may be an exception, cf.~][]{2015ApJ...802...84Y}. Deep soft X-ray observations of a large subset ($\sim 25\%$) of MSP binaries in the rotation-powered state exhibit strong DP orbitally-modulated X-ray fluxes, most often centered at IC, with BW B1957+20 unique by being centered at SC (cf.~Table~\ref{BWRBtable}). \added{This DP modulation is somewhat stable orbit-to-orbit in many systems, within $\sim 30\%$, although some sources, e.g. XSS J12270-4859 do exhibit more significant variations, across disparate observation epochs \citep[][]{2015MNRAS.454.2190D}.} \replaced{However, s}{S}ome other BWs with shallower observations also show hints in photon counts for SC-centered DP emission \citep[e.g. J1810+1744,][]{2014ApJ...783...69G}, suggesting that the preponderance of IC-centered DP emission may be an observational selection bias for brighter RBs and transitional systems. Observations by NuSTAR \citep{2013ApJ...770..103H} of J1023+0038 in its rotation-powered state also reveal \added{that} the nonthermal orbital modulation extends to at least $79$ keV \citep{2014ApJ...797..111L}, with hints for IC-centered DP emission even in the $3$--$79$ keV harder X-ray band \citep{2014ApJ...791...77T}. Such hard X-ray phenomenology is also present in other spiders (M. Roberts, private communication).  \added{For those sources that exhibit DP morphology, the global minimum to maximum count rate ratio in the light curve typically ranges from about a factor $< 2$ for J1723--2837 \citep{2014ApJ...781L..21H} up to $\sim 7$ for J2129--0429 \citep{2015arXiv150207208R}. The interpretation of the off-peak background count rate is unclear; this persistent flux component is subject to contamination and confusion. The local minimum dip in the light curve, which defines the DP morphology, typically ranges $65-85\%$ of the global maximum and may be at some small phase-offset from IC or SC. Although there is some statistical uncertainty, peaks are generally not identical with the leading peak often more prominent than the trailing peak. Peak-to-peak separation is confined to a rather narrow range of $0.2-0.35$ in normalized phase for the sources in Table~\ref{BWRBtable} while peak full-widths are generally around $0.1$. Interestingly, J2129--0429 which exhibits one of the most well-defined DP morphologies is also close to edge-on with $i \approx 80^\circ$ \citep{2016ApJ...816...74B}.}

 Simple occultation by the companion \added{of the emission region} as invoked for J1023+0038 by \cite{2011ApJ...742...97B} cannot naturally explain the DP light curve structure centered around IC as observed by \cite{2010ApJ...722...88A} and \cite{2014ApJ...791...77T}, \added{since this is $0.5$ out-of-phase of where the local minimum dip should be by occlusion.} Moreover, many of these systems have binary inclinations well away from edge-on \citep[$34^\circ$ to $53^\circ$ for J1023+0038,][]{2009Sci...324.1411A}, requiring the emission region (and intrabinary shock) relatively close to the companion in the occultation model\replaced{, which cannot naturally explain the large $>50\%$ orbital fraction of radio shrouding of the MSP in a rotation-powered state.}{. Although the X-ray emitting and radio eclipse regions need not be coincident, the large $>50\%$ orbital fraction of radio shrouding of the MSP suggests the plasma is not well-confined near the companion. However it is difficult to envision a plausible and relatively stable hydrodynamic scenario where a shock exists near the companion $L_1$ point but other plasma is shrouding the pulsar $\gtrsim 50\%$ of the orbit but \emph{generally not} at pulsar IC for such low $\lesssim 55^\circ$ inclinations. Moreover, for an X-ray emission region close to the companion, occlusion also innately leads to a DP structure that has a peak separation of $\sim 0.5$ that is too wide for any observed BW or RB.} Unlike eccentric TeV binaries, the DP light curves in circularized BWs and RBs also cannot be explained by dynamical changes of shock radius and particle cooling between periastron and apastron \citep[e.g.,][]{1997ApJ...477..439T}.
  
 We argue in this paper that geometric Doppler boosting of emission along an intrabinary shock, either bowed toward or away from the companion, can naturally explain the DP light curve structure centered at SC or IC, respectively. Then, \textit{the phase centering of the DP structure is a key discriminant of the shock orientation and system state}. In addition, the light curve structure serves as a probe of shock geometry, particle acceleration, and shock mixing. The bulk Lorentz factor that predicates the Doppler boosting is critically dependent on the level of mixing between the relativistic e$^+$e$^-$ wind and the shock-heated ionized companion matter, that is, the baryon loading of the flow. For a striped wind of magnetization $\sigma$ where the shock approximately lies around the line joining the two stars, the striped pulsar wind field orientation relative to the shock normal is critical for particle acceleration \citep[e.g.,][]{2011ApJ...741...39S,2012ApJ...745...63S}. For a striped wind that is envisioned as parallel slabs of alternating field orientation, the shock geometry is quasi-perpendicular at the nose with the highest compression ratio, transforming smoothly to quasi-parallel at the flanges with a lower compression ratio. This spatial dependence of the compression ratio, relativistic shock obliquity, along with higher particle resident time near the stagnation point (the fast cooling locale in Figure~\ref{geometry_schematic}), should inherently influence the local particle acceleration, cooling and emergent radiation depending on what shock locales the observer line-of-sight samples as a function of orbital phase. However, a detailed exploration in a self-consistent geometry with a transport model for leptons along the shock is deferred to a future paper. For our present study, we focus on the gross DP structure of the light curves in different geometries that can easily be adapted for different sources and energies.
 
 It can be shown that the equatorial upstream wind magnetic field magnitude $B_{\rm w}$, dominated by the toroidal component at large \replaced{distances}{cylindrical radii} $r_{\rm s} \gg R_{\rm LC}$ from the pulsar, is
\begin{equation}
B_{\rm w} \approx \left( \frac{3 \dot{E}_{\rm SD}}{2 c} \right)^{1/2} \frac{1}{r_{\rm s}}  = 22 \left( \frac{ \dot{E}_{\rm SD}}{10^{35} \, \, {\rm erg \, s^{-1}} } \right)^{1/2} \left( \frac{10^{11} \, \, \rm cm}{r_{\rm s}} \right) \quad \rm G.
\label{upstreamB}
\end{equation}
\added{This relatively large magnetic field advocates synchrotron cooling as a significant energy loss mechanism for electrons. A rudimentary estimate for the pulsar contribution to the electron/positron number density near the shock may be found by assuming isotropic particle outflow from the MSP at a multiplicity ${\cal M}_\pm$ of the Goldreich-Julian rate $\dot{N}_{\rm GJ}$ \citep{1969ApJ...157..869G} from the pulsar polar caps,
\begin{equation}
\dot{N}_{\rm GJ} \approx \frac{2 c A_{\rm cap} |\rho_{\rm GJ}|}{e} \approx \frac{\sqrt{6 c}}{e} \dot{E}_{\rm SD}^{1/2} \qquad \mbox{s}^{-1},
\label{GJchargerate}
\end{equation}
where $|\rho_{\rm GJ}| = |\boldsymbol{\Omega \cdot B}|/(2 \pi c) \sim B/(c P_{\rm MSP})$ the Goldreich-Julian charge density and $A_{\rm cap} \approx 2\pi R_{\rm MSP}^2 (1- \sqrt{1- R_{\rm MSP}/R_{\rm LC}})$ is the approximate pulsar polar cap area for an aligned rotator. Then for a secondary pair multiplicity $\cal{M}_\pm$, the pulsar contribution to the number density at distance $10^{11}$ cm is 
\begin{equation}
n_{\rm e,MSP} = \frac{ {\cal{M}}_\pm \dot{N}_{\rm GJ}}{ (4 \pi c r_s^2)}\approx 4 \times 10^{-2} \,  {\cal{M}}_\pm \left( \frac{\dot{E}_{\rm SD}}{10^{35} \, \, \rm erg \, s^{-1} }  \right)^{1/2} \left(\frac{10^{11} \, \, \rm cm}{r_s}\right)^2  \qquad \rm cm^{-3}.
\label{neMSP}
\end{equation}
 For MSPs, the secondary multiplicity from pair cascade codes is typically $ {\cal{M}}_\pm \sim 10^2$ -- $10^4$ of the primary polar cap outflow rate \citep{2011ApJ...743..181H, 2015ApJ...810..144T, 2015ApJ...807..130V} while constraints from young PWNe studies \citep{2003ApJ...593.1013S} or the Double Pulsar \citep{2012ApJ...747...89B} suggest $ {\cal{M}}_\pm \sim 10^3-10^5$. Thus for BWs, the typical pulsar contribution probably does not exceed $\sim 10^3$ cm$^{-3}$ unless the pair wind is highly anisotropic in the plane of the orbit. For \replaced{rotationally-powered RBs}{IC-centered spiders} and transitional systems where the shock may be much closer to the MSP, the pulsar pair density can be profoundly larger by a factor up to $(a/R_{LC})^2 \lesssim 10^8$ and may be a significant influence for the radio eclipses and radiation physics. }

For a well-defined MHD shock to develop, the magnetization must attain $\sigma \ll 1$ upstream of the shock, either by shock-mediated reconnection \citep{2011ApJ...741...39S} very near the shock precursor, or other kinetic-scale dissipation processes far upstream.  Neglecting any baryonic mass loading, the condition \replaced{$\sigma \lesssim 1$}{$\sigma = B^2/(4 \pi \, n_{\rm e,MSP} \langle \gamma_{\rm w} \rangle m_e c^2) \lesssim 1$ with Eqs.~(\ref{upstreamB})--(\ref{neMSP})} implies a mean Lorentz factor  $\langle \gamma_{\rm w} \rangle$ for an isotropic pair wind,
\begin{equation}
\langle \gamma_{\rm w} \rangle \gtrsim \left(\frac{3 \dot{E}_{\rm SD}}{2 c} \right)^{1/2} \frac{e}{{\cal M}_\pm 2 m_e c^2} \approx \frac{7 \times 10^8}{{\cal M}_\pm} \left( \frac{ \dot{E}_{\rm SD}}{10^{35} \, \, {\rm erg \, s^{-1}} } \right)^{1/2},
\end{equation}
where ${\cal M}_\pm \gg 1$ is the pulsar pair multiplicity of the Goldreich-Julian rate, cf. Eq.~(\ref{GJchargerate}). Following attaining $\sigma \ll 1$, the magnetic field in the shocked pulsar wind field $B_{\rm s}$ then scales as $B_{\rm s} \sim 3 \sqrt{\sigma} B_{\rm w}$ in the ultrarelativistic perpendicular shock limit \citep{1984ApJ...283..694K}. However, the magnetic dissipation processes upstream may convert or destroy the striped wind morphology such that the shock may be quasi-parallel in the proper frame. A containment argument, based on the observed X-ray power law provides a rudimentary lower bound on $B_{\rm s}$ -- the Larmor radius $r_{\rm L}$ of electrons in the shock must be smaller than about $1\%$ of the orbital length scale $r_{\rm L} \lesssim 0.01 a \sim 10^9$ cm. Then, assuming emission at the critical synchrotron \replaced{frequency}{dimensionless energy $\epsilon_c = 3 B_s/(2 B_{\rm cr}) \gamma_{\rm e}^2$ with $B_{\rm cr} \approx 4.414 \times 10^{13}$ G and electron Lorentz factor $\gamma_{\rm e}$}, for an observed power law extending to energy $\epsilon_{X, \rm max}$ in units of $m_e c^2$,
\begin{equation}
B_{\rm s} \gtrsim B_{\rm s, min} \approx 4.4 \, \epsilon_{X, \rm max}^{1/3} \left(\frac{10^9 \, \,\rm cm}{r_{\rm L}}\right)^{2/3}  \quad {\rm G},
\label{lowB}
\end{equation}
where we have neglected factors of roughly unity associated with Doppler shift of energies corresponding to mildly relativistic bulk speeds along the shock. Therefore power laws extending up to $ \epsilon_{X, \rm max} \approx 0.15 \approx 80$ keV$/(m_e c^2)$ observed by NuSTAR for J1023+0038 advance $2 \lesssim B_{\rm s} \lesssim B_{\rm w} \sim 200$ G in the relativistic magnetized shock if $r_s \sim 10^{10}$ cm, which implies radiating electron Lorentz factors of order $10^{5}$--$10^6$\added{, i.e. well-above a thermal population. A more loose assumption of $r_{\rm L} \sim a$ still results in $B_{\rm s} \gtrsim 10^{-1}$ G, still considerably higher than those in PWNe. Therefore this synchrotron component extends into the UV/optical/IR and lower energies, but such a power-law extrapolation yields expected fluxes well-below the sensitivity of any facility.} For other spiders where observations at energies above the classical soft X-ray band are not available, the field magnitude is still greater than about one Gauss, orders of magnitude larger than those in plerions. We consider implications of these bounds on the shock in \S \ref{mixingsec}.

\subsubsection{Radio Phenomenology} 

Orbital eclipses of the MSP's radio pulsations are a common feature in many BWs and RBs in the rotation-powered state.   Observed orbital eclipse fractions $f_E$ are ordinarily $ f_E \sim 5-15\%$ for BWs, and typically much larger for RBs, increasing in low radio frequency bands. For example, PSR J1023+0038 eclipses for less than $5\%$ at $3$ GHz to over $\sim 60\%$ of an orbit at 150 MHz \citep{2009Sci...324.1411A,2013arXiv1311.5161A}. Some BWs also have extensive eclipses. There appears to be a dichotomy in the relative stability of eclipses -- for some BWs like B1957+20 eclipses near SC are generally stable orbit-to-orbit, while sporadic mini-eclipses are seen in some other systems particularly those systems with larger eclipse fractions \citep[e.g.,][]{2009Sci...324.1411A,2016ApJ...823..105D}. However even in these erratic systems with mini-eclipses, the pulsar is generally unshrouded at IC in relevant bands. A standard decomposition of $f_E$ into symmetric and antisymmetric parts about SC is attainable as a function of observer frequency $\nu$. Frequency dependence of the eclipse fraction asymmetry is standard, with larger asymmetry in ingress-egress delays at lower observing frequencies, e.g. PSR B1957+20  \citep{1991ApJ...380..557R, 2001MNRAS.321..576S} and J1023+0038 \citep{2009Sci...324.1411A,2013arXiv1311.5161A}. At the highest radio frequencies $\nu$, the antisymmetric part of $f_E$ is typically small compared to the symmetric part. 

 For B1957+20 and other systems, the symmetric part of these eclipses encompass inferred length scales that are significantly larger than $R_*$ for a fully Roche lobe-filled companion, even for $\sin i \approx 1$. No eclipses by the companion are expected if $i < 90^\circ - \arcsin(R_{*}/a)$, but many systems with eclipses have well-constrained inclinations and companion sizes which violate this inequality. Therefore, eclipses must be predicated on plasma within the system and/or a secondary magnetosphere. Eclipses typically exhibit large plasma dispersion measures before the coherence in the timing solution of pulsations is lost, likely due to absorption rather than scattering \citep{2015ApJ...800L..12R}; continuum eclipses of the pulsar are also seen in some systems at low frequencies e.g., for BW B1957+20 \citep{1992ApJ...384L..47F} and RB J2215+5135 \citep{2016MNRAS.459.2681B} with a scaling $f_E \propto \nu^{-0.4}$. 
 
There are a panoply of potential eclipse mechanisms \citep[cf.][]{1989Natur.337..236M, 1991ApJ...370L..27E,1993ApJ...406..629G, 1994ApJ...422..304T} depending on physical parameters realized in the intervening plasma. Cyclotron absorption has been posited in B1957+20 \citep{2000ApJ...541..335K} but relatively little Faraday rotation is seen, consistent with a $1-10$ G mean magnetic field magnitude in the eclipsing medium \citep{1990ApJ...351..642F}, not inconsistent with Eq.~(\ref{lowB}) since the eclipsing medium consists of the ionized companion wind as well. Moreover, it is now known that the companion in B1957+20 is likely non-degenerate \citep{2007MNRAS.379.1117R}. Excess delays, consistent with plasma dispersion, generally show that the average free electron column density rises sharply from $\langle n_e \rangle d \sim 10^{15}$ cm$^{-2}$ to $10^{18}$ cm$^{-2}$ at phases deep into the eclipse \citep{1991ApJ...380..557R, 2001MNRAS.321..576S} for BWs, for $d\sim a$ the line-of-sight column depth, but it is anticipated that there is also clumping near the shock contact discontinuity. This $\langle n_e \rangle$ is much higher than implied by Eq.~(\ref{neMSP}), therefore the companion wind must have some influence. Whatever the mechanisms for eclipses, the momentum flux balance between the pulsar wind and a companion wind or magnetosphere defines a geometric volume of plasma through which the MSP is eclipsed, bounded by the shock surface (gray curves depicted in Figure~\ref{geometry_schematic}). 
 
 Consequently, we advance that the dichotomy of eclipse phenomenology is the orientation of the shock surface germane to the X-ray light-curve phasing in Table~\ref{BWRBtable}. For the SC-centered DP phase centering where the shock is bowed around the companion, as for BW 1957+20, the relative stability and small $f_E$ are consistent with this picture. Contrastingly, for IC-centered X-ray phasing where the shock is orientated around the pulsar, larger and more erratic eclipses are expected where the companion wind can enshroud the pulsar, and is necessarily turbulent for the obligatory angular momentum loss. The radio optical depth, as well as the shock orientation depend on the companion wind mass loss rate. This can be very low or substantial through evaporation or quasi-Roche lobe overflow \citep[e.g.,][]{2016ApJ...816...74B}, respectively, but is poorly understood. 
 
 For the IC-centered scenario, canonical Roche lobe overflow at the characteristic ion sound speed cannot be a wind source since the circularization radius $R_{\rm circ}$ must be larger than the shock radius $R_0$ (measured from the MSP), or the system will be predisposed to a disk-state \citep{2002apa..book.....F}. Moreover, for the radio pulsar state, $R_0$ must exceed the light cylinder scale, that is, $R_0 > \mbox{Max}(R_{\rm circ}, R_{\rm LC})$. This then favors an evaporatively-driven quasi-Roche lobe overflow supersonic wind model for rotation-powered states. The mass loss must be low enough to escape IR/optical detection. The scenario is somewhat fine-tuned such that the companion wind is fast enough to inhibit a disk, while dense enough such that angular momentum losses owing to turbulence are sufficient for gravitational influences to overpower the pulsar wind. Such turbulence may also be driven by the radio absorption that predicates the eclipsing mechanism. Accordingly, $R_0/{\cal{R}} \sim (\dot{m}_g c^2/\dot{E}_{\rm SD})^2 \gg 1 $ where ${\cal{R}} = 2 G M_{\rm MSP}/c^2$ is the Schwarzschild radius of the MSP and $\dot{m}_g$ is the gravitationally-captured wind's mass rate. However there are stability concerns -- the pulsar termination shock that arrests accretion flow and shrouds the pulsar and delineates the eclipsing medium may only be pushed out to a modest $2-100$ multiples of pulsar light cylinder radius $R_{\rm LC} \ll a$ \citep{2005ApJ...620..390E,2014ApJ...795...72L} unless a feedback mechanism is operating. We show in an upcoming paper that if there is a feedback mechanism operating in RBs for the mass-loss rate from the companion, then such autoregulation may permit the shock to be stable much farther from the light cylinder out to orbital length scales for rotation-powered disk-free states (Wadiasingh 2017, in prep). A detailed discussion of the poorly-understood nuances of the irradiated companion mass loss, stability, and eclipse mechanisms is beyond the scope of this paper.

\subsection{Geometric Constraints by Radio Eclipses}
\label{sec22}

Here we explore what constraints on the shock parameters can be gleaned from just the geometry of eclipses as a function of inclination. This requires an \emph{a priori} model for the shock geometry. For cases where the shock is orientated around the companion, two formalisms have been invoked for the free electron density underpinning the eclipses: an optically thin low-density plasma tail from the companion that spans several orbital semi-major axis length scales $a$, and an optically-thick model of much higher local free electron density in the intrabinary shock and shocked companion wind \citep{1989ApJ...342..934R,1991A&A...241L..25R, 1991ApJ...381L..21T}. We adopt the optically-thick formalism, as this leads to constraints that are upper limits on $R_0$ for a given $i$ which we apply to B1957+20 as a test case in \S~\ref{eclipsesB1957}. That is, for a model shock surface geometry, the transition from optically thick to thin may be parameterized in terms $\theta_{\rm max,R}$ for a given radio band. This is conceptually similar to radius-to-frequency mapping used in the study of pulsed emission in radio pulsars \citep{1970Natur.225..612K,1978ApJ...222.1006C}. Small values of $\theta_{\rm max,R}$, corresponding to high radio frequencies, sample regions of the shock closer to the shock head. Asymmetry of eclipses is interpreted as Coriolis influences on the shock, skewing it by an angle or sweeping-back a cometary tail; the latter is explored in the Appendices. 

At this stage, we do not attempt to self-consistently model the shock geometry and parameters from, for instance, generic covariant MHD jump conditions \citep{2004ApJ...600..485D}. \added{Instead we wish to constrain geometric shock parameters using radio eclipses as a function of binary inclination $i$. This not only is of some utility to synthetic X-ray light curves that follow in \S\ref{radiation}, but also informs on the ratio of wind ram pressures. Moreover, although the geometric model we present is somewhat degenerate on the parameter $\theta_{\rm max, R}$, this motivates more systematic radio eclipse population studies of BWs and RBs.} 
 
  The general problem of an arbitrarily-shaped region occulting a source involves computational geometry techniques that may require inefficient ray casting or tessellating grids. To make the problem more analytically expedient\deleted{ as well as transparent}, we assume an azimuthally symmetric form for the intrabinary shock with radial function $\{ R(\theta) \,| \, \theta \in(0,\theta_\infty)\}$ along the axis of symmetry of the bow-shaped shock, with $R(\theta = 0) = R_0$, and generalize this approach to approximate the swept-back tails due to orbital motion. Azimuthal symmetry of the shock is expected to be an acceptable approximation for the intrabinary shock locale in the vicinity of the shock stagnation point, although overall the shock angle may be skewed relative to the line joining the two stars yielding the extended-egress delayed-ingress radio eclipse phenomenology. This approximation to the shock structure should especially be good when restricted to the ingress portion of eclipses about SC that are sharp and well-defined, unlike those that are more diffuse at egress and that are likely contaminated by plasma from the cometary-tail.  \deleted{The MSP's pulsed radio emission which originates within the light cylinder is assumed to be point-like, a good approximation since the pulsar magnetosphere is small compared to the orbital separation $a$, [Equation removed]. This small distance scale relative to $a$, is also approximately the MSP striped relativistic MHD wind wavelength length scale (all distances hereafter are specified in units where $a=1$ unless otherwise noted).} In Appendix \ref{appendix_A}, we develop an analytical formalism for eclipses of a point source by an arbitrary azimuthally-symmetric surface orbiting the source around a common barycenter, as well as parameterize the asymmetry of eclipses due to the swept-back tail for finite flow velocities and wind accelerations.

 In the highly-radiative supersonic limit, azimuthally-symmetric analytic purely-hydrodynamic bow shock forms that assume nonrelativistic, momentum-dominated winds neglecting gravity, through the balance of ram pressures, can be found in \cite{1996ApJ...459L..31W} and \cite{1996ApJ...469..729C}. Here the companion shock, contact discontinuity, and shocked/deflected pulsar wind are roughly spatially coincident compared with the length scale $a$, although not necessarily well-mixed, and internal pressure contributions to the momentum-flux tensor are neglected. \added{Internal pressure contributions generally increase the geometric thickness of the shock, particularly near the stagnation point where they are non-negligible, a complication that is largely peripheral to this Section.} If the flow is not \added{highly} supersonic, the analytic forms are significantly narrower than simulations of hydrodynamic shocks with finite Mach number \citep[e.g.,][]{2003A&A...397..913V}. Relaxing the highly-radiative limit or introducing mass loading, parameterizations for azimuthally-symmetric shocks can also be found in \cite{2009ApJ...703...89G} or \cite{2015MNRAS.454.3886M}, and generically result in increasing the shock opening angle. 
 
Since the physics and geometry of the pulsar wind and induced companion wind or magnetosphere are largely unknown, we utilize a generic two-isotropic-colliding-winds analytic solution \cite{1996ApJ...469..729C} for the intrabinary shock geometry that surveys varied shock asymptotic angles through a simple parameterization. This is readily apparent from the gray contours in Figure~\ref{geometry_schematic} and qualitatively resembles hydrodynamic simulations of \cite{1991ApJ...381L..21T} for B1957+20. As explored in \S\ref{J1023_radio}, the geometric form is immaterial for radio eclipses for the case where the shock is orientated around the pulsar. However, for eclipses where the shock is around the companion, the prescribed geometry is consequential for the parameter constraints. For this reason, we consider an alternative parallel-wind geometry of \cite{1996ApJ...459L..31W}, a standard bow shock, in Appendix~\ref{parawindappendix} which has significantly narrower asymptotic shock angles and is the $R_0 \ll 1$ limit of the two-wind solution near the shock head. We limit ourselves to axisymmetric forms for simplicity -- prescriptions for the shock geometry that include nonaxisymmetric distortions by Coriolis effects are found in \cite{2008MNRAS.388.1047P}.

The radial function for the colliding isotropic winds parametrizing the shock geometry is
 \begin{eqnarray}
 R_{\rm iso} (\theta) &=& \sin \theta_1 \csc (\theta + \theta_1) \qquad \rm{with} \qquad  \theta_1 \cot \theta_1 = 1 + \eta_{\rm w}(\theta \cot \theta - 1) ,
 \label{shockform_canto}
 \end{eqnarray}
and $ \theta \in(0,\theta_\infty)$ the polar angle defining the shock, with zero taken as the line separating the two stars and $\theta_\infty$ the asymptotic shock angle,
\begin{equation}
 \theta_\infty - \tan \theta_\infty = \frac{\pi}{1-\eta_{\rm w}},
 \label{tinfty}
\end{equation}
where $\eta_{\rm w}$ is the ratio of the two wind ram pressures and $\theta_1$ is an implicitly defined function of $\theta$ and $\eta_{\rm w}$. Explicitly, $R_0$ is related to $\eta_{\rm w}$ by
\begin{equation}
 R_0 =\frac{\sqrt{\eta_{\rm w}}}{1 + \sqrt{\eta_{\rm w}}}.
 \label{R0eta_canto}
\end{equation}
We caution that the physical interpretation of $\eta_{\rm w}$ may be misleading, especially for scenarios where the shock wraps around the pulsar where gravitational influences and unknown wind anisotropies are salient. Otherwise, when the shock is orientated around the companion, as for B1957+20 in \S\ref{eclipsesB1957}, $\eta_{\rm w}$ may be connected to the pulsar's parameters if the wind of the companion is induced, e.g., Eqs.~(10)--(11) in \cite{1990ApJ...358..561H} and Eq.~(\ref{wind_energetics_1957}) in the next Section.

In \S\ref{radiation}, we only employ the geometry of these analytical shocks rather than their physical velocity and density profiles, since the companion and pulsar shock components may not be well-mixed. Generically for such a pressure-confined flow, surface mass density for the bow shocks is highest near the stagnation point and slowly decreases farther down, while tangential velocity increases approximately linearly with the shock polar angle parameter $\theta$. These are readily apparent from a Taylor series expansion of analytical expressions in the thin-shell limit. At the stagnation point, the tangential velocity approaches a small value, if internal pressure contributions are small -- this is indeed the behavior observed in nonrelativistic hydrodynamic simulations \citep[e.g., Figure 3 of][]{2014ApJ...789...87R}. The tangential velocity $v_{\rm s}$ and surface density $\Sigma_{\rm e}$ variation with $\theta$ for hydrodynamic bow shocks can be shown to follow $v_{\rm s} \propto \theta$ and $\Sigma_{\rm e} = \Sigma_0 (1 + w \theta^2)$ for $\theta \lesssim 1$ with $w <0$, where $|w| \ll 1$ is a constant, but the exact normalization of these generic forms depends critically on the wind pressures and velocities that are unknown. This generic form, generalized to relativistic velocities, is utilized in \S\ref{radiation} for the semi-analytic DP light curve synthesis. However, the hydrodynamic density profile is probably an inaccurate proxy for the spatial distribution of particle acceleration and cooling; thus a more general procedure is also described in \S\ref{partaccdiag}.

 \subsubsection{Application to PSR B1957+20, an SC-centered Spider}
 \label{eclipsesB1957}

\begin{figure}[t]
\centering
\includegraphics[scale=0.178]{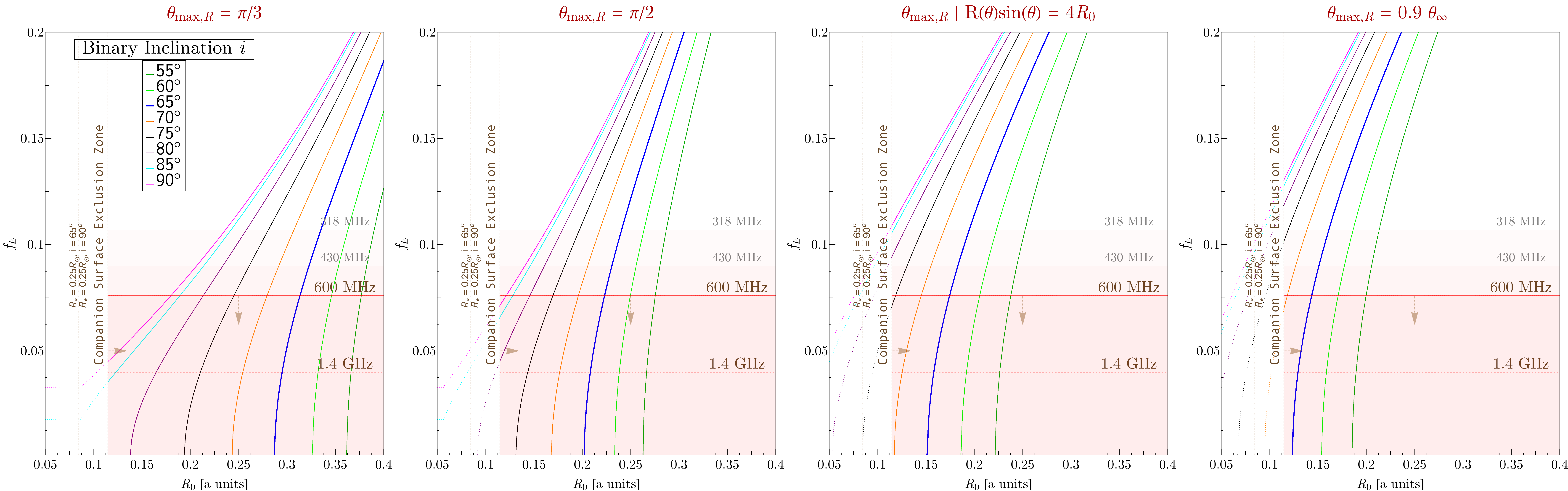}
\caption{ \replaced{Curves}{Panels of fixed $\theta_{\rm max,R}$ depicting computed curves} for the one-to-one coupling between eclipse fraction $f_E$ and the shock stagnation point $R_0$ for PSR B1957+20 in the axisymmetric optically-thick shock scenario with a range of orbital inclinations $i$, as developed analytically in Appendix~\ref{appendix_A}. \added{The panels depict successively higher values of $\theta_{\rm max,R}$ from left to right, with the  two rightmost panels prescribing values of $\theta_{\rm max,R}$ that depend on $R_0$ as stated.} \replaced{The darker gray band around the blue curve illustrates the}{The bold blue curve highlights the best-fit} optical light curve modeling observational result $ i = 65^\circ \pm 2^\circ$ in \cite{2007MNRAS.379.1117R}, while \replaced{the curves at $i = 55^\circ$ and $75^\circ$ and $85^\circ$ are}{other curves survey the} viable lower and upper systematic bounds as presented in \cite{2011ApJ...728...95V}. Note that $\zeta \approx i \approx 85^\circ$ is also found from favored outer-gap $\gamma$-ray pulsation fitting in \cite{2014ApJS..213....6J}\deleted{, with lower values for unfavored low-altitude models}.  \deleted{The left and right panels calculate the eclipse fraction for Type I and Type II shock geometric scenarios, respectively.} The mass ratio is fixed to $q = 69.2$ for all curves \replaced{while the brown region}{while the excluded region of $R_0$} indicate the companion and volumetric Roche lobe $R_{\rm vL}(q)$ radii \citep{1983ApJ...268..368E}. Radio observations from \cite{1991ApJ...380..557R} are illustrated with the horizontal \deleted{red} lines. \replaced{The $1.4$ GHz observation is largely symmetric about SC and most constraining for the upper limit of $R_0$.}{The red region below $600$ MHz, where eclipses are to some degree symmetric, isolates roughly the region of validity of the axisymmetric calculation.}
  \label{fig_R0_eclipses_fe}  } 
\end{figure}  
  
 Using the method developed in Appendix~\ref{appendix_A}, we compute the axisymmetric eclipse fraction $f_E$ for the shock geometry in Eq.~(\ref{shockform_canto}) using Eq.~(\ref{fE_equation}). This is conditional on the crucial parameter $\theta_{\rm max,R}$, where the medium transitions from optically thick-to-thin for a given observing frequency. Note that $R_0$ and $i$ are the geometric parameters that are independent of observer frequency, therefore $\theta_{\rm max,R}$ is the only parameter in the model that connects to the frequency dependence of symmetric eclipses. 
  
 In Figure~\ref{fig_R0_eclipses_fe}, we display the eclipse fraction $f_E$ dependence on $R_0$ of PSR B1957+20 with a fixed mass ratio $q = 69.2$ for various inclination angles consistent with companion light curve models \citep{2007MNRAS.379.1117R,2011ApJ...728...95V} as well as $\zeta \approx i \approx 85^\circ$ found from \cite{2014ApJS..213....6J}. The axisymmetric computations for constraining $R_0$ should be more accurate for the eclipses at $\nu \gtrsim 600$ MHz that are largely symmetric about SC; this is indicated by the red region in the panels.  We neglect the minor degenerate observational coupling between mass ratio and inclination angle due to uncertainties in the optical mass function. The four panels depict  successively larger values of the parameter $\theta_{\rm max,R}$ left to right. In the leftmost two panels the value of $\theta_{\rm max,R} = \{\pi/3, \pi/2\}$ is independent of $R_0$, while the rightmost two panels impose values that depend on $R_0$, e.g. through Eq.~(\ref{tinfty}). In particular, the prescription $R(\theta) \sin \theta = 4 R_0$ selects the value of $\theta_{\rm max,R}$ for a given $R_0$ such that the transverse shock length scale is $4 R_0$. On such a scale far from the shock head, hydrodynamic instabilities will likely develop that may break the assumptions in our rudimentary model. The rightmost panel chooses an exceptionally large value of $\theta_{\rm max,R} = 0.9 \theta_\infty$ that serves as an extreme limit for what $\theta_{\rm max,R}$ may be and represents a very substantial occluding volume. Such large values of $\theta_{\rm max,R} > \pi/2$ include regions well beyond the head of the shock, and should not be axisymmetric due to Coriolis effects and instabilities. Despite that well-defined and sharp symmetric eclipses are not expected from such large values of $\theta_{\rm max,R}$, these values are included for completeness. Similarly, values much smaller than $ \theta_{\rm max,R} < \pi/3$ require rather flat shocks for a fixed $f_E$ tending to $R_0 \rightarrow 0.5$ well past the $L_1$ point, also a rather unreasonable scenario that requires $\eta_{\rm w} \sim 1$ and is in tension with the X-ray DP light curve peak separation in \S\ref{radiation}. 
 
\begin{figure}[t]
\centering
\includegraphics[scale=0.23]{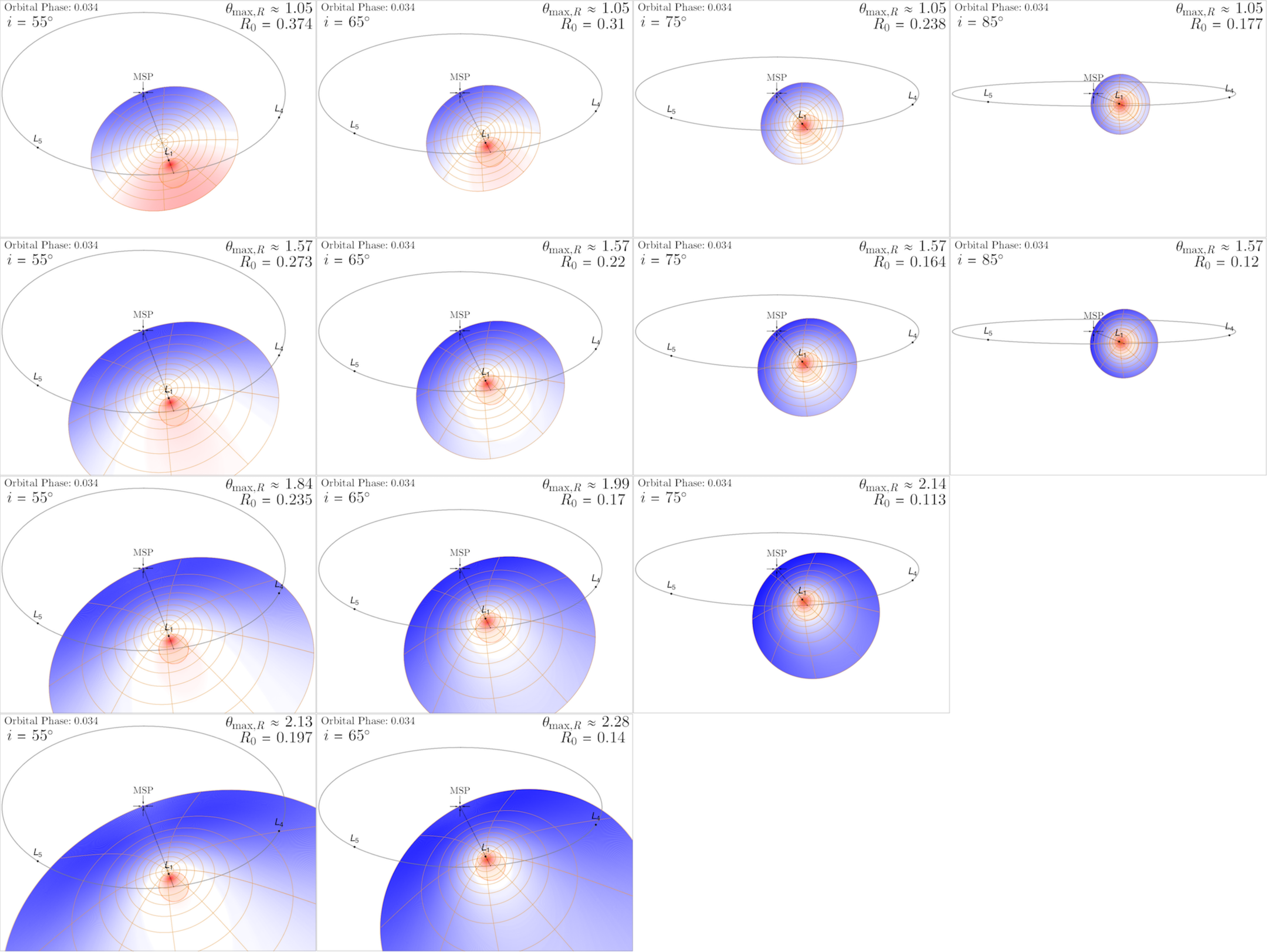}
\caption{Schematic representations of the PSR B1957+20 system, to scale, projected on the plane of the sky for inclinations \replaced{$i = 65^\circ$ (upper panels) and $85^\circ$ (lower panels)}{$55^\circ-85^\circ$ (columns)}, $q = 69.2$, and orbital phase \replaced{$\approx 0.15$}{0.034} from SC\added{ at eclipse ingress/egress}. \replaced{For all four panels, the $R_0$ is chosen such that the symmetric orbital eclipse fraction $f_E$ of the MSP is approximately $10\%$}{Rows of insets indicate several fixed $\theta_{\rm max,R}$, the same values as those in Figure~\ref{fig_R0_eclipses_fe}, with $R_0$ chosen such that $f_E \approx 7\%$}. The spherical companion is illustrated for size $R_*= 0.9 R_{\rm vL} \approx 0.1$ with the blue dot representing the pulsar. \added{Panels where $R_0 < R_*$ is obligatory are omitted.}  \deleted{(Left panels) A Type I shock scenario is depicted near the companion for $R_0 = 0.325$ (upper panel) and $0.16$ (lower panel) with a tail of length $L_{\rm z} = 3 R_0$ past the companion position outward from the pulsar position. (Right panels) A Type II scheme with $R_0 = 0.235$ and $0.16$ extending to one-half the hemisphere of the irradiated companion.} For \replaced{both}{all} panels, the color coding emphasizes, schematically, the locales of first-order Doppler boosting. A velocity profile of $v_{\rm s} \propto \theta$ tangent to the shock is imposed; the blue or red coloring accentuate those regions of the shock where the $v_{\rm s}$ along the shock is toward or away the observer line-of-sight, respectively, with intensity of coloring scaled with projected velocity component magnitude from Eq.~(\ref{DopplerV}). \added{This coloring qualitatively demonstrates the geometrical contribution to the emissivity integral in \S\ref{radiation} that culminates in DP light curves.} \deleted{The Lagrange points of the binary are also depicted.}  \label{fig_1957_cartoon}    }
 \end{figure}

For fixed inclination and $f_E$, it is clear from Figure~\ref{fig_R0_eclipses_fe} that larger values of $\theta_{\rm max,R}$ are compatible with smaller values of $R_0$.  For large inclinations near edge-on, there are clearly constraints on  $\theta_{\rm max,R}$; in particular for $i = 85^\circ$,  $\theta_{\rm max,R} \lesssim \pi/2$ for $R_0 > R_*$. On the other hand, for $i = 65^\circ$, the constraint is looser with $\theta_{\rm max,R} > \{\pi/3,\pi/2, R(\theta) \sin \theta = 4 R_0, 0.9 \theta_\infty\}$ corresponding to limits $R_0 \lesssim \{0.31, 0.235, 0.17,0.14 \}$, respectively, for $f_E \lesssim 7\%$. The upper limits on $R_0$ are modestly more stringent for $\nu > 600$ MHz for the same range of $\theta_{\rm max, R}$. Therefore we conclude that for $i = 65^\circ$, $R_0 \approx 0.17-0.3$ with a canonical value of $R_0 \approx 0.235$ for  $\theta_{\rm max,R} = \pi/2$ which defines the shock head. This latter value of $R_0 \approx 0.2$ is employed in \S\ref{SR1957} and seems plausibly compatible with the observed X-ray DP light curve; too large an $R_0$ leads to DP light curves that have too-wide of a peak separation as will become apparent in due course. Some geometric realizations for $f_E \approx 7 \%$, i.e. eclipses at $\nu \approx 600$ MHz, are illustrated in Figure~\ref{fig_1957_cartoon} with corresponding numerical values of $\theta_{\rm max,R} = \{\pi/3,\pi/2, R(\theta) \sin \theta = 4 R_0, 0.9 \theta_\infty\}$ (columns) and $R_0$. Some geometric trends are clearly evident in  Figure~\ref{fig_1957_cartoon}, for instance larger $i$ requiring smaller $R_0$ for the same $f_E$. Similarly, larger $\theta_{\rm R,max}$ for fixed $i$ allows for lower $R_0$.
 
 The upper limits found for $R_0$ allow for an estimate of the companion wind pressure if $\dot{E}_{\rm SD}$ is known.  We may express $\eta_{\rm w}$, with ${\cal{P}}$ the wind pressure due the companion, as
 \begin{eqnarray}
 \eta_{\rm w} &=&  \frac{{\cal{P}}}{\delta \Omega \dot{E}_{\rm SD}/(4 \pi c) } \sim \frac{0.5 (4 \pi) R_*^2 \sigma_B T_{\rm cold}^4/c + \xi \dot{E}_{\rm SD}/c }{\delta \Omega \, \, \dot{E}_{\rm SD}/(4 \pi c) }\ll 1,
 \label{wind_energetics_1957}
 \end{eqnarray}
where $T_{\rm cold}$ is the isotropic unheated temperature of the stellar companion (i.e., the intrinsic unirradiated radiation pressure from the secondary), $\delta \Omega/(4\pi)$ is the fractional \textit{isotropic} pulsar wind solid angle subtended by the shock. If a strong magnetosphere is not the principal source of ram pressure against the pulsar wind, then the parameter $\xi$ embodies the fractional energetic efficiency, in units of $\dot{E}_{\rm SD}$, of the induced companion wind generated by unspecified processes. The $R_0$ found above are compatible with the thermally-driven wind or companion magnetosphere scenarios of \cite{1990ApJ...358..561H}. Numerically, for typical values $R_0 = 0.2 - 0.3$ consistent with radio eclipses, the ratio of ram pressures $\eta_{\rm w}$ is between $ 6$ to $18 \%$ by inverting Eq.~(\ref{R0eta_canto}). From Eq.~(\ref{wind_energetics_1957}), we can form an estimate of the energetic efficiency $\xi$ of the induced wind, if the intrinsic or induced magnetic field of the companion is small,
\begin{equation}
\xi = \frac{1}{4 \pi} \left( \frac{ R_0}{1-R_0}\right)^2 \delta \Omega - \frac{2 \pi R_*^2 \sigma_B T_{\rm cold}^4 }{\dot{E}_{\rm SD}} \approx  \frac{1}{4 \pi} \left( \frac{ R_0}{1-R_0}\right)^2 \delta \Omega,
\end{equation}
where the ratio of cold intrinsic stellar to pulsar power can be neglected in BWs and RBs, since it of the order $\sim 10^{-4} -10^{-6}$, and thus $\xi$ is a simple function of stagnation point $R_0$. This equation is unphysical for $R_0 \rightarrow 0.5$ and should not be used in this limit. The solid angle fraction for the canonical shock head $\theta \in (0,\pi/2)$ which may participate in the heating of the companion can be routinely found $\delta \Omega/(4 \pi) \approx  3 R_0^2(1+ 2 R_0)/4$ for $R_0 \ll 1$ from Eq.~(\ref{shockform_canto}). Whence, the efficiency of the induced wind $\xi$ is of the order $0.1-5\%$, depending upon the inferred shock standoff $R_0$. This efficiency is similar in order-of-magnitude to the hemispherical quiescent induced photospheric heating fraction, i.e. $2 \pi R_*^2 \sigma_B T_{\rm hot}^4/\dot{E}_{\rm SD}\sim 0.4 \%$ for B1957+20 where $T_{\rm hot} \approx 8000$ K, as one would expect in a model where the companion's gas pressure, from a thermally-driven evaporative wind, balances the striped cold pulsar wind. 

An order-of-magnitude upper limit to the companion surface magnetic field $B_*$ may be found by assuming companion pressure is entirely due to a magnetosphere at the stagnation point \citep[e.g., Eq.~(23) of][]{1990ApJ...358..561H},
\begin{equation}
B_* \lesssim \, 4 \times 10^3 \, \left( \frac{\dot{E}_{\rm SD} }{10^{35} \,\,  {\rm erg \, s^{-1}}} \right)^{1/2} \left( \frac{10^{11} \, \, {\rm cm}}{a} \right) \left( \frac{0.05}{R_*/a} \right)^3 \left( \frac{R_0}{0.25} \right)^3 \left( \frac{0.75}{1 -  R_0} \right) \, \, {\rm G} ,
\end{equation}
which is relatively small compared to typical surface fields found in degenerate cores, and comparable to kilogauss fields found in T Tauri stars \citep{2007ApJ...664..975J}. In principle, such fields may be detectable in future, but currently requires relatively bright (magnitude $\lesssim 12-14$, C. Johns-Krull, private communication) companions for a sufficient signal-to-noise with high-precision IR/optical spectroscopy of Zeeman broadening on a model atmosphere. Any optical field constraint would be useful in constraining the physics of the pulsar and induced companion winds, as well as to appraise the \cite{1992ApJ...385..621A} model for a tidally-driven convective dynamo. Indeed, a strong companion magnetosphere would also dramatically alter shock MHD jump conditions, impact the relevant acceleration mechanisms in the shock, and influence particle heating of the companion.
 
\begin{figure}[t]
\centering
\includegraphics[scale=0.25]{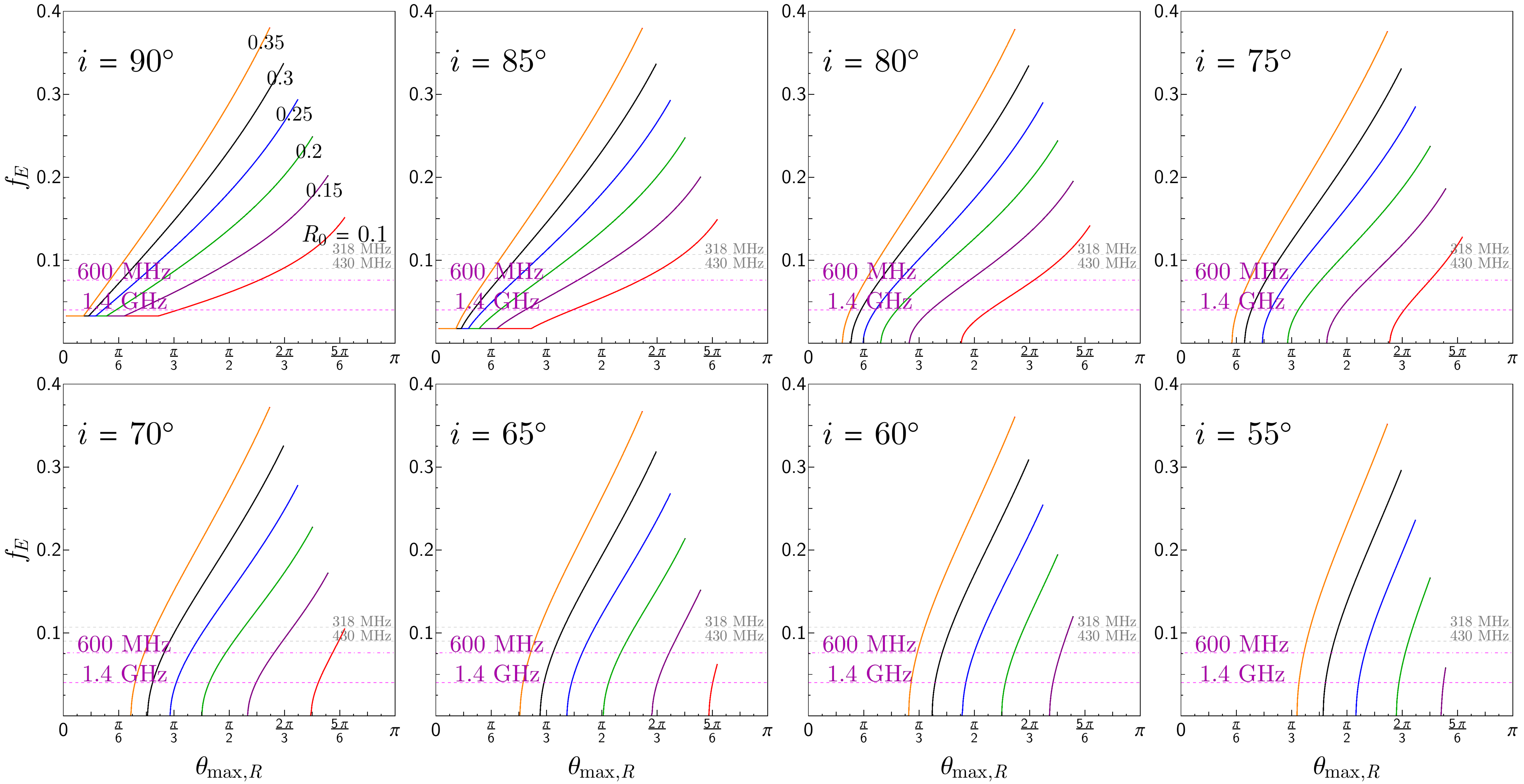}
\caption{ \added{Growth of $f_E$ as a function of $\theta_{\rm max,R} \in \{0,\theta_\infty\}$ illustrated for six values of $R_0$ between $0.1-0.35$ at fixed $i$ for the scenario where the shock is around the companion. The curves terminate at $\theta_\infty$, resulting in a maximum $f_E < 0.5$ for a prescribed $R_0$ and $i$. The flat portions at low $\theta_{\rm max,R}$ for $i=85^\circ$ and $90^\circ$ are due to the companion of $R_* \approx 0.1$ eclipsing the MSP.  Radio observations from \cite{1991ApJ...380..557R} are illustrated with the horizontal dashed lines. Interpretations are discussed in the text.}
\label{fig_thetamax_eclipses_fe}   } 
\end{figure}  

Observe in Figure~\ref{fig_1957_cartoon} that the spatial region of the shock the MSP is eclipsed by is markedly different for $55^\circ$ and $85^\circ$, with the former only sampling regions of the confined flow peripheral to the shock head. This segues to the inclination-dependent numerical computation, using Eq~(\ref{fE_equation}) of the growth rate of the eclipse fraction $f_E$ as function of $\theta_{\rm max,R}$, depicted in Figure~\ref{fig_thetamax_eclipses_fe}. The appropriate interpretation of Figure~\ref{fig_thetamax_eclipses_fe} should be restricted to the symmetric part of eclipses, i.e. below about $f_E \lesssim 0.1$. The curves terminate at $\theta_\infty$, and we note the computed $f_E < 0.5$ is finite and bounded even for the unbound axisymmetric shock geometry, i.e. $\theta_{\rm max,R} \rightarrow \theta_{\infty}$. For high inclinations, where the eclipses largely sample the shock head, the growth rate is approximately linear with $\theta_{\rm max,R}$ contrasting the nonlinear growth rates for lower inclinations which sample the periphery and tail of the shock. If the optical depth $\tau$ due to scattering or absorption by a cross section $\sigma_\nu \propto \nu^{-m}$ is given by $ 1 \ll \tau \propto \sigma_\nu \Sigma_{\rm e,R}$ for column density $\Sigma_{\rm e,R}$, then the spatial variation of the column density can be probed. Given growth curves $f_E \equiv g(\theta)$, for an observed frequency dependence in radio eclipses $f_E \sim \nu(\theta)^{-n}$, as observed for B1957+20 and RB J2215+5135 with $f_E \propto \nu^{-0.4}$, then $\nu (\theta) \sim g^{-1/n}$ and the spatial distribution of the integrated column for a given optical depth is proportional to $\Sigma_{\rm e, R} (\theta) \sim g(\theta)^{-m/n}$. For instance, if $g(\theta) \sim \theta^l$, $n=0.4$ and $m=2$ for free-free absorption in the Rayleigh-Jeans limit, then $d \log \Sigma_{\rm e, R} (\theta) / d\log \theta \sim -5 l$, an inclination-dependent line-of-sight column density that can attain a rather steep and sharp profile. This motivates frequency-dependent radio population studies of similarly eclipsing BWs and RBs for inclination-dependent trends of $f_E (\nu)$. This is of generally low utility to the X-ray Doppler-boosted emission in \S \ref{radiation} since the two populations of electrons are not necessarily concordant, but important for constraining the nature and content of the companion wind by exploring the parameter space of Figure~\ref{fig_thetamax_eclipses_fe}.

 In Appendix~\ref{parawindappendix}, we consider the simpler alternative parallel-wind bow shock geometry of \cite{1996ApJ...459L..31W} and compute correspondent $f_E$ growth curves for B1957+20 and other SC-centered spiders. Such a parallel-wind geometry is self-similar in the $R_0 \ll 1$ limit of Eq~(\ref{shockform_canto}) with much smaller shock opening angles. In this geometry $R_0$ has no impact on the shock opening angle but only serves as a scaling parameter. Since the companion and pulsar winds are not anticipated to be isotropic but somewhat stronger about the line joining the two stars in the orbital plane, this geometry is a realization where wind anisotropies obligate a much smaller shock opening angle. Such wind anisotropy is expected for an anisotropically-irradiated companion whose evaporatively-driven wind is only influential on the day-side hemisphere of the companion. For this auxiliary geometry, it is found that the constraints on $R_0$ are systematically larger than the isotropic-winds geometry, and less sensitive to the value of $\theta_{\rm max,R}$ since $f_E$ plateaus in the far-downstream tail of the shock. Since the shock angle is small, an additional upgrade to ballistic tail sweepback and eclipse asymmetry is also explored. This parameterization of eclipse asymmetry in terms of outward wind speed serves as an additional motivation for future radio characterization.

 \subsubsection{Application to PSR J1023+0038 and Other IC-centered Spiders}
  \label{J1023_radio}
  
 \begin{figure}[t]
\plotone{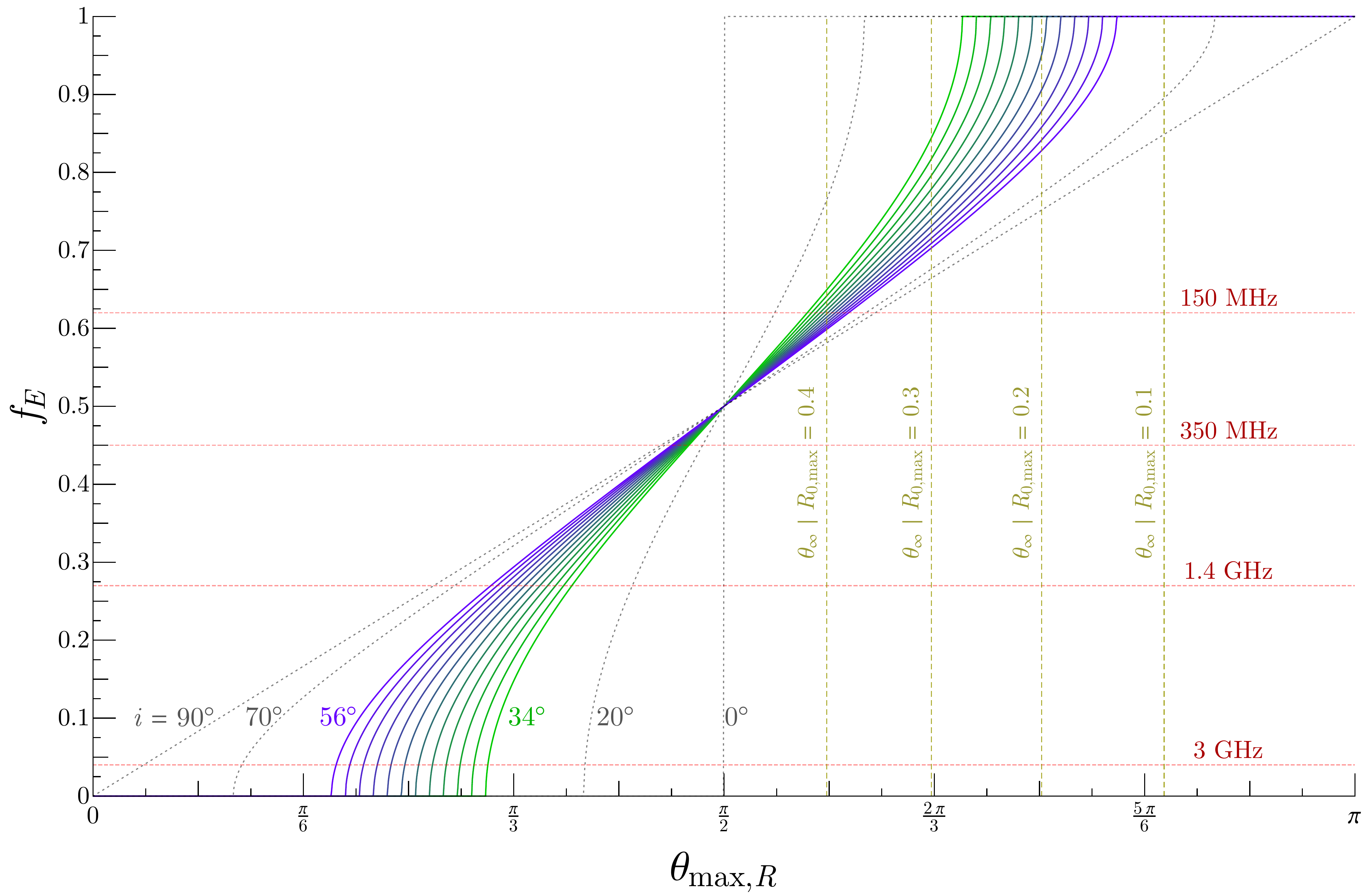}
\caption{Analytical eclipse fraction $f_E$ growth curves for any DP IC-centered system where the shock surrounds the pulsar, given by Eq.~(\ref{RBshrouding}), are shown highlighting the constrained inclination of J1023+0038 in the green to purple curves. The red horizontal lines  illustrate typically observed asymmetric total $f_E$ from \citet{2009Sci...324.1411A,2013arXiv1311.5161A}. The dark yellow vertical lines show the maximum $R_0$ that may accommodate the asymptotic shock opening angle \replaced{in the Type II shock scenario}{$\theta_\infty$ for the specific geometry of Eq~(\ref{shockform_canto})}.}
\label{J1023_fe}      
\end{figure}  

For scenarios where the \deleted{optically-thick} shock surrounds the pulsar and the perpendicular component of the shock $R(\theta)\sin \theta$ is a monotonically rising function of $\theta$, as is the case for all shocks in the limit of small sweepback, the eclipse fraction is independent of shock geometry for any given binary inclination and only depends on the shock's largest attained polar angle $\theta_{\rm max, R}$. For cases where the shock is swept back due to Coriolis effects which may introduce $\phi$ dependences and break the axisymmetry of the shock, the axisymmetric estimate below is a lower limit on the shrouding fraction (see Appendix \ref{appendix_A}). \added{For the eclipsing medium, the only assumption here is that it lies beyond the MSP termination shock, i.e. the geometric complement of the cavity excavated by the pulsar wind. Therefore, shrouding at IC is suppressed and only anticipated at inclinations away from edge-on in this model. The stand-off distance $R_0$ is measured from the pulsar, rather than the companion, in this context.}

If the shock is at an angle $\vartheta_{\rm sb}$ with respect to the orbital angular momentum vector, the eclipse fraction $f_E$ is related to $\theta_{\rm max}$ by routine spherical trigonometry, 
\begin{equation}
\cos \theta_{\rm max, R} = \cos i \cos \vartheta_{\rm sb}+ \sin i \sin \vartheta_{\rm sb} \cos (\pi f_E).
\end{equation}
We adopt the case $\vartheta_{\rm sb} = \pi/2$ corresponding to the shock axis of symmetry lying in the orbital plane. \added{The choice $\vartheta_{\rm sb} = \pi/2$ implicitly assumes $\zeta \approx i$ and the absence of jet-like anisotropy in the MSP wind \citep[e.g.,][]{1990ApJ...349..538C,1996A&A...311..172L,1999A&A...349.1017B}, a reasonable conjecture at this juncture.} \replaced{yielding}{This yields}
\begin{equation}
f_E = \frac{1}{\pi}\arccos \left( \frac{\cos \theta_{\rm max, R}}{\sin i} \right)
\label{RBshrouding}
\end{equation}
 for $\pi/2 - i \leq \theta_{\rm max, R} \leq \pi/2 + i$ with the lower and upper limits corresponding to $f_E = 0$ and $1$ respectively\added{, in contrast to Figure~\ref{fig_thetamax_eclipses_fe} which can never exceed $0.5$}.  Thus radio eclipses in MSP binaries where X-ray emission is IC-centered constrain the shrouding by the maximum shock polar angle $\theta_{\rm max, R}$ in a model-independent manner, and values of the polar angle are associated with a physical integrated line-of-sight density profile\deleted{as for B1957+20 in Figure~\ref{fig_thetamax_eclipses_sigmaE} for a fixed observing frequency}. Note that this formalism allows for axisymmetric shocks that are skewed at an angle $\Delta \phi_s$ relative to the line joining the two stars\deleted{, as is anticipated due to the finite specific angular momentum of matter falling toward the MSP and asymmetry of radio eclipses, assuming efficient angular momentum loss so that a disk-state is not realized}. If $\phi_{\rm in}$ and $\phi_{\rm eg}$ are the orbital phases of ingress and egress eclipses relative to SC, respectively, then the shock asymmetry is directly measurable at a given frequency (corresponding to a $\theta_{\rm max, R}$ in this model) as the antisymmetric part $ (1/2) (\phi_{\rm eg} - \phi_{\rm in}) \approx \Delta \phi_s/(2 \pi)$.
  
The contours corresponding to Eq.~(\ref{RBshrouding}) are illustrated in Figure~\ref{J1023_fe} with approximate J1023+0038 observational radio eclipse fractions from \cite{2009Sci...324.1411A,2013arXiv1311.5161A} for a range of inclinations. For \replaced{Type I}{parallel-wind} shocks, large values of $\theta_{\rm max, R} > \pi/2$ correspond to unphysically long tails $L_{\rm z} \gg 4 R_0$ where hydrodynamic instabilities are expected to be influential. Moreover, as we show in \S\ref{J1023_SR}, such a \added{narrow} geometry is disfavored over an \replaced{Type II-like scenario which, for a given $R_0$, is limited by a maximum $\theta_\infty$}{isotropic colliding-winds scenario which generally have larger shock opening angles limited by the} asymptotic bow shock opening angle defined by Eq.~(\ref{tinfty}). The maximum values of $R_0$ corresponding to \replaced{these asymptotic opening angles}{$\theta_\infty$} are shown in Figure~\ref{J1023_fe}, thus, one may constrain the upper limit of the shock opening angle based on the largest stable eclipse fractions observed. For J1023+0038, from eclipses at $150$ MHz of $f_{E,150 \rm MHz} \gtrsim 0.6$ this estimate yields, $R_0 \lesssim 0.4$. We caution that in such IC-centered spiders, the naive physical interpretation of $R_0$ constraints in terms of wind ram pressures given by Eqs.~(\ref{shockform_canto})--(\ref{R0eta_canto}) is erroneous due the commanding gravitational influence of the MSP past the $L_1$ point, and depends on the specifics of angular momentum loss of the companion baryonic wind. \added{Instead, $R_0$ may be envisaged as a convenient parameterization of the shock opening angle.}

\section{The Downstream Bulk Lorentz Factor $\Gamma$, Shock Mixing, and Baryon Loading}
\label{mixingsec}

For non-relativistic shocks in Eq.~(\ref{shockform_canto})\added{ and Eq.~(\ref{shockform_wilkin})}, the tangential velocity flow increases approximately linearly with the shock polar angle $\theta$ or symmetry axis $z$ for points close to the stagnation point where the shock is very nearly a spherical cap. Lacking a self-consistent relativistic MHD shock geometry, for simplicity we extend this flow scaling relation to the mildly relativistic regime \replaced{by}{assuming a scaling for} the bulk specific relativistic momentum,
\begin{equation}
p_\Gamma \equiv \Gamma \beta_\Gamma  = \left(\Gamma \beta \right)_{\rm max} \left( \frac{\theta}{\theta_{\rm max, X}}\right)  ,
\label{pmax}
\end{equation}
with $\theta_{\rm max, X}$ corresponding to a characteristic angular scale where the \replaced{shock}{shocked pulsar wind} is pressure-confined in this manner \added{and defines the region of interest where particle acceleration and synchrotron processes operate. In this rudimentary model the choice of $\theta_{\rm max, X}$, which is independent of the radio counterpart $\theta_{\rm max, R}$, is set to a benchmark value $\pi/2$ in \S\ref{radiation} to encompass the head of the shock. The specific momentum $\beta \Gamma$ is simply the spatial part of the 4-velocity that appears in the pressureless relativistic fluid continuity equation or energy density tensor. The fluid speed and Lorentz factors along the shock are a function of $\theta$, viz.}
\begin{equation}
\beta (\theta) = \frac{ \left(\Gamma \beta \right)_{\rm max} \theta}{\sqrt{ \left(\Gamma \beta \right)_{\rm max}^2 \theta^2 + \theta_{\rm max, X}^2}}  \quad, \quad \Gamma(\theta) = \frac{1}{\sqrt{1-\beta (\theta) ^2}}.
\label{betamax}
\end{equation}
We adapt this \emph{ad hoc} \added{spatially-dependent flow speed} prescription Eq.~(\ref{pmax})--(\ref{betamax}) for the calculation\added{and parameterization} of Doppler factors to synthesize light curves in \S \ref{radiation} \added{(this is a key ingredient, cf. Eq.~(\ref{deltaD}))}. This linear spatial dependence of specific momentum is a general characteristic of pressure-confined relativistic flows \citep[e.g.,][]{2006MNRAS.367..375B,2007MNRAS.380...51K, 2009MNRAS.394.1182K} \added{whose bulk flow speeds exceed relativistic gas adiabatic sound speed  $c/\sqrt{3}$}. The physical interpretation of the bulk Lorentz factors $\Gamma$ depends on the baryon loading of the shocked pulsar wind by the shocked companion component across the contact discontinuity, and couples to the companion mass loss rate in a nontrivial manner; \added{a loading $n_e/n_i \sim 10^3$ is sufficient for ions to be influential where $n_{e,i}$ are electron and ion number densities. For the energy budget $\dot{E}_{\rm SD}$, ``well-mixed" ion-dominated hypothesis such as that of \cite{2014A&A...561A.116B} requires a substantially lower companion mass loss rate than their assumed $\dot{m} \approx 6.3 \times 10^{14}$ g s$^{-1}$ to yield bulk Lorentz factors large enough to yield DP X-ray modulation as observed, if attributed to Doppler boosting in B1957+20  (cf. \S\ref{SR1957}).} The relative amount of mixing is also important for the normalization of spectra to the energy budget $\dot{E}_{\rm SD}$ since it augments the leptonic population in the shocked pulsar wind, and is consequential for spectroscopically resolved light curves in \S\ref{partaccdiag} but largely unimportant for the geometric considerations in \S\ref{radiation} that prescribe values of $\left(\Gamma \beta \right)_{\rm max}$ and other parameters in a fixed energy bin. \added{Notice that the level of mixing will also influence the geometric thickness across the contact discontinuity of the leptonic pulsar and baryonic companion winds. This is because in a kinetic picture shock thickness is proportional to the mean Larmor radius of the particles, while in a macroscopic fluid picture, the geometric thickness of the shock structure is related to the internal pressure terms which are anticipated to be larger for the nonrelativistic baryonic component versus the momentum-dominated pulsar leptonic component. Therefore one may expect mixing to increase the geometric thickness and volume where leptons radiate.}

We anticipate a partitioned shock morphology in spiders, with a shocked companion wind and pulsar wind, separated by a contact discontinuity. From flux-freezing arguments in an idealized steady-state laminar MHD framework, there is no charged particle transport across the contact discontinuity except perhaps at the stagnation point. This lack of transport seems to be in tension with the assumptions of \cite{2016arXiv160603518R}.  RMHD instabilities and kinetic-scale processes alter this overly-\replaced{simple}{simplified} picture for mixing. Following \cite{2015MNRAS.454.3886M}, an estimate of mixing due to kinetic-scale effects may be discerned by consideration of cross-field diffusion of charges. In the Bohm limit of strong turbulence and wave-particle interactions, the perpendicular diffusion coefficient $\kappa_\perp$ proceeds at the gyroscale attaining $\kappa_\perp \lesssim \kappa_\parallel = r_{\rm L} v/3$. Although such strong turbulence is not anticipated in the shocked companion wind, this estimate constitutes a lower limit for the cross-field diffusion timescale $\tau_{{\cal D}, \perp}$ of protons,
\begin{equation}
\tau_{{\cal D}, \perp} \gtrsim  \frac{\ell^2}{2 \kappa_\parallel} = 1.4 \times 10^4 \left( \frac{\ell}{10^9 \, \, \rm cm} \right)^2 \left(\frac{ B_{\rm s}}{1 \, \, \rm G} \right) \left(\frac{ v_{\rm p} }{10^9 \, \, \rm cm \, s^{-1}} \right)^{-2} \quad {\rm s} ,
\label{protontau}
\end{equation}
where we have taken $v_{\rm p} = 10^9$ cm s$^{-1} \gg a/P_{\rm b}$ for the proton speed and assumed $\ell = 0.3 R_0 \sim 10^{9}$ cm, the length scale corresponding to the separation between the shock components, roughly that recovered from RMHD simulations \citep{2005A&A...434..189B}. The value of $\tau_{{\cal D}, \perp}$ attained in Eq.~(\ref{protontau}) is much greater that the typical advection timescale of ions of $10-100$ s in the shocked companion wind, thus mass loading due to diffusion is unlikely to be a significant influence unless $\sigma$ is significantly smaller and \added{the wind is} particle dominated. Eq.~(\ref{protontau}) adapted to pairs in the shocked pulsar wind is a factor of $1836$ larger, thus less constraining. The onset of significant baryon loading due to kinetic effects then corresponds to $\langle B_{\rm s} \rangle \lesssim 10^{-3}$ and $\sigma \lesssim 10^{-9}$ in the \cite{1984ApJ...283..694K} prescription for oblique shocks, values that are largely excluded since they obligate a Lorentz factor of leptons in the shocked pulsar wind $\gamma_e \gtrsim 10^8$ (which violate the Hillas bound for containment in the shock) to emit at the critical synchrotron frequency corresponding to the $>10$ keV NuSTAR band. The contribution of mixing due to instabilities is difficult to quantify and depend on $\sigma$ in a nontrivial manner, but less germane for $\sigma \ll 1$ as they are likely to be hydrodynamic in nature and occur far from the head of the shock in the tail, at distances scaling with the natural driving scale $R_0$.

Additionally, using the magnetic field constraint Eq.~(\ref{lowB}) and the pair wind density Eq.~(\ref{neMSP}), it is simple to form an upper limit for the shocked magnetization parameter $\sigma_{\rm shocked}$ if there is no baryon loading,
 \begin{equation}
 \sigma_{\rm shocked} \lesssim  \frac{B_{\rm s, min}^2}{4 \pi n_{\rm e, MSP} \langle \gamma_e \rangle m_e c^2} \approx \frac{3 \times 10^6}{ {\cal M}_\pm \langle \gamma_e \rangle} \left( \frac{B_{\rm s, min}}{1 \, \, \rm G}  \right)^2 \left( \frac{r_{\rm s}}{10^{11} \, \, \rm cm} \right)^2 \left( \frac{10^{35} \,\, \rm erg \, s^{-1}}{\dot{E}_{\rm SD} }\right)^{1/2}.
 \end{equation}
Therefore, modest values of ${\cal M}_\pm$ and  $\langle \gamma_e \rangle$ are sufficient to yield $ \sigma_{\rm shocked} \lesssim 1$ which implies that mixing is not compulsory for efficient leptonic acceleration, but probably occurs at some low level. Neutral hydrogen may still cross fields and mix; however, the prolific soft X-ray emission from the shock should ionize any such neutrals close to the shock so that the mass loading length scale is much larger than $R_0$. Therefore, the mixing of shock components is insignificant unless small-scale instabilities are prolific near the head of the shock or there are unaccounted for magnetic fields quasi-parallel to the shock normal. \replaced{The deduction that such mixing is not significant is also consistent with MSP binaries being significant astrophysical sources of energetic cosmic-ray positrons at Earth \citep{2015ApJ...807..130V}.}{Mixing with the companion wind electron/ion plasma also dilutes the energy budget for the shock-accelerated energetic positrons that escape the system. Hence, models that posit MSP binary shocks as significant astrophysical sources of energetic cosmic-ray positrons at Earth assume low mixing \citep[e.g.,][]{2015ApJ...807..130V}.} Unless the unknown mass loss rate of the companion is small, low levels of mixing are also necessary for the mildly-relativistic bulk Lorentz factors ascribed by Eq.~(\ref{pmax})--(\ref{betamax}) and in \S\ref{radiation} \added{since mixing will baryon load the bulk flow, with large speeds demanding unreasonably large fractions of the total energy budget}. An observational constraint of mixing on the shocked pulsar wind and underlying leptonic particle distribution also follows from the method outlined in \S\ref{partaccdiag}. 

\section{Geometric Doppler-boosted Orbitally Modulated Synchrotron Emission}
\label{radiation}

\subsection{Formalism}
\label{formalism}

\input{SR_formalismv_submit}

\section{Conclusion}
\label{conclu}

 In this paper we have focused on the radio and X-ray phenomenology of MSP binaries constructing geometric models for X-ray double-peaked light curves and radio eclipses. We have shown that in the \deleted{optically-thick} thin-shell model, one may constrain the shock parameters as a function of binary inclination using radio eclipses and X-ray light curves.  \added{The two different modes for the orbital phase-centering of double-peaked X-ray light curves are interpreted as owing to the relative shock orientation, reflecting the ratio of MSP to companion wind ram pressures. When this ratio is much larger than unity, orbital radio eclipse fractions of the MSP $f_E$ may be low and X-ray DP light curves are centered at superior conjunction of the pulsar. In contrast, when this ratio is small, the phase-centering is at inferior conjunction and $f_E$ is large. This shock orientation along with large $f_E$ advances the scenario where the pulsar is enshrouded by the shock somewhere past the $L_1$ point. Synthetic X-ray light curves are a complex function of orbital inclination, shock geometry, bulk Lorentz factor, and emissivity distribution. In the power-law regime, the light curve morphology is energy-dependent if and only if the power-law index of the steady-state particle distribution is spatially dependent along the shock. This then is a powerful probe of particle acceleration in relativistic oblique shocks.} 
 
 We have argued that the shock in BWs and RBs have two components that are not well-mixed unless the \replaced{mass rate}{companion mass loss rate} is low, consistent with the reasoning that relatively low baryon loading is necessary for the mildly relativistic bulk Lorentz factors attained in the shocked pulsar wind component that generates the DP Doppler-boosted light curves. The inferior conjunction phase-centered double-peaked light curves imply a relatively stable shock at orbital length scales in our model.  Accordingly, some mechanism is necessary to stabilize the shock from gravitational influences for these cases to prevent the systems from readily transitioning to an LMXB state.  
 
 The geometric explorations of the shock presented here will be used in a future paper for modeling high-energy emission and transport processes. The quantitative agreement of the synthetic light curves with observations strongly encourages further development of the model to include considerations of diffusive and convective transport in the shock environs. There are implications for orbitally-modulated inverse Compton emission in the \emph{Fermi} LAT and TeV bands, depending on the shock geometry and target photons from the shock and companion.  Model fitting explorations in the near future of DP X-ray light curves may be able constrain the shock physics given orbital parameters, or constrain orbital parameters such as binary inclination (and consequently MSP mass) using a model for the shock geometry and the particle distribution in different regions.  Energy-dependent light curves by NICER or NuSTAR may also probe the underlying relativistic shock acceleration's spatial dependence in the future. Our study lastly motivates population studies of radio eclipse phenomenology in the MeerKAT/SKA era, where the number of MSPs discovered will increase by severalfold \citep{2015aska.confE..40K}.

\acknowledgments 

We thank the anonymous referee for aiding to improve the flow and structure of the manuscript. Z.W. acknowledges helpful discussions with Cees Bassa, Julia Deneva, Guillaume Dubus, Jason Hessels, Christopher Johns-Krull, \added{Patrick Kilian}, Edison Liang, Alessandro Papitto, \added{Martin Pohl}, Scott Ransom, Mallory Roberts, Bronek Rudak, Ben Stappers, and Kent Wood. \added{C.V. and Z.W. acknowledge Tunde Ayorinde's aid in some cross-checks.} C.V. \& Z.W. are supported by the South African National Research Foundation (NRF). The work of M.B. is supported by the South African Research Chairs Initiative of the Department of Science and Technology and the NRF. Any opinion, finding and conclusion or recommendation expressed in this material is that of the authors and the NRF does not accept any liability in this regard.  A.K.H. and M.G.B. acknowledge support from the NASA Astrophysics Theory Program. A.K.H., Z.W., and C.V. also acknowledge support from the {\textit{Fermi}} Guest Investigator Cycle 8 Grant. 

\appendix

\input{Eclipses_Appendix_input}

{\added{
\section{Parallel Wind Bow Shock}
\label{parawindappendix}

\begin{figure}[t]
\centering
\includegraphics[scale=0.178]{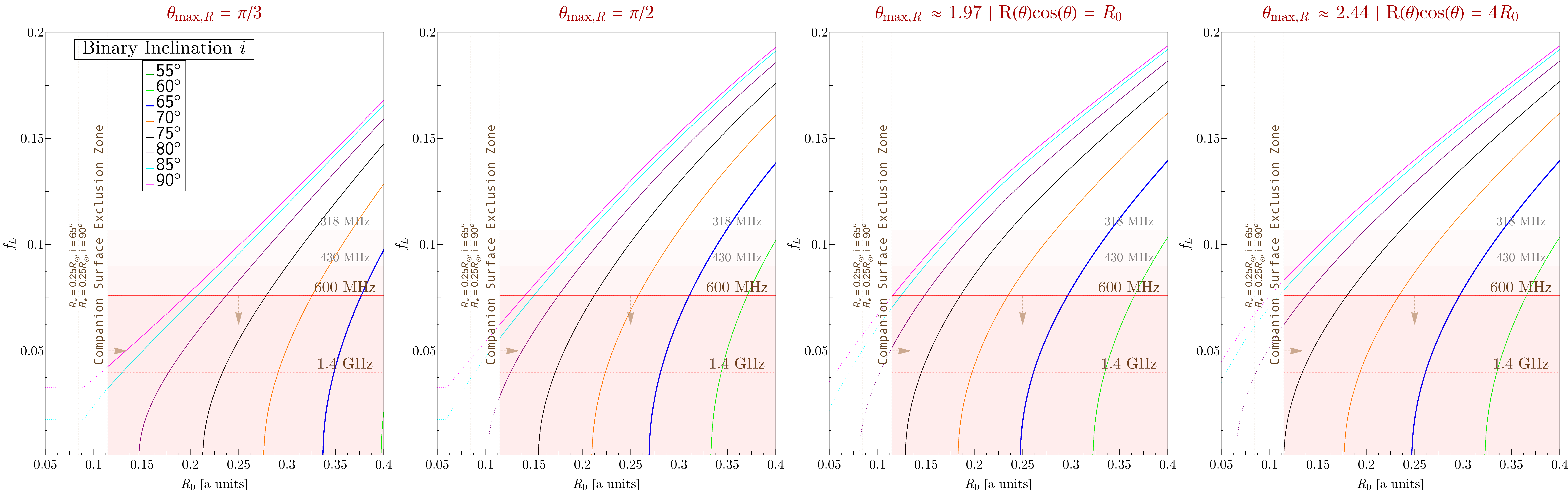}
\caption{\added{A similar suite of axisymmetric computations as that of Figure~\ref{fig_R0_eclipses_fe} but for the parallel-wind shock geometry, with different but appropriate choices $\theta_{\rm max,R}$ in the rightmost two panels corresponding to shock tail lengths past the companion of $R_0$ and $4 R_0$.}
  \label{fig_R0_eclipses_fe_W}  } 
\end{figure}  

An analytic alternative to the colliding isotropic winds shock geometry of Eq.~(\ref{shockform_canto}) is that of a parallel-isotropic wind interaction, typically invoked in the context of bow shock nebulae \citep{1996ApJ...459L..31W}. The radial function, defined on $\theta \in (0,\pi)$, is manifestly scale-invariant
\begin{equation}
\frac{R_{\rm para} (\theta)}{R_0} =  \csc \theta  \sqrt{3(1- \theta \cot \theta)} \, .
\label{shockform_wilkin}
\end{equation}
The geometry also constitutes the $R_0 \ll 1$ or $R_0^2 \approx \eta_{\rm w} \ll 1$  limit of Eq.~(\ref{shockform_canto}) for the head of the shock, and may have an arbitrarily long tail as $\theta \rightarrow \pi$. 

This geometry may be more relevant to SC-centered BWs such as B1957+20 where the companion pressure support from an irradiated evaporative wind is anisotropic and weaker on the night side. The net effect of such anisotropies would narrow the shock opening angle versus that of Eq.~(\ref{shockform_canto}) for the same stagnation point $R_0$ value. The two geometries converge only for $R_0$ values well below the companion radius,  $R_* \sim 0.1$. Therefore for $R_0 \geq R_*$, the geometry Eq~(\ref{shockform_wilkin}) is much narrower than that of Eq.~(\ref{shockform_canto}). Moreover, the tail region $\theta \gtrsim \pi/2$ is not asymptotic to any finite $\theta$, in dramatic contrast to the isotropic-winds scenario; the ratio of perpendicular to longitudinal shock components (with respect to the shock symmetry axis) tends to zero as $\theta \rightarrow \pi$.

\subsection{B1957+20 Radio Eclipses}
\label{parawindfe}

\begin{figure}[t]
\centering
\includegraphics[scale=0.25]{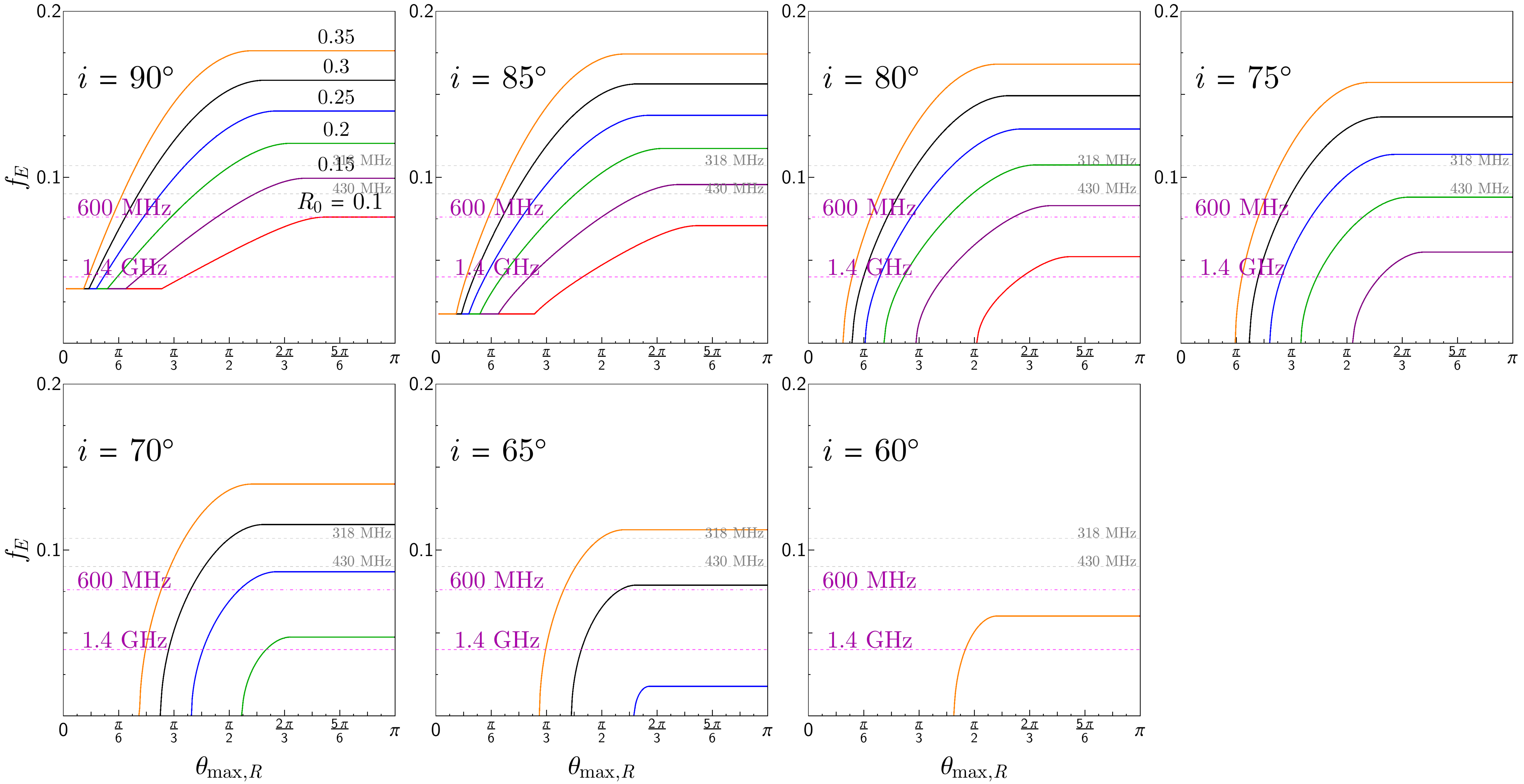}
\caption{ \added{Eclipse fraction growth rates for the parallel-wind shock geometry, contrasting the isotropic-winds case in Figure~\ref{fig_thetamax_eclipses_fe}. The case $i=55^\circ$ requires $R_0 > 0.35$ for non-zero $f_E$ for all $\theta_{\rm max,R}$ and is omitted.}
   \label{fig_thetamax_eclipses_feW}   } 
\end{figure}  

\begin{figure}[t]
\plotone{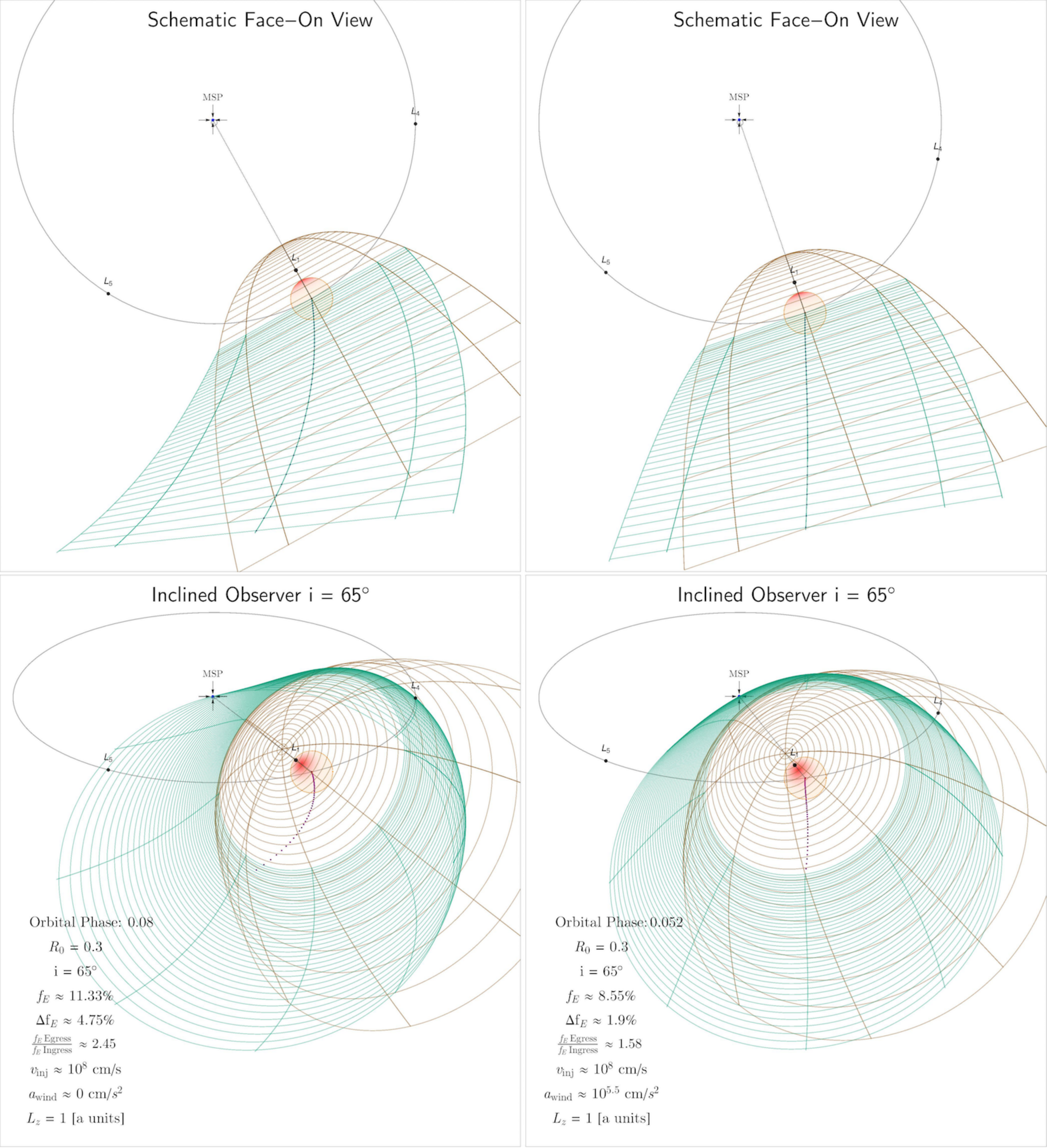}
\caption{The effect of constant velocity and acceleration on the \replaced{Type I}{parallel-wind} swept-back shock tail, out to unity semi-major distance $a$, in the PSR B1957+20 system due to orbital motion and Coriolis effects. The system angular momentum vector direction is pointing out of the plane in the top two panels, i.e. the system is rotating counterclockwise. The depicted orbital phase is at eclipse egress. For all panels, the dark green coloring mimics the loci of points corresponding to the contact discontinuity injected supersonically at velocity $v_{\rm inj} = 10^8$ cm s${}^{-1}$ at the companion position, while the brown colored lines illustrate a symmetric unswept high-speed shock component. The left panels highlight the case of zero wind acceleration, while the right panels accelerate the shock at a constant rate $a_{\rm wind} = 10^{5.5}$ cm s${}^{-2}$ (outward, parallel to the line joining the two stars). The constant acceleration, contrasted to zero acceleration, decreases the total eclipse fraction from $11.33\%$ to $8.55\%$, eclipse asymmetry from $\Delta f_E = 4.75\%$ to $1.9\%$ and egress-ingress ratio from $2.45$ to $1.58$. The purple points denote the geometric center of the projected ellipses from Eq.~(\ref{ellipseeq}).  \label{Ballistic_grid}    }
\end{figure}

\begin{figure}[t]
\plotone{ballistic_shock_densityplot_i65_85v2.pdf}
\caption{ The asymmetry of eclipses given by the proxy $\Delta f_E$ for the B1957+20 system due to orbital sweep-back of a \replaced{Type~I}{parallel-wind} shock, as a function of injected velocity, with $a_{\rm wind} = 0$, at the companion and shock length $L_{\rm z} = | R(\theta) \cos \theta | $ downstream from the companion to a maximum of $4$--$5 R_0$. The white region corresponds to the portion of the parameter space not allowed by the radio results of \cite{1991ApJ...380..557R}; here either the asymmetry of eclipses or the total eclipse fraction are too large even at the lowest frequencies. The orbital speed of the companion is shown as the blue line and varies with inclination due to observed pulsar mass function for a fixed mass ratio $q = 69.2$. \added{Larger speeds or shorter tails reduce total eclipse asymmetry.} Horizontal cuts at a fixed $v_{\rm inj}$ in the figure \replaced{can}{may} be interpreted as the frequency dependence of the eclipse asymmetry, with lower frequencies sampling larger $L_z$. }
\label{fig_ballistic_densityplot_dfe}      
\end{figure}  

\begin{figure}[t]
\plotone{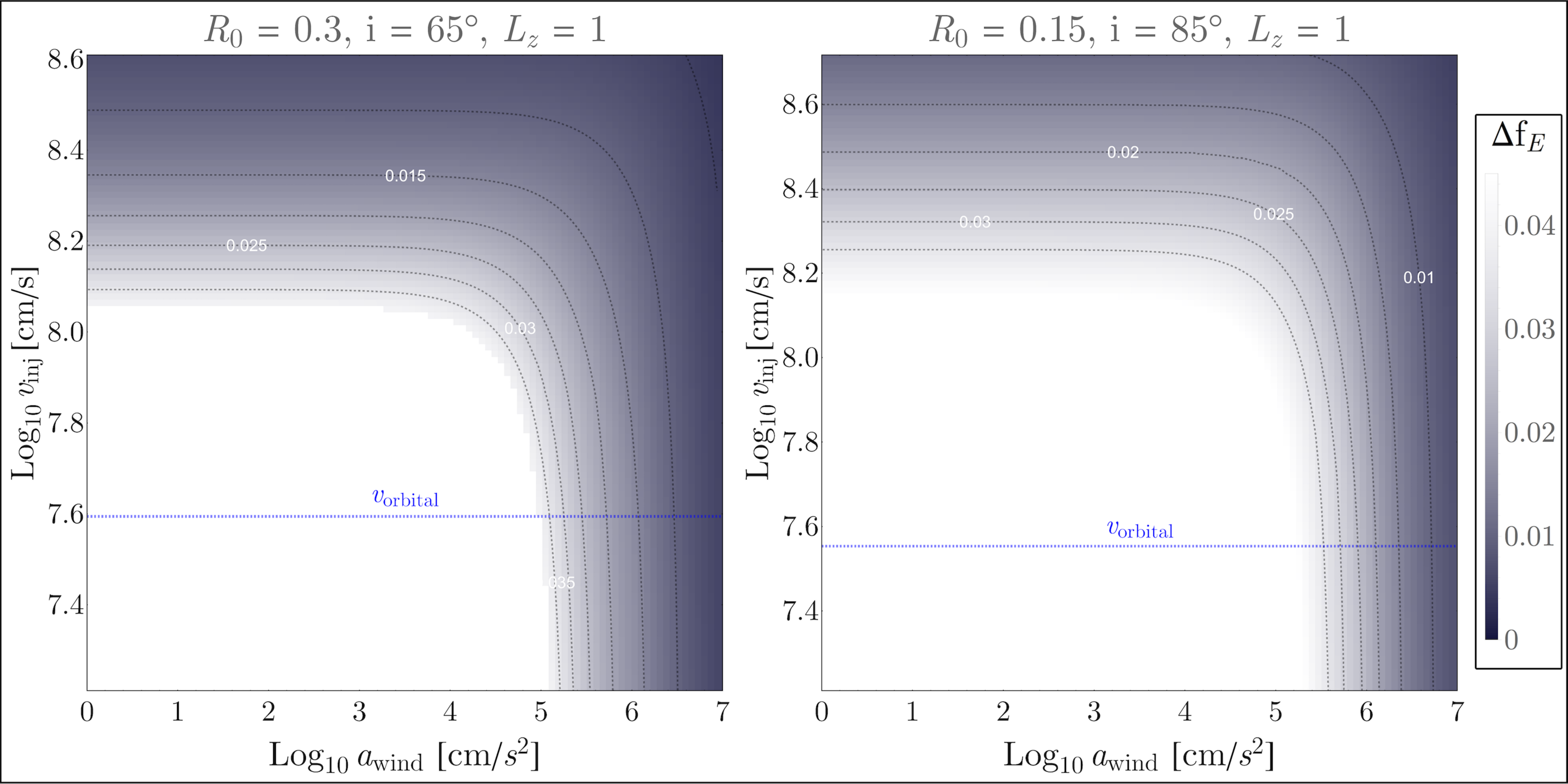}
\caption{ The asymmetry of eclipses given by the proxy $\Delta f_E$ for the B1957+20 system due to orbital sweep-back of a \replaced{Type~I}{parallel-wind} shock, as a function of injected velocity and constant wind acceleration $a_{\rm wind} $ for a fixed shock length $L_{\rm z} = 1 $ downstream from the companion. Horizontal cuts depict the eclipse asymmetry $\Delta f_E$ varying the $a_{\rm wind} $ at a fixed injection velocity, while vertical cuts correspond to $\Delta f_E$ as a function of $v_{\rm inj}$.   The white region corresponds to the portion of the parameter space not allowed by the radio results of \cite{1991ApJ...380..557R}; here either the asymmetry of eclipses or the total eclipse fraction are too large even at the lowest frequencies. \added{Larger speeds and accelerations reduce eclipse asymmetry.} The orbital speed of the companion is shown as the blue line and varies with inclination due to observed pulsar mass function for a fixed mass ratio $q = 69.2$. }
\label{fig_ballistic_avesc_densityplot_dfe}   
\end{figure}

When the shock enshrouds the companion, scale-invariance of Eq.~(\ref{shockform_wilkin}) is broken for eclipses and $f_E$ is dependent on $R_0$, as for the two isotropic-winds geometry in Figures~\ref{fig_R0_eclipses_fe}--\ref{fig_thetamax_eclipses_fe}. We compute the parallel-wind geometry counterparts of these Figures as applied to B1957+20, mirroring \S\ref{eclipsesB1957} with similar caveats. 

Observe that in Figure~\ref{fig_R0_eclipses_fe_W}, the four panels again depict successively larger values of the parameter $\theta_{\rm max,R}$ left-to-right. However, unlike its counterpart Figure~\ref{fig_R0_eclipses_fe}, the two rightmost panels prescribe values of $\theta_{\rm max,R}$ independent of $R_0$ owing to the scale-invariance of Eq.~(\ref{shockform_wilkin}). The values $\theta_{\rm max,R} \approx \{1.97, 2.44\}$, independent of $R_0$, prescribe shocks with tails of longitudinal lengths $| R(\theta) \cos \theta | = L_{\rm z} = \{R_0,4R_0\}$. The narrowness of the shock geometry also imparts a much slower growth of $f_E$ as a function of $R_0$, and implies larger values of $R_0$ for the same $f_E$ than the isotropic-winds case. As for the isotropic-winds case, there are constraints on $\theta_{\rm max,R}$. For $i = 85^\circ$, the implication is unchanged: $\theta_{\rm max,R} \lesssim \pi/2$ for $R_0 > R_*$. Observe that for $i = 65^\circ$ and $f_E \lesssim 7\%$, the inequality $\theta_{\rm max,R} > \{\pi/3,\pi/2, 1.97, 2.44\}$ corresponds to limits $R_0 \lesssim \{0.37, 0.325, 0.3, 0.3 \}$ and is of lower sensitivity to the value of $\theta_{\rm max,R}$ than the isotropic-winds geometry. The origin of this behavior is readily apparent from inspection of the growth curves in Figure~\ref{fig_thetamax_eclipses_feW} -- there is a plateauing of $f_E$ for large values of $\theta_{\rm max,R}$, following a steep rise through the head of the shock. This result intrinsic to the geometry and owing to narrowing of the shock in the tail. Low inclinations are therefore untenable with this shock geometry, with $i \lesssim 55^\circ$ requiring $R_0$ values that may exceed $0.5$ to attain $f_E \gtrsim 7\%$.

The aforementioned exploration has been in the context of symmetric eclipses in the axisymmetric limit for the head of the shock. Using the method outlined in \S\ref{appendix_tail} we may explore the asymmetry of eclipses for the long tails of the parallel-wind geometry. We assume this asymmetry arises from the relatively slow and ablated shocked companion wind with eclipsing locales being far behind the companion in the downstream tail of the shocked companion flow.  The shock is assumed to be azimuthally symmetric at every instantaneous orbital phase at the companion position, an expedient approximation that is accurate when the transverse velocity component of the shock is much smaller than the parallel component $v_{\perp} \ll v_{\rm inj}$, which exceeds the escape velocity $v_{\rm inj} \gtrsim v_{\rm esc} \sim 10^7$ cm s$^{-1}$. To simulate the tail sweepback, a supersonic flow of velocity $v_{\rm inj} \gg v_{\rm esc}$ is prescribed for the downstream shocked companion wind, accelerated by radiation pressure from the pulsar wind by parameter $a_{\rm wind}$ parallel to the line joining the two stars, neglecting gravitational forces. The flow must be supersonic in this ballistic model, so that the internal hydrodynamics of the flow can be neglected, e.g., \cite{1989ApJ...342..934R}, and we go beyond previous analyses by considering general inclination angles. Using the analytics developed in \S \ref{appendix_tail}, the ingress and egress eclipse fractions can be computed for a given $R_0$, inclination $i$, tail length $L_{\rm z}$ measured from the companion position, $v_{\rm inj}$, and $a_{\rm wind}$. In Figure~\ref{Ballistic_grid}, we depict two different cases of constant velocity and constant acceleration, and it is evident that acceleration of the flow can have dramatic consequences on the geometry and eclipse asymmetry depending on the parameters $L_z, a_{\rm wind}$, and $v_{\rm inj}$, which unfortunately, are poorly constrained observationally due to degeneracies in these parameters for a given asymmetric eclipse.

From observations of B1957+20 in \cite{1991ApJ...380..557R}, the total eclipse fraction is constrained to be conservatively $ 0.04 \lesssim f_E \lesssim 0.15$, while the egress minus ingress eclipse difference about SC, $0 \lesssim \Delta f_E \lesssim 0.05 $, with largest asymmetry for the lowest observing frequencies. In Figures~\ref{fig_ballistic_densityplot_dfe} and \ref{fig_ballistic_avesc_densityplot_dfe} we explore the parameter space for B1957+20 computing $\Delta f_E$ as a function of $v_{\rm inj}$ versus $L_{\rm z} = | R(\theta) \cos \theta | $ or $a_{\rm wind}$, respectively, excluding portions of the parameter space where the total eclipse fraction or asymmetry is disallowed by observations. In these figures, $R_0$ is modified upon changing $i$ to keep the total eclipse fraction similar. For fixed $L_{\rm z}$, the asymmetry decreases with increasing $v_{\rm inj}$ and the flattening in the asymmetry at larger $L_{\rm z}$ is due to the long shock tail. If the spatial steady-state density monotonically decreases with $L_{\rm z}$ downstream, then larger distances naturally produce larger eclipse asymmetry sampled at lower observing frequencies. Horizontal cuts in Figure~\ref{fig_ballistic_densityplot_dfe} trace the frequency dependence of the eclipse asymmetry in this simplified model of eclipse mapping. For a modest tail length of $L_{\rm z} \sim 1$ that participates in the eclipses, it is evident that $v_{\rm inj} \gtrsim 10^8$ cm s$^{-1}$ although smaller values approaching the escape speed $\sim 10^7$ cm s$^{-1}$ are allowed if the acceleration $a_{\rm wind} \gtrsim 10^5$ cm s$^{-2}$. The allowed parameter space is generally larger for lower $i$. There are obviously several confounding variables and large degeneracies in the parameter space for such non-axisymmetric eclipses at a fixed observational frequency that may produce different combinations of $f_E$ and $\Delta f_E$. This motivates radio population studies of eclipsing MSP binaries and eclipse phenomenology to unearth commonalities and constrain the parameter space of the two-wind interaction. 

\subsection{B1957+20 Synthetic X-ray Light Curves}
\label{parawindSR}

Mirroring Figure~\ref{fig_SR_emission1}, we compute light curves using the parallel-wind geometry for the head of the shock $\theta_{\rm max, X} = \pi/2$ in Figure~\ref{fig_SR_emission1W}. We choose $R_0 = \{0.16, 0.325\}$  consistent with eclipses for the head of the shock at $i = \{85^\circ,65^\circ \}$, respectively, as well as the limiting case $R_0 = R_*$. Unlike for the isotropic-winds geometry, the value of $R_0$ in the scale-invariant parallel-wind geometry does not regulate peak separation or width, only the influence of shadowing governed by the scale $R_* \approx 0.1$. The light curves are qualitatively similar to that of the isotropic-winds geometry with no change in the major conclusions: shadowing is a minor influence and $\beta_{\rm max} > 0.5$ mildly relativistic and supersonic bulk speeds are required to generate DP morphology. The narrower shock geometry does result in smaller peak separation and larger peak width, particular for the lower inclination $i=65^\circ$ similar to small $R_0$ behavior in Figure~\ref{fig_phaseplots}. It may be viable to observationally differentiate between isotropic- and parallel-wind geometries in the future with a transport model that reduces the number of free parameters.

 \begin{figure}[t]
\plottwo{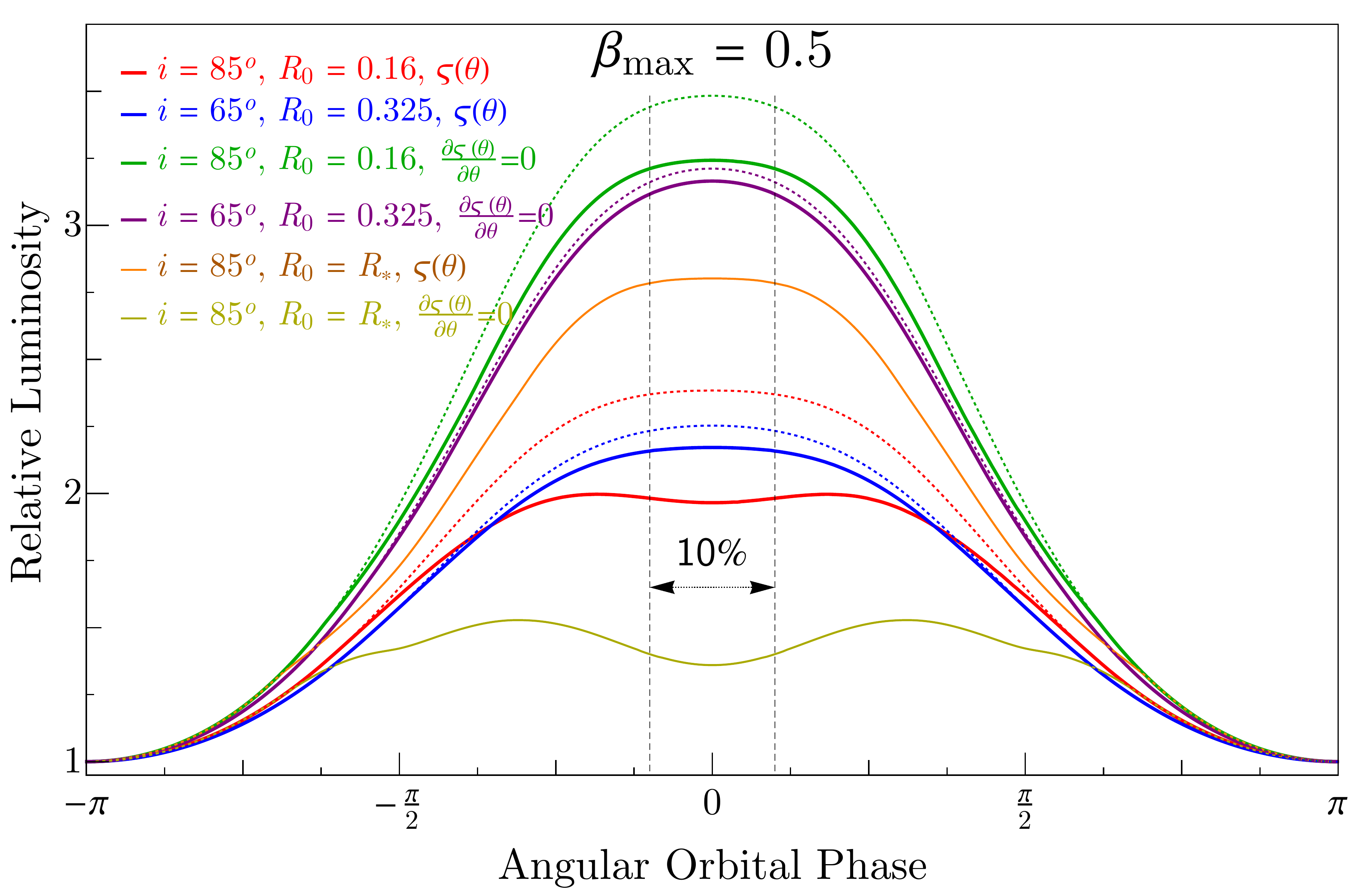}{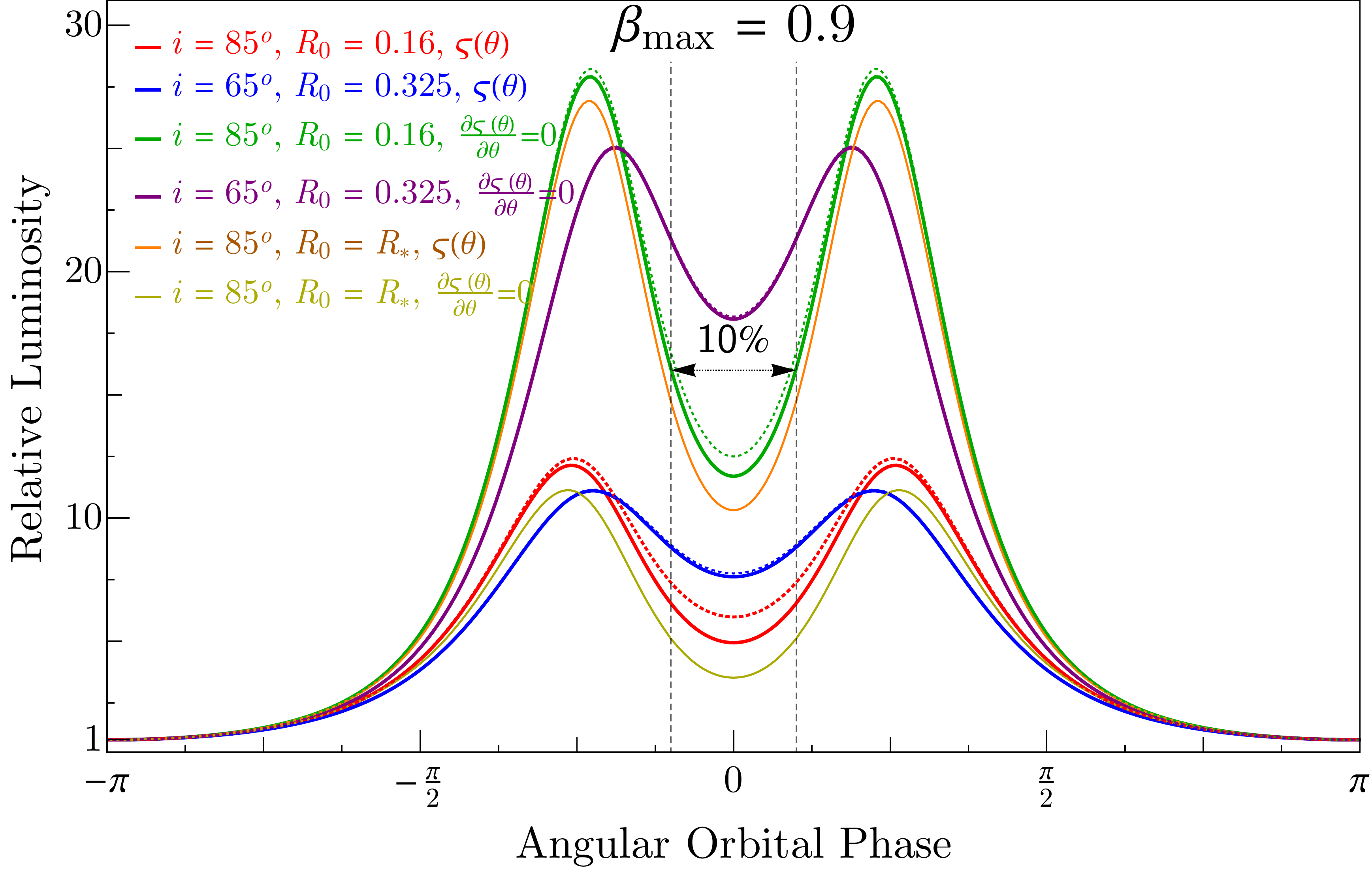}
\caption{\added{Parallel-wind shock geometry synchrotron light curves contrasting those in Figure~\ref{fig_SR_emission1}. Note the different value of $R_0$ for $i=65^\circ$ required for consistency with radio eclipses. }
 \label{fig_SR_emission1W}}  
\end{figure}  
}}

\bibliographystyle{aasjournal}
\bibliography{refs_BWRBs}

\end{document}

%% file: geo_intro.tex
\section{INTRODUCTION}

\added{The old population of rapidly-spinning neutron stars, generally known as the millisecond pulsars (MSPs), are frequently found as binaries. In the standard ``recycling'' evolutionary scenario, MSPs attain their short rotation periods through angular momentum transfer by accretion from a main-sequence companion in a low-mass X-ray binary phase \citep{1982Natur.300..728A}. Depending on the initial conditions, such evolution can yield an MSP binary with low-mass companion ($\ll 1.4 M_\odot$) in a circular orbit with a short orbital period $<1$ day. }\replaced{Millisecond pulsar (MSP)}{This small subset of radio and $\gamma$-ray MSP} binaries in tight circular orbits with low-mass companions are useful astrophysical laboratories for the physics of pulsar winds and relativistic shock acceleration. They are relevant to not only striped pulsar winds but also to the physics of Poynting-flux-dominated relativistic outflows in active galactic nuclei and gamma-ray bursts. \replaced{High-energy}{Observer-dependent high-energy} light curves and spectra advance constraints on the underpinning physical phenomena due to observer-dependence of sampling, via orbital modulations, of the emission region in a viewing geometry constrained by radio and optical determinations of the binary mass functions.  Prior to the launch of the \textit{Fermi} Large Area Telescope \citep[LAT;][]{2009ApJ...697.1071A} $\gamma$-ray observatory, only three such \added{``black widow" low-mass radio MSP} binaries were known in the Galactic Field. The first of these \replaced{was}{is} the original ``black widow" B1957+20 \citep{1988Natur.333..237F}, a $1.6$ ms MSP orbited by a stellar companion of mass $M_{\rm c} \gtrsim 0.02 M_\odot$ with a binary period of $9.17$ hours. It is now well-established that old ``recycled'' MSPs emit $\gamma$-rays up to several GeV with a \replaced{similar spectrum}{spectrum similar} to many young pulsars \citep{2013ApJS..208...17A} and that prolific e$^+$e$^-$ pair cascades \citep{1971ApJ...164..529S} must occur in the MSP magnetospheres \citep{2009ApJ...707..800V}; some of these pairs are advected into the relativistic pulsar wind that then interacts with the companion star and \added{its} wind. \added{Accelerated leptons from MSP binaries may also significantly contribute to the anomalous rise in the Galactic energetic positron fraction observed in low-Earth orbit \citep{2015ApJ...807..130V} by the Alpha Magnetic Spectrometer (AMS-02), {\textit{PAMELA}}, and \emph{Fermi} LAT instruments.}  Follow-up observations, predominantly in the radio band, of \textit{Fermi} LAT unidentified sources \replaced{has}{have} expanded the binary population to over 30\footnote{https://confluence.slac.stanford.edu/display/GLAMCOG/Public+List+of+LAT-Detected+Gamma-Ray+Pulsars} in the Field, bringing their total number to over 70 known when including those residing in globular clusters \citep{2005AJ....129.1993M} \added{with the caveat that the companion mass is unclear in many of these systems}. 

Precision radio timing of the binary MSPs accounting for orbital Doppler wobbles yields the pulsar binary mass function and\replaced{a precise}{an} estimate of the minimum mass of the companion and semi-major axis of the orbit, typically $\sim 10^{11}$ cm\added{, based on inclination $\sin i \leq 1$ and under the reasonable assumption that the MSP mass is at least the canonical $1.4 M_\odot$}. Empirically, the known population of rotation-powered MSP \added{low-mass short-period} binaries are loosely segregated based on the minimum companion mass $M_{\rm c}$ \citep{2011AIPC.1357..127R}: black widows (BWs) with minimum companion masses $M_{\rm c} \lesssim 0.05 M_\odot$ that may be degenerate, and the rarer redbacks (RBs) with non-degenerate companions $M_{\rm c} \gtrsim 0.1 M_\odot$. These MSP binaries, colloquially termed ``spider'' binaries, are ancient with characteristic ages $>$Gyr. They ``devour'' and \replaced{destroying}{destroy} their low-mass companion by accretion followed by ablation and mass loss exacerbated by the pulsar wind. Recently, some RBs have been observed to transition between a rotation-powered pulsar state and a low-mass X-ray binary accretion state \citep{2009Sci...324.1411A,2013Natur.501..517P}, confirming the association between these source classes. The behavior of these transitional objects is complex, exhibiting poorly understood transitions in X-ray luminosity and accretion states \citep{2014ApJ...795...72L,2015ApJ...806..148B}. The focus of this paper is modeling the conceptually clearer rotation-powered state of BWs and RBs where the total energy budget is constrained by the pulsar rotational energy loss rate $\dot{E}_{\rm SD}$.
 
In the standard narrative envisioned for BWs and RBs, an intense $\dot{E}_{\rm SD} \sim 10^{34 } - 10^{35}$ erg s$^{-1}$ MSP pair plasma ``striped" wind reprocesses in an intrabinary shock that irradiates the tidally-locked companion, preferentially heating \replaced{one}{the facing} side. The companion exhibits orbitally-modulated optical light curves \replaced{in many systems implying viewing geometries deviating from face-on}{interpreted as a convolution of ellipsoidal variations due to the companion being nearly Roche lobe-filling and anisotropic photospheric emission due to pulsar irradiation}. Besides heating by irradiation, particle acceleration beyond TeV energies can take place in the relativistic magnetized intrabinary shock \citep{1990ApJ...358..561H,1993ApJ...403..249A}, whose pressure support derives from either the companion wind or a magnetosphere. The companion wind matter also generates radio eclipses of the MSP at orbital phases where the companion is between the pulsar and observer. In this picture, eclipses in the radio pulsations of the MSP in \replaced{many}{some} systems can assist in constraining the shock and wind geometry as we demonstrate in this paper. A photometric light curve model of the orbital optical variations of the companion can be used to constrain the system inclination and irradiation efficiency \citep[e.g.,][]{2013ApJ...769..108B}. If radial velocities of the companion can be measured spectroscopically, the companion mass function constrains orbital parameters in the usual way. Combining the radio pulsar mass function, optical companion mass function and modeled system inclination then yields the complete orbital solution of the system, including the neutron star mass. This procedure has been applied to a handful of systems yielding \deleted{extremely} heavy neutron star masses \replaced{$\gtrsim 1.6-2 M_\odot$}{$M_{\rm MSP}$ well-above the canonical $1.4 M_\odot$} \citep[e.g.,][]{2011ApJ...728...95V}. Such massive stars constrain theories of the nuclear equation of state, marking these systems as attractive targets for the Neutron star Interior Composition Explorer \cite[NICER, ][]{2014SPIE.9144E..20A} due for launch in\deleted{February} 2017 with an energy range of $0.2$ -- $12$ keV and sensitivity about twice that of \emph{XMM-Newton}.

\replaced{The careful study}{The scrutiny} of BWs and RBs can help uncover the largely unknown physics of pulsar winds in MSPs, and aid in understanding where the transition occurs from a magnetic flux-dominated to a particle-dominated flow. Unlike the Crab Nebula whose termination shock or inner knot is $\sim 10^{15} -10^{17}$ cm away from the pulsar, the companions in BWs provide a fixed target at a distance only $\sim 10^{11}$ cm from the pulsar with much higher magnetic fields realized than in PWNe shocks. \added{Indeed, the ``clean" nature of the circular orbit, tidally-locked companion, steady well-constrained pulsar spin-down energy budget, and multiwavelength observations establish these systems as useful probes for studying the physics of pulsar winds and shock acceleration.} Kinetic-scale magnetic dissipation \citep[i.e., shock-driven reconnection, e.g.,][]{2011ApJ...726...75S, 2011ApJ...741...39S, 2016arXiv160305731L} in the shocked pulsar wind is a probable acceleration process for leptons if the pulsar wind magnetization $\sigma$, the ratio of magnetic to pair plasma particle kinetic energy density, is larger than unity -- however too large a $\sigma$\ may preclude the existence of the observed shock. Conversely, if $\sigma$ is small, a more conventional diffusive shock acceleration (DSA) may be the energization mechanism but is likely less efficient due to the oblique shock geometry. Moreover, leptons also may or may not be accelerated in the far upstream pulsar wind, although this scenario is under contention \citep{2001ApJ...547..437L, 2012Natur.482..507A, 2016ApJ...823...39Z}. There might be feedback between the intrabinary shock and the upstream wind content, as well \citep{2012AIPC.1505..402D}. \deleted{Accelerated leptons from MSP binaries may also significantly contribute to the anomalous rise in the Galactic energetic positron fraction observed in low-Earth orbit \citep{2015ApJ...807..130V} by the Alpha Magnetic Spectrometer (AMS-02), {\textit{PAMELA}}, and \emph{Fermi} LAT instruments.} 

Unlike massive TeV binaries such as B1259--63, the intrabinary shock in some BWs envelopes the companion rather than the pulsar since the pulsar wind ram pressure dominates\added{ that of the companion wind}.  \deleted{Indeed, the ``clean" nature of the circular orbit, tidally-locked companion, steady well-constrained pulsar spin-down energy budget, and multiwavelength observations establish these systems as useful probes for studying the physics of pulsar winds and shock acceleration.} Although many physical processes employed in BW and RB models mirror those invoked in massive TeV binaries  \citep{1997ApJ...477..439T, 2013A&ARv..21...64D}  the shock and orbital geometries are qualitatively different\added{,} as depicted in Figure~\ref{geometry_schematic} \added{and elaborated in \S\ref{sec21}}.  However, as we argue in this paper, many RBs and transitional systems support the interpretation of being ``inverted", where the interaction shock orientation is reversed. It then bows around the pulsar outside the light cylinder of radius $R_{\rm LC} = c P_{\rm MSP}/(2 \pi)$, where $P_{\rm MSP}$ is the MSP spin period, rather than around the companion. Such inversions can be envisaged as a state preceding or following accretion in transitional systems, \added{with gravitational influences of the MSP significantly affecting the companion wind}. The shock \replaced{geometry and location of its stagnation point can}{geometry may} significantly impact models of orbitally-modulated high-energy emission as well as the shrouding of the MSP in radio.

In this paper, we construct semi-analytical geometric models for radio eclipses and the Doppler-boosted orbitally-modulated X-ray light curves to constrain the geometry and orientation of the intrabinary shock. Previous analyses of MSP ``spider" binaries have largely focused on BW B1957+20, and principally its radio aspects \citep[e.g.,][]{1989ApJ...342..934R,1991ApJ...381L..21T}, leaving the double-peaked (DP) X-ray orbital modulation found in many BWs and RBs (see \S \ref{sec21} and Table \ref{BWRBtable}) unmodeled. We note \added{that} the parallel and independent work by \cite{2016arXiv160603518R} for the DP modulation shares some conclusions of this work. We focus on BW B1957+20 and RB J1023+0038 as representative systems, but our framework is generically applicable to other MSP binaries, setting the stage for a future population analysis as well as aiding future models of particle transport in the shock. Such advances will go beyond previous analyses invoking inverse Compton by \cite{2014A&A...561A.116B} and \cite{2012ApJ...761..181W}, and aid target selection for orbitally-modulated high-energy emission for {\textit{Fermi}} LAT and the planned \v{C}erenkov Telescope Array (CTA). In \S\ref{sec2} we present interpretations of recent observational developments and construct machinery to constrain the intrabinary shock with a simple geometric model for frequency-dependent radio eclipses. We explore the implications of recent observations and results for shock mixing in \S\ref{mixingsec}.  In \S\ref{radiation} we develop a semi-analytical model for the Doppler-boosted orbitally-modulated X-ray emission, developing synthetic light curves that will be useful for a planned model fitting study.  \added{Our conclusions follow in \S\ref{conclu}.}

%% file: doublepeak_list.tex
\begin{deluxetable}{lccc}
\setlength{\tabcolsep}{0.2in}
\tablecaption{Rotation-Powered BWs and RBs with DP X-Ray Light Curve Morphology}
\tablecolumns{4}
\tablehead{ \colhead{ Name } & \colhead{Type}  &\colhead{DP Phase Centering} &\colhead{Refs.} }
\startdata
B1957+20 & BW & SC & [1] \\
J0024--7204W & RB & IC & [2] \\
J1023+0038 & RB & IC & [3] \\
J12270--4859 & RB & IC & [4] \\
J1723--2837 & RB & IC & [5] \\
J2039--5618 & RB & IC & [6] \\
J2129--0429 & RB & IC & [7] \\
J2215+5135 & RB & IC & [8] \\
J2339.6--0532 & BW & IC & [9] \\ 
\enddata
\label{BWRBtable}
\tablecomments{Current list of MSP binaries in the rotation-powered state for which DP X-ray emission attributed to Doppler-boosting has been observed. IC and SC denote inferior and superior conjunction of the pulsar, respectively.}
\tablerefs{
[1] \cite{2003Sci...299.1372S, 2012ApJ...760...92H}
[2] \cite{2005ApJ...630.1029B}
[3] \cite{2010ApJ...722...88A,2010ApJ...724L.207T,2011ApJ...742...97B, 2014ApJ...789...40B, 2014ApJ...791...77T}
[4] \cite{2014ApJ...789...40B, 2015MNRAS.454.2190D}
[5] \cite{2014ApJ...781....6B,2014ApJ...781L..21H}
[6] \cite{2015ApJ...812L..24R,2015ApJ...814...88S}
[7] \cite{2015arXiv150207208R,2015ApJ...801L..27H}
[8]  \cite{2014ApJ...783...69G,2015ApJ...809L..10R}
[9] \cite{2011ApJ...743L..26R, 2015ApJ...802...84Y} 
}
\end{deluxetable}

%% file: SR_formalismv_submit.tex
We assume local quasi-isotropy in pitch angle of the differential electron number density distribution along the shock in the comoving frame of the bulk flow in a thin emitting region such that $\bar{n}_e (\boldsymbol{\bar{r}},\gamma_{\rm e}) \rightarrow \Sigma_{\rm e} (\theta, \phi, \gamma_{\rm e})/\Delta (\theta, \phi) $ at each lab-frame coordinate pair $\{\theta, \phi\}$ on the shock, where $\Delta(\theta, \phi)$ is the shock thickness and $\Sigma_{\rm e}$ the differential electron surface density distribution. In reality, the local electron distribution accelerated in relativistic shocks by DSA or prolific magnetic reconnection is expected to be highly anisotropic relative to the mean field direction in the comoving frame. However, for locales near the stagnation point where the residence timescale is large compared to the adiabatic and convection timescales, the flow is expected to be well-isotropized with the field randomly oriented, especially if magnetic dissipation is the principal particle acceleration mechanism. Moreover, for locales where the convection speed is large, Doppler boosting transforms any emission small-scale anisotropy into a narrow cone along the shock, the geometry of which dominates in the observer frame over any smaller-scale anisotropy within the cone due to the underlying electron pitch angle distribution. 

The differential volumetric emissivity, in the comoving frame (denoted by bars over variables), is defined by
\begin{equation}
\bar{j}_{\epsilon_f} =  \frac{d \bar{E}}{d\bar{V} d\bar{t} d\bar{\epsilon}_f d\bar{\Omega}_f}  = m_e c^2 \bar{\epsilon}_f \bar{\dot{n}}_{\rm SR}^{\rm iso},
\end{equation}
where $\bar{\dot{n}}_{\rm SR}^{\rm iso}$ is the isotropic differential synchrotron photon production rate at each interaction point. The variables $V$, $t$, $\epsilon_f$ and $\Omega_f$ constitute the volume, time, outgoing photon energy, and outgoing photon solid angle, respectively,\added{ with photon energies in units of $m_e c^2$.}

{\deleted{
The differential emissivity averaged over an isotropic distribution of pitch angles in the comoving frame at an interaction point, first derived by \cite{1986A&A...164L..16C}, is given by
$\bar{j}_{\epsilon_f} = \frac{3 \sqrt{3}}{2 \pi} \sigma_T c \, U_{\rm cr} \left( \frac{\bar{B}_s}{B_{\rm cr}} \right) \int_1^\infty d\gamma_{\rm e} \bar{n}_e (\gamma_{\rm e}) \,  {\cal R} \left( \frac{\bar{\epsilon}_f}{\bar{\epsilon}_c} \right) ,$%
where photon energies are in units of $m_e c^2$, $\bar{\epsilon}_c = 3 \bar{B}_s/(2 B_{\rm cr}) \gamma_{\rm e}^2$, $U_{\rm cr} = B_{\rm cr}^2/(8\pi)$, $B_{\rm cr} \approx 4.414 \times 10^{13}$ G the quantum critical or Schwinger field, and,
${\cal{R}}(\chi) = \frac{\pi}{2} \chi \left[ W_{0, \frac{4}{3}} (\chi)  W_{0, \frac{1}{3}} (\chi)  - W_{\frac{1}{2}, \frac{5}{6}} (\chi)  W_{-\frac{1}{2}, \frac{5}{6}} (\chi)  \right].$%
Here $W_{\kappa, \mu}$ are Whittaker functions \citep{1965hmfw.book.....A}, and ${\cal R} \propto \chi^{1/3}$ for small arguments $\chi \ll 1$, and exponentially cut-off at large arguments. These two asymptotic domains can be employed to define approximations to ${\cal R}(\chi )$ that facilitate numerical computations of the emissivity.  However, here, the orbital light curves are mostly computed in the domain of power-law spectra, for which the detailed form of ${\cal R}(\chi )$ serves only to determine an overall normalization factor, which cancels out of flux ratio computations.}}

 For a power-law distribution of electrons with index $p$ between Lorentz factors $\gamma_{\rm min}$ and $\gamma_{\rm max}$ \added{and spatial range $\theta \in (0,\theta_{\rm max, X}$}, denoted by the Heaviside step function $\Theta$, such that $\bar{n}_e ( \gamma_{\rm e},\theta) = \Sigma_{\rm e} (\theta, \phi)/ \Delta (\theta, \phi)  (\gamma_{\rm e}/\gamma_{\rm min})^{-p} \allowbreak \times \Theta(\gamma_{\rm e}\; \gamma_{\rm min}, \gamma_{\rm max}) \Theta(\theta; 0, \theta_{\rm max,X})$, the emissivity \replaced{far from the integration endpoints in Eq.~(\ref{fulljSR}),}{in the synchrotron power-law regime} $ \gamma_{\rm min} \ll \sqrt{\bar{\epsilon}_f B_{\rm cr}/\bar{B}_s} \ll \gamma_{\rm max}$\added{,} is given by \citet{2009herb.book.....D},
\begin{equation}
\bar{j}_{\bar{\epsilon}_f} \approx  {\cal C}(p) \left(\frac{3}{2}\right)^{(p+1)/2} \frac{\Sigma_{\rm e}(\theta, \phi)}{\Delta(\theta, \phi)} \sigma_T c \,U_{\rm cr}  \left(\frac{\bar{B}_s(\theta, \phi) }{B_{\rm cr}} \right)^{(p+1)/2} \bar{\epsilon}_f^{-(p-1)/2} ,
\label{SR_j}
\end{equation}
where $\bar{B}_s (\theta, \phi)$ is the post-shock turbulent magnetic field magnitude in the comoving frame, \added{ $U_{\rm cr} = B_{\rm cr}^2/(8\pi)$, $B_{\rm cr} \approx 4.414 \times 10^{13}$ G the quantum critical or Schwinger field,} and
\begin{equation}
 {\cal C}(p) = \frac{2^{(p-1)/2}  \sqrt{3}  }{32 \sqrt{\pi}} \left[ \Gamma\left( \frac{p}{4}+\frac{7}{4} \right)   \right]^{-1} \Gamma\left( \frac{p}{4}-\frac{1}{12} \right)  \Gamma\left( \frac{p}{4}+\frac{19}{12} \right)   \Gamma\left( \frac{p}{4}+\frac{1}{4} \right) .
 \end{equation}
 In the idealized MHD formulation where the bulk flow is force-free and there is no magnetic field perpendicular to the flow, we note that Lorentz transforming from the lab frame to the comoving blob frame preserves the parallel component $\langle B_s \rangle$. \deleted{In principle, the particle index may vary at points along the shock such that $p \rightarrow p(\theta, \phi)$, steepening in locales where acceleration is inefficient, but this a priori character is poorly constrained theoretically. For expediency we consider it constant for the head of the bow shock under consideration in this section. Unless the shock is highly oblique at the stagnation point, particle acceleration along the shock should decrease and cease at locales farther removed from the nose, demanding that the injection spectrum product $\Sigma_{\rm e} \gamma_{\rm e}^{-p}$ for a large enough given $\gamma_{\rm e}$, be a decreasing function of $\theta$. This injection spectrum may be convolved into a spatially-dependent particle transport model along the shock, which we defer to a future study.}

To suitably develop light curves that would correspond to an observable, we form the differential synchrotron luminosity, $L_{\rm SR}(\Omega_f, \epsilon_f )\equiv d{\cal L}_{\rm SR}/(d\Omega_f d\epsilon_f)$.  This is the total source luminosity ${\cal L}_{\rm SR}$, or volume-integrated emissivity, binned in solid angle \added{and energy} elements.  We take advantage of the simple Lorentz transformation property of the spectral emissivity $\bar {j}_{\epsilon_f}$, namely that $j_{\epsilon_f}/{\epsilon_f^2}$ is invariant under boosts. 
Forming the differential luminosity in the observer frame, the result is an integral over the shock volume that is essentially a reconstruction of 
Eq.~(10) of \cite{2015A&A...581A..27D} in the optically thin limit:
\begin{equation}
L_{\rm SR}(\Omega_f , \epsilon_f) \equiv {{ d{\cal L}_{\rm SR}}\over {d\epsilon_f d\Omega_f}} 
= \int dV \epsilon_f^2 \, \left( {{  {j}_{\epsilon_f} }\over { \epsilon_f^2 }} \right)
= \int dV \delta_{\rm D}^2  \bar {j}_{\epsilon_f} ,
 \label{SR_transform} 
 \end{equation}
with the emissivity and photon energies being computed using the Doppler-shifted photon energy
\begin{equation}
\bar{\epsilon}_f = \epsilon_f / \delta_{\rm D} .  
\end{equation}
Observe that the factor $\delta_{\rm D}^2$ in Eq.~(\ref{SR_transform}), rather than $\delta_{\rm D}^3$ found routinely in relativistic jet contexts, arises because the integration over the emitting volume element is chosen to be in the observer's frame rather than the comoving frame. 
This orbitally-modulated Doppler factor $\delta_{\rm D}$ is calculated for each point along the shock and dependent on the prescribed local bulk speed $\beta_\Gamma (\theta)$  and bulk Lorentz factor $\Gamma (\theta)$ along the shock, defined by
\begin{equation}
\delta_{\rm D} (\theta, \phi, \Omega_b t, i)  = \frac{1}{\Gamma(\theta) \left(1 - \beta_\Gamma (\theta) \boldsymbol{\hat{n}}_v \boldsymbol{\cdot} \boldsymbol{\hat{u}}^\prime \right)},
\label{deltaD}
\end{equation}
where $\boldsymbol{\hat{u}}^\prime$ is the unit tangent vector along the polar direction of the shock in the inclined and orbital phase-rotated coordinate system (cf. Appendix~\ref{appendix_A} for definitions and conventions). Thus, we have
\begin{eqnarray}
 \boldsymbol{u} &\equiv& \frac{ d R_x} {d \theta } \hat{\boldsymbol{x}}  + \frac{d R_y}{ d \theta }\hat{\boldsymbol{y}} + \frac{d R_z}{d \theta }\hat{\boldsymbol{z}} \quad ,\quad \boldsymbol{\hat{u}} = \frac{\boldsymbol{u}}{u} \\ 
\boldsymbol{\hat{u}}^\prime &=& \Lambda_i \Lambda_{\Omega_b t_0} \boldsymbol{\hat{u}}.
\label{DopplerV}
\end{eqnarray}
In addition, $ \boldsymbol{\hat{n}}_v $ is the unit vector in the direction of the observer, applicable to all orbital phases. Throughout, we employ the scaling Eq.~(\ref{pmax})--(\ref{betamax}), and associated $\beta_{\rm max}$ to prescribe the bulk flow speed along the shock.

 The convolution of different Doppler factors realized in different portions of the integration volume near the shock is what defines the relative sharpness of peaks in the ensuing figures of orbital modulations of the flux. For the assumed symmetry, the integration Eq.~(\ref{SR_transform}) is two-dimensional in variables $\theta$ and $\phi$ for each point along the intrabinary shock. For a shock that is thin compared to the orbital length scale $a$, we may write the lab frame volume element as
\begin{equation}
dV \approx \Delta (\theta, \phi) dA = a^2 \Delta (\theta, \phi) R(\theta) \sin \theta \sqrt{R(\theta)^2 + (dR/d\theta)^2} d\theta \, d\phi.
\end{equation}
\deleted{This differential volume element differs for Type I and II geometries through its dependence on $R(\theta)$}. The factor $\Delta(\theta, \phi)$ cancels with the factor in Eq.~(\ref{SR_j}), and the net result is that the orbital-modulated differential luminosity or intensity is proportional to
\begin{eqnarray}
L_{\rm SR} &\propto& \epsilon_f^{-(p-1)/2}  \int d\theta \, d\phi \, R(\theta) \sin \theta \sqrt{R(\theta)^2 + (dR/d\theta)^2} \, \, \delta_{\rm D}^{2+(p-1)/2} \varsigma (\theta, \phi) \label{LSRpropto} \\
\varsigma (\theta, \phi) &=& \Sigma_{\rm e}(\theta, \phi) \left[ \bar{B}_s (\theta, \phi) \right]^{(p+1)/2} 
\end{eqnarray}
when $p$ is a constant along the region of interest. Crucially, the flux ratio of Eq.~(\ref{LSRpropto}) generates energy-independent light curves when $p$ is spatially independent.  In this regime, the ensuing large curves are largely regulated by geometric influences of the observer and ascribed velocity profile Eq.~(\ref{pmax}) \added{through Eq.~(\ref{deltaD})}.  It is quickly discerned that the maximum amplitude of orbital modulation is attained when $\boldsymbol{\hat{n}}_v \boldsymbol{\cdot} \boldsymbol{\hat{u}}^\prime  \approx 1/\beta_\Gamma$ realizing a flux enhancement of order $\lesssim 2 \Gamma_{\rm max}^{2+(p-1)/2}$ modulo \added{weighting of the emissivity across the full shell and observer impact angles with respect to the tangential velocity component in Eq.~(\ref{LSRpropto}) that moderates this upper limit}. \deleted{If $p$ is spatially dependent, the overall spectral index may deviate from the usual $(p-1)/2$ value and a pure power law in an orbital phase-dependent manner.}

\begin{figure}[t]
\centering
\includegraphics[scale=0.55]{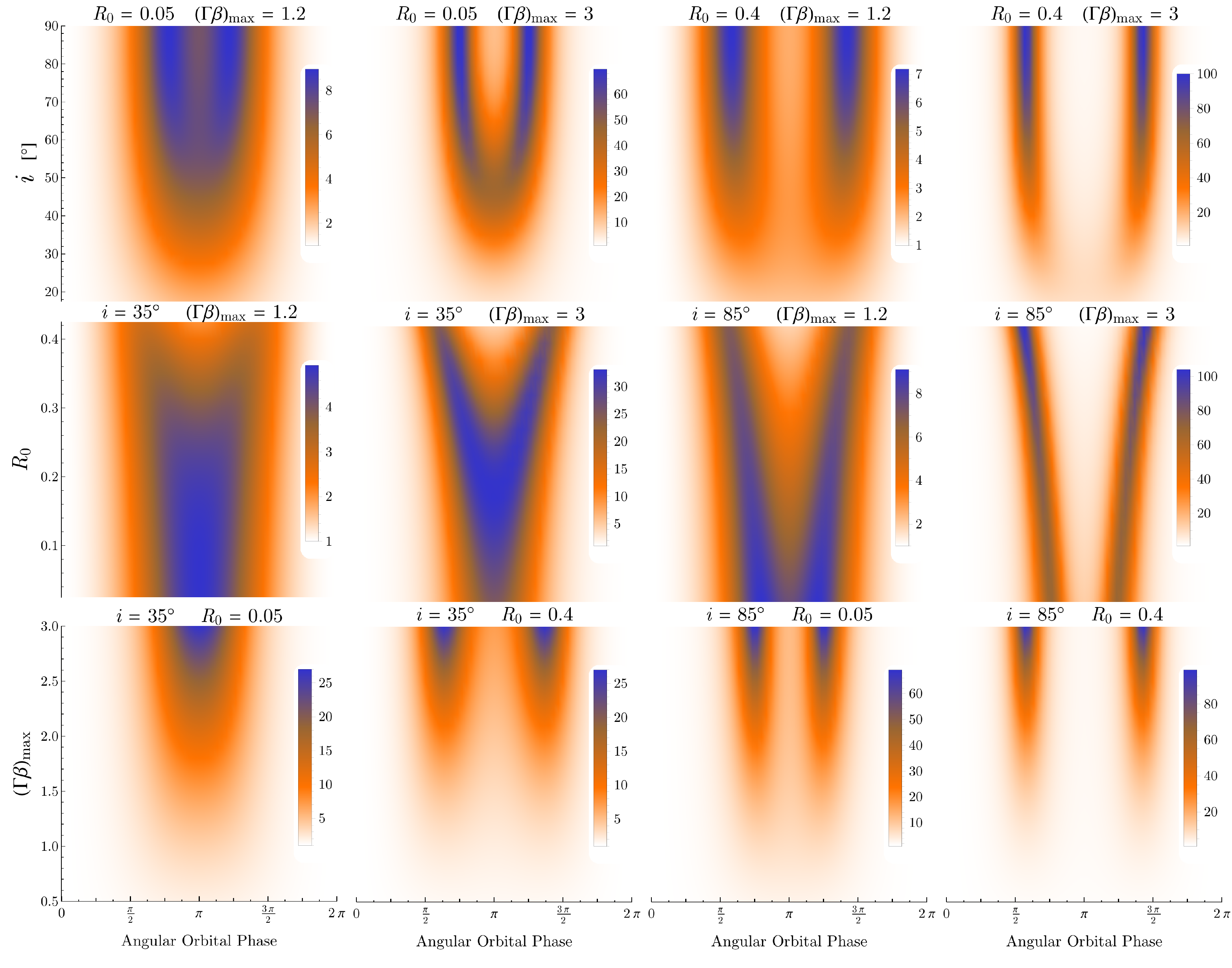}
\caption{ \added{Partial atlas of IC-centered light curves for synchrotron emission, normalized to unity at SC (white coloring), computed for $p=2$, $\theta_{\rm max,X} = \pi/2$ and $\varsigma =$ constant using Eq.~(\ref{LSRpropto}). $R_0$ is measured from the MSP. The top, middle and bottom rows depict the $i$, $R_0$ and $(\Gamma \beta)_{\rm max}$ dependency on the synthetic light curves, respectively. The columns contrast four pairings of parameters somewhat extreme in range. No shadowing by a companion is included.}
  \label{fig_phaseplots}  } 
\end{figure}

\added{The accelerated/cooled electron power-law and magnetic field spatial distribution in the thin shell, encapsulated in $\varsigma$, is poorly understood. Geometric and Doppler boosting influences must dominate in the integral Eq.~(\ref{LSRpropto}) far from the shock nose since this is a necessary condition for the existence and generation of DP light curves. Therefore the weighting $\varsigma$ is expected to be a modest nuance on the light curve morphology, pending a future proper treatment of self-consistent particle transport. Even so, the underlying particle acceleration mechanism is unknown which couples to the form of $\varsigma$ in a transport model, and the unknown spatial distribution of $p$. Crucially, if $p$ is spatially dependent, the overall spectral index may deviate from the usual $(p-1)/2$ value and a pure power law in an orbital phase-dependent manner, a profound consequence discussed in \S\ref{partaccdiag}. 

To isolate the geometric aspects of the model, we take $\partial \varsigma/\partial \theta = \partial \varsigma/\partial \phi = 0$ with $p=2$ and routinely integrate Eq.~(\ref{LSRpropto}) for the isotropic-winds geometry of Eq.~(\ref{shockform_canto}) for a single thin shell in the IC-centered scenario, IC corresponding to angular phase $\pi$. For such a shock orientation, geometric occlusion by the companion is a negligible influence. A phase plots atlas of light curves, depicted in Figure~\ref{fig_phaseplots} normalized to unity at SC, establishes several features as a function of parameters $R_0, (\Gamma \beta)_{\rm max}$, and $i$, with $\theta_{\rm max,X} = \pi/2$. Observe that with a single shell, only a single broad peak or DPs may be produced, as discernible in the ``U" or ``V" patterns in the top two rows. The boost parameter $ (\Gamma \beta)_{\rm max}$ straightforwardly regulates the relative sharpness and width of the peaks. The characteristic shock asymptotic opening angle, parameterized by $R_0$, regulates peak separation at fixed $i$ with larger shock opening angles (corresponding to large $R_0$) yielding wider peaks. The characteristic shock opening angle is also regulated by the weighting $\varsigma$. Transient changes in X-ray peak separation therefore may be interpreted as changes in the effective shock opening angle with respect to the observer or an equivalent spatial change in $\varsigma$, since emission near the shock nose is not Doppler-boosted. There is degeneracy in peak separation with $i$ and $R_0$ highlighting the need for multiwavelength constraints on $i$. The other degeneracy for the peak width between $R_0$ and $(\Gamma \beta)_{\rm max}$ at fixed $i$ may be resolved in a future particle transport and mixing model, since $(\Gamma \beta)_{\rm max}$ cannot be arbitrarily large by energy budget considerations. If there is universality among BWs and RBs in the shock structure or opening angle, correlations between X-ray peak separation and $i$ may become evident across a population of BWs or RBs.

Note that the assumption of pitch angle isotropy also suppresses a phase-dependent polarization signature. A polarization-dependent calculation follows routinely from the development for the synchrotron emissivity, but requires a specification of the unknown magnetic structure and level of turbulence. Future proposed X-ray polarimetry instruments \citetext{PolSTAR: \citealp{2016APh....75....8K}; XIPE: \citealp{2013ExA....36..523S}; IXPE: \citealp{2016SPIE.9905E..17W}} may be able to discriminate among types of acceleration mechanisms from orbital phase-dependent Stokes parameters. High levels of linear polarization, for example, may suggest radiative losses in an ordered magnetic field rather than turbulent reconnection, with orbital phase-dependent polarization yielding tomographic information of the magnetic field geometry on the shock.
}

 \subsection{Application to PSR B1957+20, an SC-centered System}
 \label{SR1957}
 
 \begin{figure}[t]
\plottwo{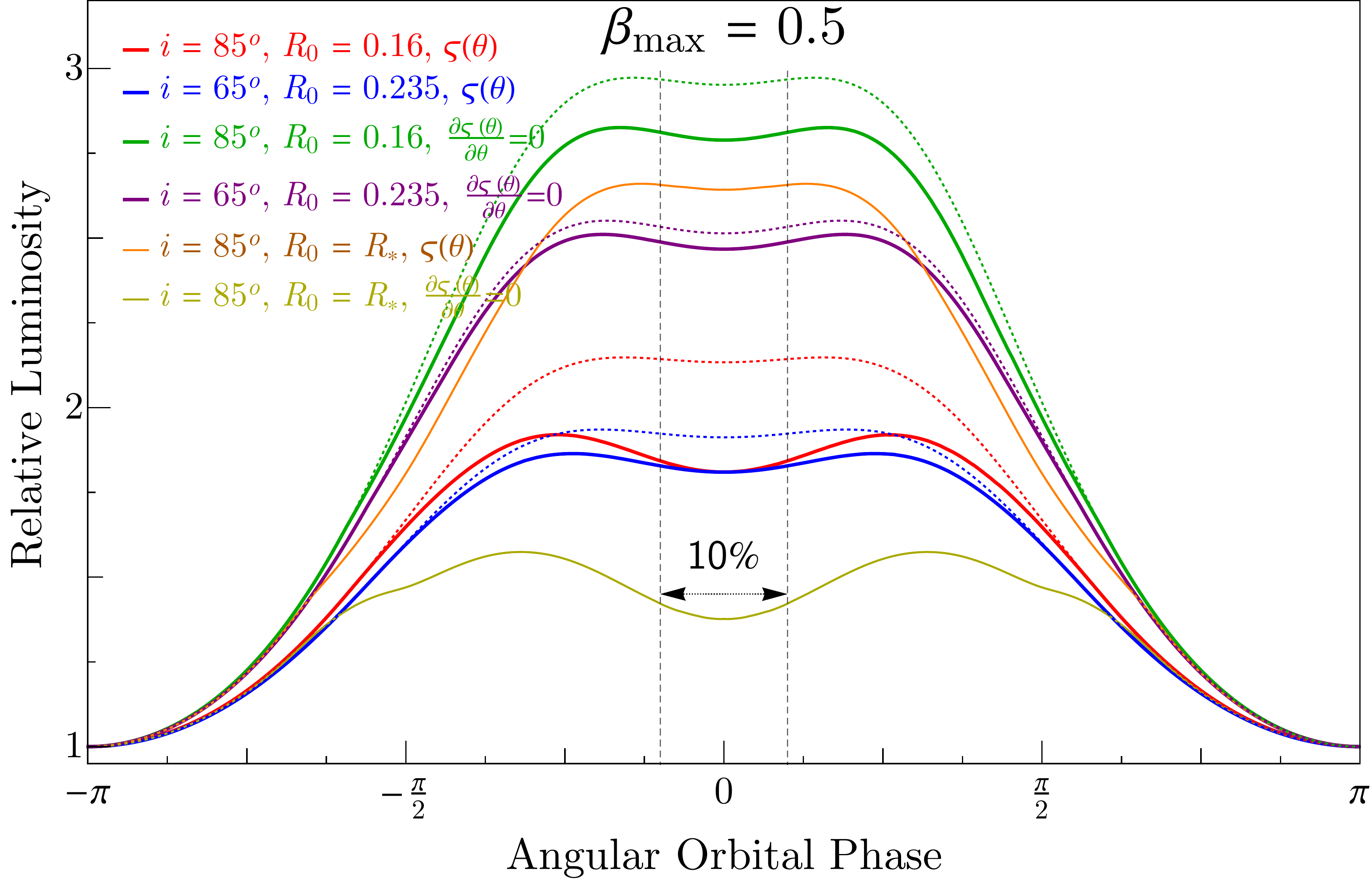}{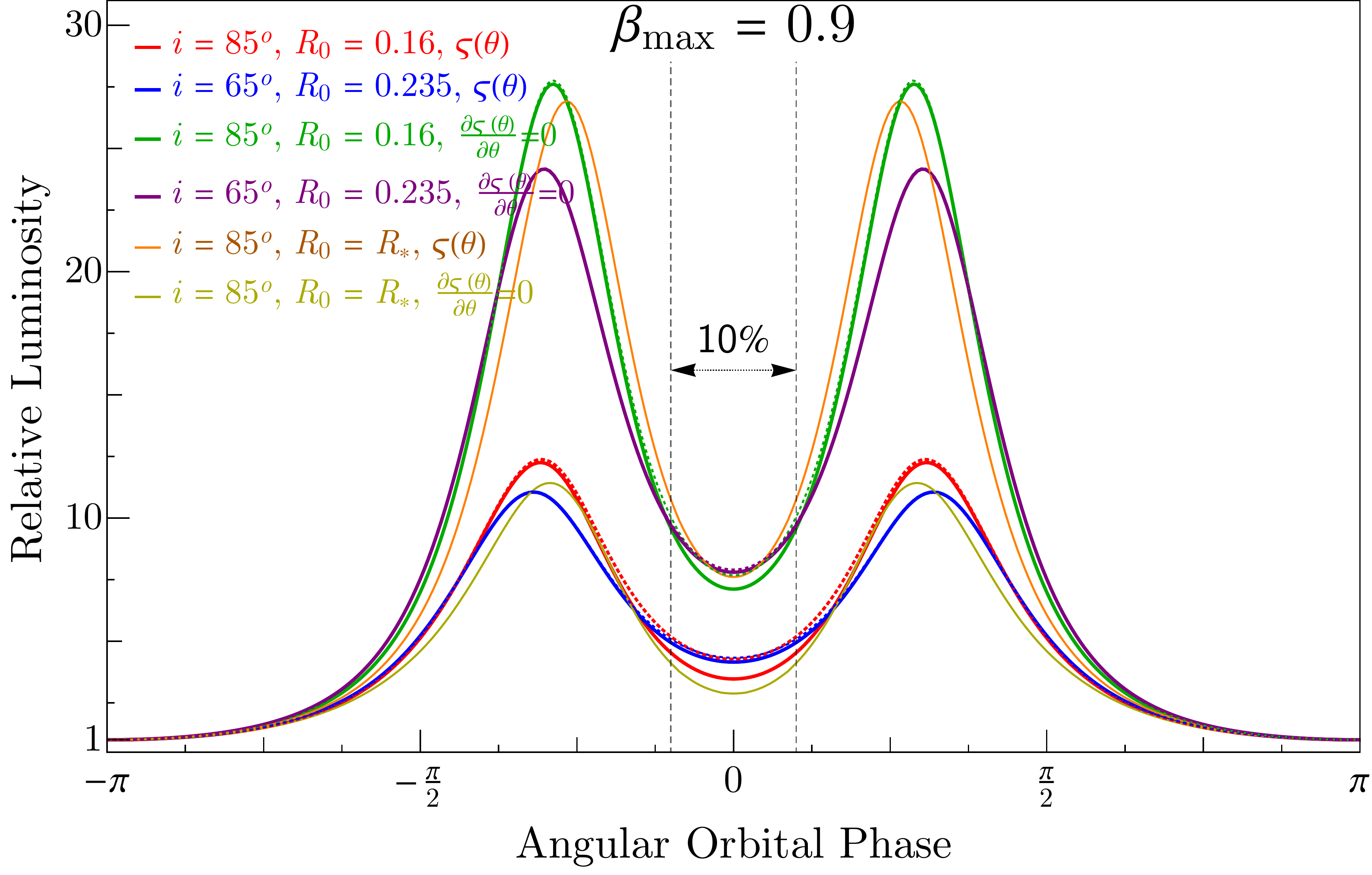}
\caption{ Orbitally-modulated synchrotron flux ratios of superior-to-inferior conjunction for \replaced{$\beta_{\rm max} = 0.5$}{$\beta_{\rm max} = \{0.5, 0.9\}$ with $\theta_{\rm max,X} = \pi/2$}, at an arbitrary energy where the power-law approximation is valid, for different inclinations and shock stand-off $R_0$ \replaced{approximately derived from}{generally consistent with} radio eclipses. Phase zero defines SC of the pulsar. The solid and dashed curves are \added{cases, where the shock is} shadowed and unshadowed by the companion \deleted{cases}, respectively. \label{fig_SR_emission1} }  
\end{figure}  

\replaced{In Figure \ref{fig_SR_emission1}}{Using Eq.~(\ref{shockform_canto})} we compute the volume-integrated emissivity ratio of superior-to-inferior conjunction \replaced{with $p=2$}{in Figure \ref{fig_SR_emission1}}, and take $ \varsigma $ constant or $\varsigma(\theta) = \varsigma_0\left[1 -(\theta/\theta_{\rm max, X})^2 \right]$ at a fixed energy $\epsilon_f$ where the particle power-law is valid, with all numerical constant factors canceling. \added{We choose $p=2$ as a benchmark value, which is somewhat harder than implied from photon indices $\Gamma_X \sim 1.5-2$ found by \cite{2012ApJ...760...92H}.} The \deleted{latter} ad hoc prescription of surface density profile $\varsigma(\theta) $ is known from hydrodynamic bow shocks, but likely does not trace the true time-averaged emitting particle distribution.\deleted{Indeed, the underlying particle distribution even in this constant $p$ approximation is likely more concentrated near the shock nose due to transport effects and efficiency considerations as discussed in \S\ref{minitransport}.} The integration is taken to $\theta_{\rm max, X} = \pi/2$, that is, we only consider the head of the bow shock that participates for the hemisphere of the companion facing the pulsar. This portion of the shock geometry is expected to be largely axisymmetric and comprise the preponderance of accelerated charges and emission. 

We employ the results of \S\ref{eclipsesB1957} to inform the choice of $R_0$ for a given inclination in the axisymmetric case. For lower maximum bulk speeds $\beta_{\rm max} = 0.5$, \added{the left panel in} Figure \ref{fig_SR_emission1}, although there is orbital modulation it is relatively modest and flat around SC. Little to no DP structure is exhibited except when the shock stagnation point is taken at the companion surface, assumed to be 90\% of the Roche Lobe radius for B1957+20, where the influence of geometric occultation of the shock by the companion (solid curves) is larger. However, this DP morphology is rather flat and would require a large fraction of emission to be concentrated near the nose to produce the observed peak-dip ratio in \cite{2012ApJ...760...92H}, \added{and even then would not produce the correct $\approx 0.2$ phase separation of peaks}. This small value of $R_0 \approx R_*$ would also appear to be in tension with the radio eclipse estimates in \S\ref{eclipsesB1957} unless an optically-thin model for eclipses is operant. For other values of $R_0$, the influence of shadowing (solid versus dashed curves) is small especially for the more moderate inclination $i=65^\circ$. \added{In general, shadowing of \emph{unboosted} optically-thin emission can only produce a single dip, not DPs; high maximum bulk velocities, greater than $c/\sqrt{3}$, are required to establish the DP structure. Therefore, we conclude that shadowing of the emission region by the companion alone cannot explain the DP light curve features, particularly peak separation and peak width.}

\replaced{Higher bulk velocities}{Larger bulk speeds} in \replaced{Figure~\ref{fig_SR_emission2}}{the right panel in Figure~\ref{fig_SR_emission1} do} produce characteristically DP synthetic light curves\deleted{ for both shock geometries}, which are in qualitative agreement with \cite{2012ApJ...760...92H} for the DP positions. As expected, modulation increases with inclination $i$ for either shock geometry. Given the large error bars on the data in \cite{2012ApJ...760...92H} and the early development stage of our model, we do not attempt any fitting. There is a complex interplay between the shock geometry, the system inclination, the value of $R_0$ (which principally moderates the influence of shadowing), the bulk Lorentz factor at points along the shock, and the surface density profile $\varsigma(\theta)$. Smaller inclinations reduce the peak separation for a given set of parameters, which can also be mimicked by altering the prescribed shock geometry for wider opening angles. \added{Shadowing is generally a negligible influence, but may modestly enhance the dip at SC for $i=85^\circ$.} \deleted{The Type II geometry produces systematically larger orbital modulations than Type I shock geometry largely due to the larger shock opening angle, but the differences are modest. The larger opening angle of the shock also widens the peak separation and increases the peak-dip ratio. Interesting, although differences in the light curve morphology between inclinations $i=65^\circ$ and $i=85^\circ$ are noticeable for the Type I geometry, they are less so for the Type II geometry.}
 
It is clear from this quantitative parameter exploration for B1957+20 that higher bulk Lorentz factors are preferred such that $\beta \approx 0.5-0.8$ is sampled by the observer line-of-sight cutting across the shock. This implies the shock components are not well-mixed unless the companion mass loss rate is lower than expected. Some concentration of emissivity near the stagnation point where Doppler-boosting is negligible, like the $\varsigma(\theta)$ prescription, is essential to mitigate the larger than observed amplitude and peak-dip ratio. 
   
  \subsection{Application to PSR J1023+0038, an IC-centered System}
 \label{J1023_SR}
 
  \begin{figure}[t]
\centering
\includegraphics[scale=0.35]{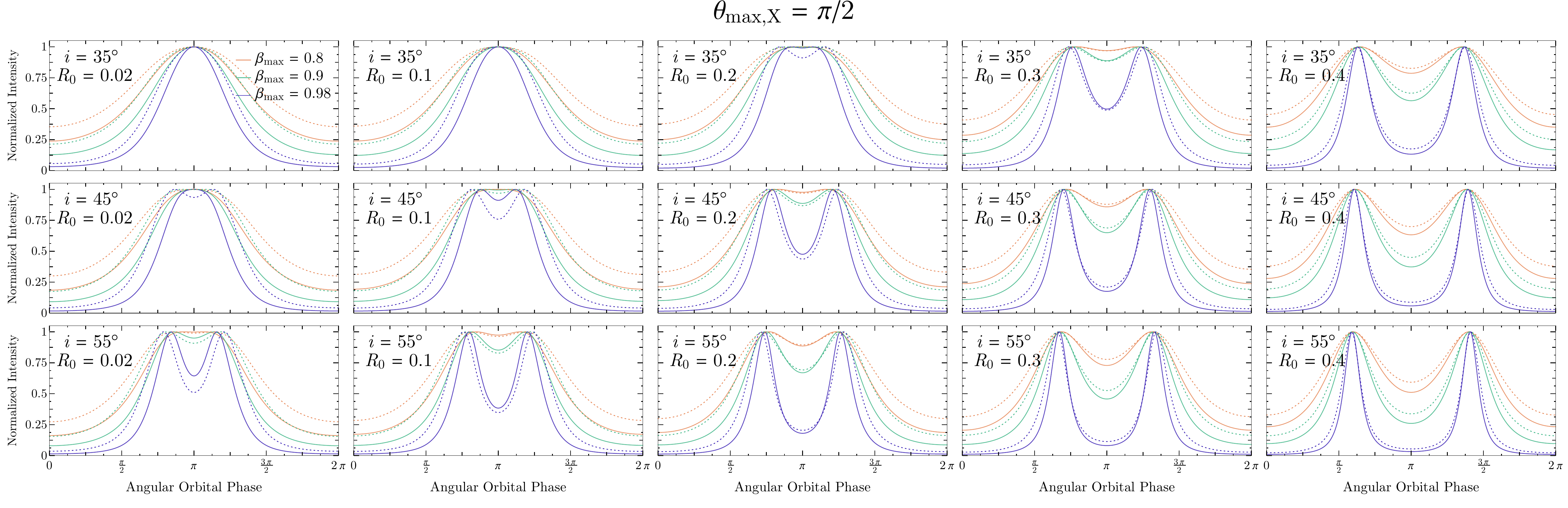}
\caption{\replaced{Type II shock s}{S}ynthetic light curves highlighting the $R_0$ and inclination dependence of the DP structure, with $\theta_{\rm max, X} = \pi/2$.  The fluxes are scaled \replaced{by a factor $\Gamma_{\rm max}^{-2.5}$}{to unity at maximum} for clarity. \added{The solid and dotted curves correspond to $\varsigma =$ constant and $\varsigma(\theta)$ prescriptions, respectively. All light curves are computed for $p=1.2$ corresponding to photon index $\Gamma_X = 1.1$ in \cite{2014ApJ...791...77T}.} Phase zero defines SC of the pulsar. \label{fig_SR_J1023_2}    }  
\end{figure}  

  \begin{figure}[t]
\centering
\includegraphics[scale=0.5]{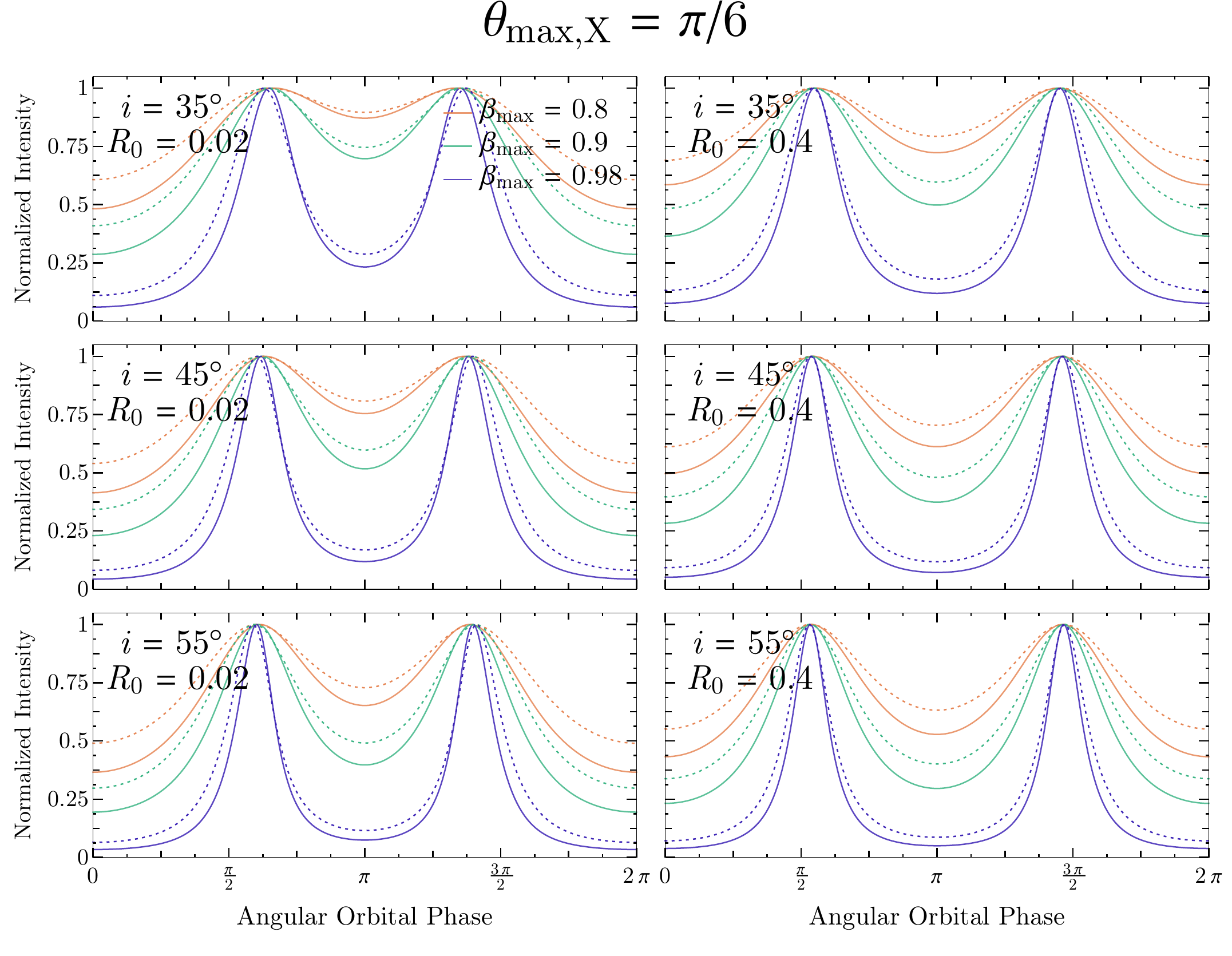}
\caption{\replaced{Type II shock s}{S}ynthetic light curves highlighting the $R_0$ and inclination dependence of the DP structure, with $\theta_{\rm max, X} = \pi/6$ that limits the emissivity to a small cap around the stagnation point. The fluxes are scaled \replaced{by a factor $\Gamma_{\rm max}^{-2.5}$}{to unity at maximum} for clarity.  \added{The solid and dotted curves correspond to $\varsigma =$ constant and $\varsigma(\theta)$ prescriptions, respectively. All light curves are computed for $p=1.2$ corresponding to photon index $\Gamma_X = 1.1$ in \cite{2014ApJ...791...77T}.} Phase zero defines SC of the pulsar. \label{fig_SR_J1023_3}    }  
\end{figure}  

\added{The observed X-ray light curve of PSR J1023+0038 is centered at pulsar IC, so that the shock must be surrounding the MSP instead of the companion. In this case, we take $R_0=0$ is at the MSP rather than at the companion.} In Figure~\ref{fig_SR_J1023_2} we compute the symmetric Doppler-boosted light curves with \replaced{$p=2$}{$p=1.2$} and $\theta_{\rm max, X} = \pi/2$, normalized to \replaced{SC (phase zero)}{peak maximum}, for a variety of inclinations and $\beta_{\rm max} = \{0.8, 0.9, 0.98\}$, the largest value corresponding to a maximum bulk Lorentz factor of about $5$. \added{This calculation constitutes a variant of Figure~\ref{fig_phaseplots}, focusing on parameters relevant to J1023+0038 restricted to $i=35-55^\circ$, and $p=1.2$ corresponding to $\Gamma_X \approx 1.1$. Moreover, the $\varsigma(\theta) = \varsigma_0\left[1 -(\theta/\theta_{\rm max, X})^2 \right]$ prescription is illustrated with dotted curves.} Since the shock surrounds the MSP, no shadowing by the companion is incorporated in the calculation since even a fully Roche lobe-filling companion for the given low inclinations $35^\circ-55^\circ$ only obscures a negligible fraction of the shock surface emissivity, and then so only near SC for emission that is Doppler de-boosted. As for B1957+20, we do not attempt any fits here, pending the development of a more complete and self-consistent model of particle transport, cooling and shock geometry, but rather explore shock geometry and physical parameters in a more generic way that will be useful to such a future study. 

\added{A wide range of shock opening angles are surveyed by the range of $R_0$ values. Smaller values of $R_0$, corresponding to narrow bow shocks, for instance of the parallel-wind type, are generally disfavored by failing to yield DP modulation even the largest prescribed value of $\beta_{\rm max} = 0.98$ except at the highest inclinations (corresponding to unusually low pulsar masses) for either $\varsigma$ constant or $\varsigma(\theta) = \varsigma_0\left[1 -(\theta/\theta_{\rm max, X})^2 \right]$. Larger values of $\beta_{\rm max}$ than considered here may allow these smaller values of $R_0$, a degeneracy in the model. The bulk Lorentz factor parameter largely controls the peak width and the depth of the dip at IC. It is evident that $R_0 \gtrsim 0.3$ produces peak separations that are generally too large; this is consistent with radio eclipses in \S\ref{J1023_radio}, where it was surmised $R_0 \lesssim 0.4$. For the intermediate inclination $i=45^\circ$ corresponding to an MSP mass of $M_{\rm MSP} \approx 1.7 M_\odot$ employing the radial velocity $K_2$ value from \cite{2015MNRAS.451.3468M}, we may constrain $R_0 \lesssim 0.3$ and $\beta_{\rm max} > 0.8$. Higher inclinations can yield DP structure with lower $\beta_{\rm max} $ values. The variation in peak separation with $R_0$ implies that decreasing or increasing peak separations may be evident in X-ray light curves of RBs promptly preceding or following an LMXB state transition, respectively.

Interestingly, for large values of $\beta_{\rm max}$, the $\varsigma(\theta)$ concentration near the stagnation point does enhance DP modulation for small values of $R_0$, but has the opposite effect for larger values. This behavior for small $R_0$ results from differential weighting and contributions of the observer line-of-sight cutting across the beamed portion of the shock, with emissivity weighting near the nose increasing the the average shock opening angle sampled by the observer. Therefore, narrow shock opening angles the light curve exhibit a degeneracy where emissivity weighting near the shock nose can effectively mimic larger opening angles, at the cost of requiring higher maximum bulk Lorentz factors. Peak separation constrains the maximum shock opening angle or $R_0$ for a given inclination. The different effect at larger $R_0$ is due to weighting the overall particle distribution leading to lower values of bulk motion, effectively lowering the impact of beaming, and hence the magnitude of modulation. Clearly, a self-consistent analysis is necessary for the emissivity distribution to disentangle model parameters.
}

\deleted{It is evident from the upper red panel of Figure~\ref{fig_SR_J1023_1} that depicts the Type I scenario that even the largest prescribed value of $\beta_{\rm max} = 0.98$ fails to yield DP modulation except at the highest inclinations (corresponding to unusually low pulsar masses) for either $\varsigma$ constant or $\varsigma(\theta) = \varsigma_0\left[1 -(\theta/\theta_{\rm max, X})^2 \right]$. Moreover, the self-similarity of Eq.~(\ref{shockI}) produces light curves that are invariant for different values of $R_0$ without shadowing. The blue panel of Figure~\ref{fig_SR_J1023_1} computes a similar suite of curves for a Type~II shock with $R_0 = 0.1$, and produces more DP modulation but again only at the largest inclinations and bulk Lorentz factors. A different $\varsigma$ that increasingly concentrates the accelerated particle distribution near the stagnation point would only diminish the DP modulation. Thus we conclude the Type~I scenario is disfavored in its parameter space for reproducing the rotation-powered state DP light curves of J1023+0038 as reported in \cite{2010ApJ...722...88A} and \cite{2014ApJ...791...77T} and similar IC-centered DP systems with moderate-to-low inclinations, and that $R_0 > 0.1$ in the Type II scenario if $i < 55^\circ$. Since the Type I and II shock geometries converge at $R_0 \ll1$, we conclude that the shock of J1023+0038 in its past rotation-powered state was at distances large compared to the pulsar light cylinder and required a stabilizing mechanism on such orbital scales.}

\deleted{From radio eclipses in \S\ref{J1023_radio}, we surmised $R_0 \lesssim 0.4$. We explore larger values of $R_0$ in Figure~\ref{fig_SR_J1023_2} for the same $\beta_{\rm max}$ values as Figure~\ref{fig_SR_J1023_1}. Although an unmodulated DC component or $\varsigma$ distribution ought to be combined with these light curves for fits, it is evident that the bottom row corresponding to $R_0 = 0.4$ produces peak separations that are generally too large. However, such large peak separation may be mitigated by a $\varsigma$ distribution that dominates emission near the shock stagnation point, or if the real shock geometry is not as flat, relative to the line joining the two stars. Significant modulation at lower inclinations still requires relatively high speeds $\beta_{\rm max}$. For the intermediate inclination $i=45^\circ$ corresponding to a MSP mass of $M_{\rm MSP} \approx 1.7 M_\odot$ emplying the radial velocity $K_2$ value from \cite{2015MNRAS.451.3468M}, we may constrain $0.2 \lesssim R_0 \lesssim 0.4$ and $\beta_{\rm max} > 0.8$. Higher inclinations can yield DP structure with lower $\beta_{\rm max} $ values. The variation peak separation with $R_0$ implies that decreasing or increasing peak separations may be evident in X-ray light curves of RBs promptly preceding or following a LMXB state transition, respectively.}

Finally, in Figure~\ref{fig_SR_J1023_3} we consider restricting $\theta_{\rm max, X} = \pi/6$, only considering emission from this small shock cap rather than the full head as in Figure~\ref{fig_SR_J1023_2}. This scenario approximates one where emission from the shock is extraordinarily concentrated near the stagnation point. It is clear that such a restriction generally results in too-wide peak separation for any inclination\added{, $\varsigma$ distribution,} or $R_0$ values, principally due to the local flatness of this region. \added{There is little change in light curve shape with different values of $R_0$. The emitting region is geometrically similar to large values of $R_0 \gtrsim 0.3$ with $\theta_{\rm max,X} = \pi/2$, highlighting a degeneracy in parameters for ``flat" shocks. However, this degenerate regime is ruled out by the large peak separation.} Therefore, the emitting region in J1023+0038 and similar IC-centered systems must include synchrotron contributions from larger $\theta_{\rm max, X}$ values where the shock significantly curves or bows, consistent with large radio eclipse fractions enshrouding the MSP. 

\added{
\subsection{Discussion: Caveats and Future Refinements}
\label{caveats}

\subsubsection{The Thin-Shell Approximation}
\label{thinshell}

An objection to the formalism we have presented may be the use of the thin-shell approximation, especially when known X-ray bow shocks in PWNe do not exhibit geometrically-thin morphologies traced by radiatively cooling electrons. Simulations, as well as observed pulsar bow shocks \cite[e.g.,][]{2014ApJ...784..154B} typically show a thickness of order $\delta R_0/R_0 \sim 30\%$ near the head of the shock with increasing thickness far from the head. The thickness is highly dependent on the physical conditions such as the in situ MHD $\sigma \lesssim 1$ as well as transport of the tracers within the medium. It is unclear if there is a mapping between those parsec-scale shocks and more compact shocks in MSP binaries for the relative geometric thickness of radiating electrons distribution. In particular, the relative ordering of Bohm diffusion, cooling, collisionsional and convective/adiabatic timescales for charges may be different. Nevertheless, in MSP binaries, the shocked pulsar wind component will have some geometrical thickness, at least on the order of the electron/positron Larmor radius but perhaps greater depending on the internal macroscopic fluid/magnetic pressure and level of mixing with ions. Hydrodynamically, the thin-shell approximation is only valid in the highly-radiative momentum-dominated limit, which is realized when $\sigma \ll1$, i.e., where pressure terms in the Euler equation are omitted compared with velocity terms. For low bulk velocities near the shock nose, this approximation is clearly overstepped. However, this low-speed component near the nose is also not Doppler-boosted, barely impacting the X-ray modulation and therefore constitutes a DC or background-level offset to the X-ray modulation. 

Since we are principally concerned with DP light curves, only the radiative portion of the electron population, whose photons are then Doppler-boosted, are consequential in flux ratio Eq.~(\ref{LSRpropto}). The spatial distribution of this leptonic population is poorly understood but may be geometrically thinner than $\delta R_0/R_0 \sim 30\%$, since the coherence of the observed DP morphology implies some coherence in bulk motion towards the observer line-of-sight at mildly relativistic speeds at orbital phases where the two peaks are observed. This is crucial and essential in the model, or else DPs are not tenable. The narrowness of the peak widths is then a constraint on the relative geometric thickness of the pressure-confined shell, since a large angular spread in velocity streamlines would smear out the DP morphology. That is, the observed peak phase width of $\sim 0.1$ implies a similar limit on the geometric thickness $\delta R_0/R_0 \lesssim 0.1 $ for the boosted component, and perhaps substantially smaller since each velocity streamline also introduces an intrinsic width in the light curves, a strong function of $(\beta \Gamma)_{\rm max}$ (cf. Figure~\ref{fig_phaseplots}). Therefore, presently, geometric thickness of the radiative population may be viewed as a dispensable supplemental complexity. Certainly, the formalism above is amenable to an ensemble of thin-shells with the cost of additional free parameters --  a multi-zone model may embrace an arbitrary assemblage of multiple single-zone thin-shells. With a single shell, quantities such as the bulk Lorentz factor and shock density profile $\Sigma_e$ are to be interpreted as ensemble averages in this ``one-zone" model. A natural extension is a two-zone model, for two different electron populations separated by the contact discontinuity where one is baryon-loaded and attains bulk velocities much lower than the relativistic adiabatic sound speed $ \lesssim c/\sqrt{3}$. Such an exploration is deferred for future work.

\subsubsection{Diagnosing Spatially-Dependent Relativistic Particle Acceleration From Light Curves}
 \label{partaccdiag}

Any spatial variation of the index $p \rightarrow p(\theta, \phi)$ induces energy dependence of the light curves in Eq.~(\ref{LSRpropto}). As discussed previously, the upstream relativistic shock geometry is quasi-perpendicular at the stagnation point relaxing to quasi-parallel at higher bulk velocity locales. Accordingly, the field compression, reconnection, the particle density, and the acceleration spectral index $p$ probably vary from locale to locale, deviating from the canonical value of $p\approx2$ in nonrelativistic DSA or $p=2.23$ for ultrarelativistic parallel shocks \citep{2000ApJ...542..235K}, steepening in locales where acceleration is inefficient. Such index variation will be convolved with spatially-dependent contributions from particle cooling. The distribution $\varsigma$ serves as an accessible diagnostic of the spatial dependence underlying the nonthermal lepton population, shock acceleration, and post-shock magnetic field. The winding of the pulsar wind in a Parker spiral must yield different field obliquities at different angles $\theta$ along the shock surface.  These obliquities will depend on the tilt of the pulsar's spin axis to the orbital plane, and importantly, on the plasma dynamics of its magnetosphere: the morphology of the field lines emanating from the pulsar surface is influenced by the MHD conductivity of the wind \citep{2014ApJ...781...46C,2014ApJ...793...97K}.  The field obliquity angle along the shock interface can have a critical impact upon the acceleration process.  First, if the field is dynamically important, it can alter the curvature of the shock surface as well as the level of field compression and the field orientation downstream.  Next, the index $p$ is sensitive to the choice of the field obliquity in mildly-relativistic shocks, and the power-law index resultant from diffusive (Fermi-like) and shock drift acceleration is generally less than $p=2$ as long as the shock interface is subluminal \citep{2012ApJ...745...63S}.  Such circumstances will likely occur at significant angles $\theta$ away from the nose or stagnation point of the shock hyperboloid.  Thus, unless the microstructure of the field at the nose is highly turbulent, the nose locale should provide a fairly steep particle distribution that flattens as $\theta$ increases, with the index eventually declining at the lateral extremities of the shock because its MHD compression ratio will decline. Kinetic turbulence will also impart a spatial dependence on the accelerated spectrum, a function of the spatially-dependent turbulent cell lengthscale \citep[e.g.,][]{2016arXiv160904851Z}. There are clearly a number of parameters involved in describing such complexity. 
 
Asymmetries and energy dependence of structure in light curves also serve as a probe of the underlying particle distribution. Coriolis effects near the stagnation point break axisymmetry but are only relevant if the slow baryonic component dominates the dynamics and geometry of the mildly relativistic shocked pulsar component. Observe that inclination of the pulsar spin axis to the normal to the orbital plane may also generate an asymmetric bow shock structure, leading to asymmetric modulation of the observed flux. Analogous to the Vela pulsar in $\gamma$-rays \citep{2010ApJ...713..154A}, energy dependence in ratio of the main peaks and widths may be evident.  Any energy dependence of light curve asymmetry therefore is a signature of differential index variation rather than simple geometric asymmetry in the shock geometry.

To constrain and probe such influences in the future demands high fidelity phase-resolved spectroscopy of light curves. There exists an integral equation for the flux for each observed energy bin constructed from Eqs.~(\ref{SR_j})--(\ref{deltaD}) that is fairly involved.  We defer such an exploration for when such data are available, and note that such an observational program on multiple systems would potentially be a remarkable astrophysical probe of particle acceleration in oblique relativistic shocks. 
 
 \subsubsection{Phase-Dependent Synchrotron Spectral Breaks and Cut-offs}
 \label{minitransport}

A secondary source of spectral index $p$ variation and energy dependence far beyond the classical X-ray band may arise from transport phenomena for the steady-state particle distribution $\bar{n}_e$.  Such influences could lead to a different origin of spectral curvature, similar in origin to the observed spectral cooling breaks in GRBs and AGNs but in an orbital-phase-dependent manner, since the selection of Doppler factors that modify the spectral elements is dependent on the observer's viewing perspective. The underlying particle distribution and transport are constrained by where such spectral breaks occur and whether they are dependent on orbital phase. Synchrotron cooling is the dominant energy loss mechanism at large Lorentz factors $\gamma_e \gtrsim 10^4$, and if the shock acceleration is efficient at the gyroscale like the nebular shock in the Crab \citep{1996ApJ...457..253D}, the radiation-reaction limited synchrotron exponential cut-off is independent of magnetic field at $\sim 160$ MeV in the comoving frame; this is the rough upper limit for energy unless extremely fast acceleration processes are occurring. Therefore, for the binary systems considered here, the convolution of Doppler shifts from different shock locales will instill only a modest smearing in this spectral turnover in the gamma-ray waveband.  Moreover, this will not impact our light curve determinations, which are germane to lower frequencies.  At the other end of our broadband spectral window, simple estimates of the synchrotron self-absorption frequency \citep{1979rpa..book.....R} indicate that it is well into the radio band. This is quickly discerned for typical number densities $\langle n_e \rangle \sim 10^{6}-10^9$ cm$^{-3}$ and length scales $R_0 \sim 10^8-10^{10}$ cm.

Although we do not attempt self-consistent transport calculations in this paper, let us briefly discuss its general aspects. The radiative efficiency is governed by the bulk convective timescale $\tau_{\rm conv} \sim R_0 a/ (\beta c)$ in comparison to the radiative timescale $\tau_{\rm rad}$ in the comoving frame,
\begin{equation}
\tau_{\rm rad} = \frac{3 m_e c}{4 \sigma_T \gamma_e u} \approx 6 \left( \frac{10^6}{\gamma_e} \right)  \left( \frac{5 \, \, \rm erg \, cm^{-3}}{u} \right) \quad \rm s \, ,
\end{equation}
where the energy density $u = u_{\rm rad} + u_{\rm B} \sim 5$ erg cm$^{-3}$ with $u_{\rm B} = \bar{B}_s^2/(8 \pi)$ and $u_{\rm rad} \approx \langle \Gamma \rangle^2 \eta \dot{E}_{\rm SD}/(4 \pi R_0^2 c)$ (efficiency $\eta <1$). The convective timescale is large near the stagnation point, steadily decreasing according to Eq.~(\ref{betamax}) and bounded from below $\tau_{\rm conv} > \tau_{\rm conv, min} = R_0 a/c$. The radiative efficiency is, crudely, $\tau_{\rm eff}/\tau_{\rm rad} =(1+\tau_{\rm rad}/\tau_{\rm conv})^{-1}$ where $\tau_{\rm eff}^{-1} = \tau_{\rm rad}^{-1} + \tau_{\rm conv}^{-1}$ is the effective timescale \citep{1997ApJ...477..439T}. The radiative efficiency $\tau_{\rm eff}/\tau_{\rm rad} \ll 1$ establishes the slow cooling locale of Figure~\ref{geometry_schematic}, while large convective timescales near the stagnation point naturally yields efficient cooling for even modest Lorentz factors well below $10^6$. The inefficient cooling in the wings of the shock may seem detrimental to the model, but observe the Doppler-boosting in flux in Eq.~(\ref{SR_transform}) grows faster than the decline in comoving radiative efficiency for orbital phases and shock locales where observer beaming is significant, assuming $n_e \sim Q_e \tau_{\rm eff}$ for a spatially-independent injection rate $Q_e$. That is, $\beta(\theta)$ in $\tau_{\rm eff} \approx \tau_{\rm conv}$ is slower growing than $\delta_D^{2+(p-1)/2}$ when $\boldsymbol{\hat{n}}_v \boldsymbol{\cdot} \boldsymbol{\hat{u}}^\prime  \approx 1/\beta$. Alternatively, one may view it as a constraint for the unknown spatial acceleration/injection rate $Q_e (\gamma_e)$, which cannot wane along the head of the shock faster than compensated by the Doppler-boosting influences. Note that the surface area element corresponding to the shock nose is also much smaller than those associated with the wings. The fast-cooled locale is also not significantly Doppler-boosted, therefore not impactful on the background-normalized light curve morphology, but does influence cumulative spectral determinations. Accordingly, at the wings of the shock, $\tau_{\rm rad} = \tau_{\rm conv} > \tau_{\rm conv, min}$ defines a break Lorentz factor below which radiative cooling is inefficient, which dominates the Doppler-boosted light curves,
\begin{eqnarray}
\gamma_{e, \rm break} &\lesssim& \frac{3 \, m_e c^2}{4 R_0 a \, \sigma_T \, u} \approx 5 \times 10^6  \left(\frac{ 3\times 10^{10} \, \rm cm}{R_0 \, a} \right)  \left(\frac{ 5 \, \rm \, erg \, cm^{-3}}{u} \right).
\end{eqnarray}
This Lorentz factor is somewhat lower than the multi-TeV-scale $\gamma_{\rm max} \sim 10^{8}$ values expected from a purely gyroscale acceleration radiation-reaction limited scenario in a $\sim 1-100$ G field. The concomitant characteristic comoving break energy $\bar{\epsilon}_{c, \rm break} = 3 \bar{B}_s/(2 B_{\rm cr}) \gamma_{e, \rm break}^2$ may be well beyond the classical soft X-ray band,
\begin{eqnarray}
\epsilon_{f, \rm break} &=& \bar{\epsilon}_{f, \rm break} \delta_{\rm D} \lesssim  \frac{27 (m_e c^2)^2 }{32 R_0^2 a^2 \sigma_T^2 u^2 } \left(\frac{\bar{B}_s}{B_{\rm cr}}\right) \delta_{\rm D} \label{coolingbreak} \\
&\approx& \delta_{\rm D} \frac{ 10  \, \rm MeV}{m_e c^2} \left(\frac{10 \, \rm G}{ \bar{B}_s} \right)^3  \left(\frac{ 3\times 10^{10} \, \rm cm}{R_0 \, a} \right)^2 \nonumber \quad  , \quad u_{\rm B} \gg u_{\rm rad}.
\end{eqnarray}
Although the characteristic comoving break energy is independent of observer perspective, the Doppler factor dependence of Eq.~(\ref{coolingbreak}) introduces an orbital and geometrical influence to the break energy at any particular emission point along the shock. Thus, the Doppler factor must be spatially convolved with the shock emission region, introducing factors of unity variation in the cooling break and smearing any sharp spectral transition.

Observe that in the context of IC-centered systems the break energy Eq.~(\ref{coolingbreak}) scales as $r_s$ using Eq.~(\ref{upstreamB}), although with a $\sigma$ dependence for the post-shock magnetic field $B_s$ that depends on the kinetic-scale physics in the shock. For J1023+0038, no phase-averaged spectral cut-off is seen by NuSTAR up to $50-80$ keV. Assuming $B_s \sim \sqrt{\sigma} B_w$ with $\sigma \approx 10^{-2}$ results in $r_s \gtrsim 10^8$ cm for the shock radius constraint in J1023+0038, and requires $\sigma < 1$ for $r_s < a \approx 10^{11}$ cm. During transient flaring optical states of the companion, the temporary intensification of a dominant Compton cooling for Lorentz factors $\gamma_e \lesssim 10^5$ will reduce the characteristic synchrotron break energy as $u_{\rm rad}^{-2} \sim T_{\rm hot}^{-8}$. Such spectroscopic features, phase-dependent, transient, or correlated may be observable for some bright MSP binaries with NuSTAR and a future next-generation hard X-ray to soft $\gamma$-ray Compton telescope \citetext{e.g., ComPair: \citealp{2015arXiv150807349M}; e-ASTROGAM: \citealp{2016arXiv161102232D}}.

}

%% file: Eclipses_Appendix_input.tex
\section{Eclipses by Surfaces in Binary Systems}
\label{appendix_A}

\subsection{General Formalism}

Although the focus of this text is on eclipses by optically-thick intrabinary shocks, the analytical formalism here can also be applied to arbitrary azimuthally symmetric surfaces that occult a point source. The case of eclipses by quasi-static Roche lobes has been treated analytically in \cite{1976ApJ...208..512C} and \cite{1959cbs..book.....K}, however the method presented here is more broadly applicable. 

Without loss of generality, we prescribe a right-handed orthonormal cartesian coordinate system, in flat spacetime, such that the barycenter is the origin and the orbital angular momentum vector of the binary is in the $\boldsymbol{\hat{z}}$ direction with binary angular frequency $\Omega_{\rm b}$.  The observer angle is defined such that $\boldsymbol{\hat{z} \cdot \hat{n}_v} = \cos i$ with $\boldsymbol{\hat{n}_v} = \cos i \boldsymbol{\hat{z}} + \sin i \boldsymbol{\hat{x}}$ the observer line-of-sight unit vector. To obtain the projection of the binary system into the plane of the sky perpendicular to $\boldsymbol{\hat{n}_v}$, we construct a new primed rotated coordinate system such that $\hat{\bf{x^\prime}}$ is parallel to $\hat{\bf{n}}_v$ at arbitrary orbital phase. If the cartesian basis $ \boldsymbol{\hat{r}} \rightarrow (\hat{\bf{x}},\hat{\bf{y}},\hat{\bf{z}})^\intercal$ defines the vector space at phase $\Omega_{\rm b} t = 0$ and inclination $i = \pi/2$, then the primed orthonormal coordinate basis for arbitrary orbital phase and inclination is constructed by two successive rotations. Here the inclination is formulated such that $i = 0$ and $i=\pi/2$ are face-on and edge-on views, respectively, and with the phase convention $\Omega_b t =0 $ prescribing the superior conjunction where the companion is between the MSP and observer in Eq.~(A3) of the orbital equations of motion. The rotations are given by
\begin{equation}
\boldsymbol{\hat{r}}^\prime = \Lambda_i \Lambda_{\Omega_b t}  \boldsymbol{\hat{r}},
\label{rotation}
\end{equation}
where
\begin{eqnarray}
\Lambda_{\Omega_b t} &=&
\left( \setlength\arraycolsep{3pt}
 \begin{array}{ccc}
\cos \left( \Omega_{\rm b} t \right) & -\sin \left( \Omega_{\rm b} t \right) & 0 \\
\sin \left( \Omega_{\rm b} t \right) & \cos \left( \Omega_{\rm b} t \right) & 0 \\
0 & 0 & 1
\end{array}
\right) \qquad \qquad
\Lambda_i = \left(
 \setlength\arraycolsep{3pt}
 \begin{array}{ccc}
\sin i & 0 & \cos i \\
0 & 1 & 0 \\
- \cos i & 0 & \sin i
\end{array}   \right)
\end{eqnarray}
define the primed coordinate basis about the barycenter such that $\hat{\bf{x^\prime}} = \hat{\bf{n}}_v$, with the span of $\hat{\bf{y^\prime}}$ and $\hat{\bf{z^\prime}} $ characterizing the plane of the sky.

The stars are treated as following unperturbed Keplerian trajectories, although a post-Newtonian generalization is straightforward, with phase zero constructed to rest on the $\hat{\bf{x}}$-axis with the following equations of motion,
\begin{eqnarray}
\textrm{MSP (primary): } \boldsymbol{r}_{\rm NS}(t) &=& -r_{\rm NS}(t) \left[ \cos\left( \Omega_{\rm b} t \right) \boldsymbol{\hat{x}} + \sin\left( \Omega_{\rm b} t \right) \boldsymbol{\hat{y}} \right] \\
\textrm{Companion (secondary): } \boldsymbol{r}_{\rm c} (t) &=& r_{\rm c}(t)\left[ \cos\left( \Omega_{\rm b} t \right) \boldsymbol{\hat{x}} + \sin\left( \Omega_{\rm b} t \right) \boldsymbol{\hat{y}} \right]. \nonumber
\end{eqnarray}
The scalars $r_i(t)$ for non-zero eccentricity depend on the orbital phase,
\begin{equation}
r_i(t) = \frac{r_i (1 - \epsilon^2)}{1 + \epsilon \cos\left( \Omega_{\rm b} t \right)} , \qquad r_i = r_{\rm c}, r_{\rm NS}
\end{equation}
where $r_{\rm c} + r_{\rm NS} \equiv 1$ (normalized to $a$) are the respective semi-major axes for each star, with $r_{\rm c} = q/( q +  1)$ and $r_{\rm NS} = 1/( q +  1)$ where $q = M_{\rm MSP}/M_{\rm c}$ is the mass ratio. For nonzero eccentricity, which is beyond the scope of this paper, an additional rotation of coordinates by the argument of periastron must be incorporated in Eq.~(\ref{rotation}).

Rotating coordinates to the primed coordinate basis by Eq.~(\ref{rotation}), the position vectors of the primary and secondary are given by
\begin{eqnarray}
\boldsymbol{r}_{\rm NS}^\prime (t) &=& r_{\rm NS}(t)\left[ - \sin i \cos\left( \Omega_{\rm b} t \right) \hat{\bf{n}}_v -\sin\left( \Omega_{\rm b} t \right) \hat{\bf{y^\prime}} + \cos i \cos\left( \Omega_{\rm b} t \right) \hat{\bf{z^\prime}}\right] \\
\boldsymbol{r}_{\rm c}^\prime(t) &=& r_{\rm c}(t)\left[\sin i \cos\left( \Omega_{\rm b} t \right)  \hat{\bf{n}}_v +\sin\left( \Omega_{\rm b} t \right) \hat{\bf{y^\prime}} - \cos i \cos\left( \Omega_{\rm b} t \right) \hat{\bf{z^\prime}}\right] \nonumber
\end{eqnarray}
with a parallel/orthographic projection on the $  {\rm Span }\{ \hat{\bf{y^\prime}} , \hat{\bf{z^\prime}} \}$ plane comprising the observer view. 

For orbital angular speeds $a \, \Omega_{\rm b} \ll c$, the coordinates of a vector $\boldsymbol{\ell}$ defining a ray leaving the primary towards the observer traversing a distance $l$ (in units of $a$) can be expressed as
\begin{equation}
\boldsymbol{\ell}^\prime = \boldsymbol{r}_{\rm NS}^\prime (t)  + l \hat{\bf{n}}_v \, ,
\end{equation}
with the distance between an interaction point at the ray and an arbitary location in the system $\boldsymbol{r}_{\rm m}^\prime$ (e.g., the secondary) given by $\left| \boldsymbol{\ell}^\prime - \boldsymbol{r}_{\rm m}^\prime \right|$. The optical depth for a specified absorption coefficient $\alpha$ is computable in the usual way through the observer line-of-sight integral at each orbital phase, $\tau = \int \alpha dl$.  Such a procedure for pulsar eclipses by specified volumes has been performed in the context of MSP binaries \citep{1989ApJ...342..934R} and other pulsar systems \citep[e.g.,][]{2005ApJ...634.1223L} but relies on a physical model underpinning the absorption coefficient, and is computationally inefficient for arbitrary volumes when the optical depth is large for the region of interest, and when the transition region of low-to-high optical depth is sharp. In the case of optically thick surfaces and volumes, we employ a geometric occultation formalism described below.
 
For a surface with azimuthal symmetry with radial function $R(\theta)$, every locus of points at fixed $\theta$ is a circle. When projected onto the $ {\rm Span } \{ \hat{\bf{y}}^\prime , \hat{\bf{z}}^\prime \}$ plane of the sky, these circles are transformed to ellipses by the elementary rotations of orbital phase and inclination angle. The definition of eclipses by a bow shock is thus transformed to the problem of determining if the projected location of the MSP is interior to any projected bow shock ellipse, for all prescribed $\theta$. \replaced{In the real system with an intrabinary shock where the flow velocity is finite, there is a $\theta_{\rm max,R}$ where the bow shock approximation no longer holds due to hydrodynamic instabilities as well as the orbital motion.}{We define $\theta_{\rm max,R}$ as the maximum polar angle where the shock surface transitions from optically-thick to thin at a given frequency.} The maximum length along the axis of symmetry of the shock tail past the companion position $L_z(\theta_{\rm max,R}) = - R(\theta_{\rm max,R}) \cos\theta_{\rm max,R}$ should be smaller than the typical orbital length scale $a$ for flow velocities that are similar in scale to the orbital speed. \added{These relatively slow flow velocities are interpreted to originate from the nonrelativistic baryonic component of the companion wind, rather than the leptonic pulsar wind, with the two winds separated by a contact discontinuity.}

The locus of points for a particular projected ellipse, given the conventions developed above, can be found to be
\begin{eqnarray}
\boldsymbol{E} (t, \theta, \phi) &=& (\alpha_y + \kappa_y \sin \phi )  \hat{\bf{y^\prime}} + (\alpha_z + \beta_z \cos \phi + \kappa_z \sin \phi) \hat{\bf{z^\prime}} ,
\end{eqnarray}
with
\begin{eqnarray}
\alpha_y &=& S(t, \theta)  \sin\left( \Omega_{\rm b} t \right) \nonumber \\
\alpha_z &=& -S(t, \theta) \cos\left( \Omega_{\rm b} t \right) \cos i \nonumber \\
\beta_z &=& R(\theta) \sin\theta \sin i \label{ellipseeq} \\
\kappa_y &=& R(\theta) \sin\theta \cos\left( \Omega_{\rm b} t \right) \nonumber \\
\kappa_z &=& R(\theta) \sin\theta \sin\left( \Omega_{\rm b} t \right) \nonumber ,
\end{eqnarray}
where $\phi \in (0,2\pi)$ is the azimuthal angle parameterizing the azimuthally-symmetric surface. The function $S(t, \theta)$ defines whether the shock surface surrounds the primary or secondary with the orbital phase convention defined above: for surrounding (bowing around) the secondary it takes the form $S_{\rm c} = r_{\rm c}(t) - R(\theta) \cos \theta$ but $S_{\rm NS} = -r_{\rm NS}(t) + R(\theta) \cos \theta$ for where the shock enshrouds the pulsar.

 With these definitions, it is evident that the center of the ellipse is then $\{\alpha_y,\alpha_z\}$. In general, the ellipses' axes of symmetry are rotated by an angle with respect to the $\hat{\bf{y}}^\prime$ or $\hat{\bf{z}}^\prime$ axes. The semi-major $E_a$ and semi-minor $E_b$ axes lengths are easily shown to be
\begin{eqnarray}
E_a &=& \left | R(\theta) \sin \theta \right | \\
E_b &=& \left | R(\theta) \sin \theta \sin i \cos (\Omega_{\rm b} t ) \right | , \nonumber
\end{eqnarray}
for a given inclination and orbital phase and, for a given $\theta$. The projected eclipsed area is  $ \propto R(\theta)^2 \sin^2 \theta \sin i \cos (\Omega_{\rm b} t ) $. Using elementary methods, the angle between an ellipse's semi-major axis and the $\hat{\bf{y}}^\prime$ direction is found to be
\begin{equation}
\cos \vartheta_E = \frac{2 \left| \cos \left( \Omega_{\rm b} t \right) \right| \cos i}{\sqrt{3 + \cos (2 i) - 2 \cos \left(2 \Omega_{\rm b} t \right) \sin^2 i}} \, ,
\end{equation}
for phases $-\pi/2 \leq  \Omega_{\rm b} t  \leq \pi/2$. 

These relations then allow us to analytically test whether the MSP's projected coordinate is eclipsed by an ellipse. Rotating and translating coordinates, we construct new axes aligned with the ellipse with origin at the ellipse center. The coordinates of the MSP are then given by
\begin{eqnarray}
G_{\rm x} (t) &=& \left[\boldsymbol{r}_{\rm NS}^\prime\boldsymbol{ \cdot}  \hat{\bf{y^\prime}} - \alpha_y \right] \cos \vartheta_E + \left[\boldsymbol{r}_{\rm NS}^\prime\boldsymbol{ \cdot}  \hat{\bf{z^\prime}} - \alpha_z \right]\sin \vartheta_E \\
G_{\rm y} (t) &=& -\left[\boldsymbol{r}_{\rm NS}^\prime\boldsymbol{ \cdot}  \hat{\bf{y^\prime}} - \alpha_y \right] \sin \vartheta_E + \left[\boldsymbol{r}_{\rm NS}^\prime\boldsymbol{ \cdot}  \hat{\bf{z^\prime}} - \alpha_z \right] \cos \vartheta_E \nonumber \, ,
\end{eqnarray}
which can be used to test if the projected position of the MSP is eclipsed. We define a unit Heaviside test function $\Theta(\Gamma)$ \replaced{where}{with argument}
\begin{equation}
\Gamma = \left(\frac{G_{\rm x}}{E_a}\right)^2 + \left(\frac{G_{\rm y}}{E_b}\right)^2 -1 \, ,
\label{heaviside_gamma}
\end{equation}
is less than zero for eclipses. Hence the position $G_{\rm x,y}$ and parameter $\Gamma$ can be associated with a physical attenuation (and $\tau$) at any given $\theta$, i.e., $e^{-\tau} \sim f(G_{\rm x},G_{\rm y})$ for some function $f$; we choose $f$ to be a unit step function in our optically thick formalism. Then, $\Theta(\Gamma)$ is unity when the pulsar is not eclipsed. The eclipsing by the entire ensemble of ellipses $j_{\rm max}$ that encompass the employed surface is a product of $\Theta$ for every $\theta$ up to $\theta_{\rm max,R}$, suitably discretized to sample the complete surface,
\begin{equation}
\Theta_{\rm tot} = \prod_{j}^{j_{\rm max}} \Theta [ \Gamma(\theta_j)].
\label{tmax_product}
\end{equation}

For the case where the surface surrounds the emission point source and is optically-thick, a lower limit of the shrouding fraction can be found by considering the condition Eq.~(\ref{tmax_product}) evaluated only at $\theta_{\rm max,R}$. This yields a lower limit because a general surface may have pockets of plasma that may occlude the emission source in a nontrivial manner. For cases where the bow shock is not swept back and the curvature is positive everywhere, the preceding construction is unnecessary, since the shrouding fraction is analytically expressible by Eq.~(\ref{RBshrouding}).

It is numerically expedient to adopt a continuous approximation of the Heaviside function, $\Theta[\Gamma(\theta_j)] \approx 1/(1+e^{- {\cal T} \Gamma(\theta_j) })$, where ${\cal T}$ is an associated pseudo optical depth that parametrizes the geometric transition thickness near the occluding surface $R(\theta)$ and approximates a sharp transition for ${\cal T} \gg 1 $ for a fixed observational frequency. This transforms the product to a sum,
\begin{equation}
\Theta_{\rm tot} \approx \exp\left[{- \sum_j^{j_{\rm max}} \log \left(1+ e^{- {\cal T} \Gamma_j} \right)} \right] \qquad {\cal T} \gg 1.
\end{equation}
 
Because of the symmetry in the problem, the full orbit and other branches for the solution of $\cos \vartheta_E$ need not be considered, and the calculation can be restricted to  one quarter of the total orbit. The total eclipse fraction $f_E$ by the surface during a total duration $2 \pi$ of an orbit is then given by the integral,
\begin{equation}
f_E = \frac{1}{\pi} \int_0^{\pi/2} d(\Omega_{\rm b} t) \left[1- \Theta_{\rm tot} (\Omega_{\rm b} t)\right]
\label{fE_equation}
\end{equation}
for a given set of Keplerian orbital parameters and radial function $R(\theta)$. 

\subsection{``Cometary" Tail Sweepback Due to Orbital Motion}
\label{appendix_tail}

If the locus of points of the shock flow is azimuthally symmetric relative to the companion, i.e., axisymmetric at each instantaneous orbital phase then it is straightforward to include the sweep-back and Coriolis forces due to orbital motion by retarding the surface at each $\theta$ by a specified time delay. We neglect Coriolis effects perpendicular to the outward direction, equivalent to the assumption that $v_\perp \ll v_{\rm inj}$ where $v_\perp$ is the bulk flow velocity perpendicular to the line connecting the two stars; this approximation, which preserves axisymmetry of the shock, is generally valid for locales distant from the stagnation point, the principal focus of this work associated with radio eclipses. Connecting orbital phases to the surface parameter $\theta$ then suffices to produce a good approximation of such swept-back shock tails, and the eclipse fraction formalism of the previous section can be expediently utilized. This ballistic assumption is valid if the shock bulk velocity is highly supersonic, as in MSP binaries, especially if the winding number of the swept-back tail that contributes to the eclipses is zero. \added{The assumption  $v_\perp \ll v_{\rm inj}$ requires $R_0 \ll 0.1$ for the isotropic wind geometry Eq.~(\ref{shockform_canto}), but is satisfied to a good degree for $\theta \gtrsim \pi/2$ for the self-similar parallel-wind geometry of Eq.~(\ref{shockform_wilkin}), with the approximation becoming increasingly better for larger $\theta$ or longer tails. Therefore, the validity of this approach is restricted to the parallel-wind geometry explored in Appendix \ref{parawindappendix}.}

The model we adopt here simply injects the shock flow at a finite velocity $v_{\rm inj}$ near the companion at $\theta=\pi/2$, assumed much greater than the escape velocity $v_{\rm inj } \gg v_{\rm esc} \sim 10^7$ cm s${}^{-1}$ which, coincidently, is also the same order of magnitude as the orbital velocity $v_{\rm orb}$, so that gravitational effects can be neglected for BWs. More complex self-consistent prescriptions for the bulk flow can also adopted in this framework, but may introduce more parameters. The time delay for nonrelativistic velocities and finite constant acceleration $a_{\rm wind}$ (in the lab nonrotating frame) due to radiation pressure from the pulsar wind is
\begin{equation}
\delta t = \frac{-v_{\rm inj} + \sqrt{2 a_{\rm wind} a L_z(\theta) +  v_{\rm inj}^2}}{a_{\rm wind}} \, ,
\end{equation}
which reduces to $a L_z / v_{\rm inj}$ for small flow accelerations. For zero acceleration, the approximate tail angle  with respect to the line connecting the two stars is then $\arctan (v_{\rm orb}/v_{\rm inj})$, good when $v_{\rm inj} \gg v_{\rm orb}$. For large $a_{\rm wind}$ or $v_{\rm inj}$, the time delay approaches zero and the symmetric shock solution is recovered.  Here $L_z (\theta)= - R(\theta) \cos\theta $ connects the time delay to points along the shock surface. Thus, generically, increasing asymmetry increases the total eclipse fraction at a given $\theta_{\rm max,R}$. \deleted{In Figure~\ref{Ballistic_grid} we show two different cases of constant velocity and constant acceleration, and it is evident that acceleration of the flow can have dramatic consequences on the geometry and eclipse asymmetry depending on the parameters $L_z, a_{\rm wind}$, and $v_{\rm inj}$, which unfortunately, are poorly constrained observationally due to degeneracies in these parameters for a given asymmetric eclipse.}\explain{Moved to \S B1.}

The computation of eclipse fraction is straightforward with the simple replacement $\Omega_{\rm b} t \rightarrow \Omega_{\rm b} (t^\prime - \delta t)$ in Eq.~(\ref{heaviside_gamma}) at each $\theta > \pi/2$ up to a maximum $L_z$ or $\theta_{\rm max,R}$ for all variables associated with the shock surface. The ingress and egress of eclipses are defined to occur for phases $\Omega_{\rm b} t^\prime \in [-\pi/2,0) $ and $\Omega_{\rm b} t^\prime \in (0,\pi/2] $ respectively. The total eclipse fraction is then $f_E = f_E ({ \rm Ingress}) + f_E ({ \rm Egress})$, and the asymmetry can be characterized by the difference $\Delta f_E \equiv f_E ({ \rm Egress})  -f_E ({ \rm Ingress})$ and the ratio of egress to ingress eclipse durations.